\documentclass[preprint]{imsart}

\RequirePackage[numbers]{natbib}
\RequirePackage[colorlinks,citecolor=blue,urlcolor=blue]{hyperref}

\usepackage{amsmath, amsthm, amsfonts, bm, multirow}
\usepackage{graphicx, caption, subcaption, comment, xcolor}
\usepackage{enumerate}

\arxiv{2505.04884}
\startlocaldefs
\theoremstyle{plain}

\newtheorem{theorem}{Theorem}[section]
\newtheorem{lemma}[theorem]{Lemma}
\newtheorem{example}{Example}[section]
\newtheorem{remark}{Remark}[section]
\theoremstyle{definition}

\theoremstyle{remark}

\DeclareMathOperator{\argmin}{arg\,min}
\DeclareMathOperator{\argmax}{arg\,max}
\DeclareMathOperator{\I}{\mathbf{I}}
\DeclareMathOperator{\OPM}{\mathbf{H}}
\DeclareMathOperator{\E}{\mathbb{E}}

\DeclareMathOperator{\HDIC}{\mathrm{HDIC}}
\endlocaldefs

\begin{document}
\begin{frontmatter}
\title{Model selection for unit-root time series with many predictors}
\runtitle{Model selection for unit-root time series}

\begin{aug}
\author[A]{\fnms{Shuo-Chieh}~\snm{Huang}\ead[label=e1]{sh1976@stat.rutgers.edu}},
\author[B]{\fnms{Ching-Kang}~\snm{Ing}\ead[label=e2]{cking@stat.nthu.edu.tw}}
\and
\author[C]{\fnms{Ruey S.}~\snm{Tsay}\ead[label=e3]{ruey.tsay@chicagobooth.edu}}
\address[A]{Department of Statistics,
Rutgers University\printead[presep={,\ }]{e1}}

\address[B]{Institute of Statistics and Data Science, National Tsing Hua University\printead[presep={,\ }]{e2}}

\address[C]{Booth School of Business, University of Chicago and Institute of Statistics and Data Science, National Tsing Hua University\printead[presep={,\ }]{e3}}

\runauthor{Huang, Ing, and Tsay}
\end{aug}

\begin{abstract}
This paper studies model selection for general unit-root time series, including the case with many exogenous predictors. We propose a new model selection algorithm, FHTD, that leverages forward stepwise regression (FSR), a high-dimensional information criterion (HDIC), a backward elimination method based on HDIC, and a data-driven thresholding (DDT) approach. Under some mild assumptions that allow for unknown locations and multiplicities of the characteristic roots on the unit circle of the time series and conditional heteroscedasticity in the predictors and errors, we establish the sure screening property of FSR and the selection consistency of FHTD. Our theoretical analysis relies on two novel technical contributions, namely a functional central limit theorem for multivariate linear processes and a uniform lower bound for the minimum eigenvalue of the sample covariance matrices, both of which are of independent interest. Simulation results corroborate the theoretical properties and show the superior performance of FHTD in model selection. We apply the proposed FHTD to model U.S. monthly housing starts and unemployment data, showcasing its practical utility.
\end{abstract}



\end{frontmatter}

\section{Introduction} \label{sec:intro}

With the widespread availability of large-scale fine-grained datasets, researchers analyzing time series data now have a plethora of predictors available for constructing informative and interpretable models. 
Regularization techniques \cite{tib1996, zou2006, candes2007, Zhang2010, Zheng2014}, which select a few relevant features in a sparse model for prediction, have thus been adapted from the independent framework to time series data \citep{medeiros2016, Han2020}. In addition, greedy forward selection algorithms
 \citep{Buhlmann2016, wang2009, Fan2008, ing2011} have also proven useful for a similar task involving dependent data \citep{ing2019p}. 

However, the aforementioned methods are generally not applicable to 
unit-root nonstationary time series, prevalent in economics, finance, and environmental sciences. To apply these methods to unit-root time series, one must carefully transform the series under study into stationary ones. 
This step often involves multiple intricate unit-root tests since the underlying unit-root structure is typically unknown. 
In addition, it becomes even more challenging to take the correct difference transforms when the data are driven by complex unit roots, such as those exhibiting persistent cyclic behavior. 
Determining the order of integration and the frequency at which the series is integrated are far from straightforward and are sometimes sensitive to model specifications.
Yet, persistent cyclic (or seasonal) time series are widely encountered in applications, such as the unemployment rate \citep{bierens2001}, spot exchange rates \citep{AlZoubi2008}, entrepreneurship series \citep{Faria2009}, firms’ capital structure \citep{AlZoubi18}, sunspot numbers \citep{Gilalana09, Maddanu22}, oil prices \citep{Gilalana14}, tourist arrivals \citep{Castro22}, and CO$_2$ concentrations \citep{Proietti22}, to name just a few examples.

In this paper, 
we study model selection for an autoregressive model with exogenous variables, known as the ARX model, 
when the dependent variable contains general unit roots and the number of exogenous variables is large. Specifically, the model employed is 
\begin{align} \label{Model1}
    (1-B)^{a}(1+B)^{b}\prod_{k=1}^{l}(1-2\cos \vartheta_{k} B +B^{2})^{d_{k}} \psi_{n}(B) y_{t,n}
   = \sum_{j=1}^{p_{n}}\sum_{l=1}^{r_{j}^{(n)}}\beta_{l,n}^{(j)}x_{t-l,j}^{(n)} + \epsilon_{t,n}, 
\end{align}
$t=1, \ldots, n$, where $n$ is the sample size, $B$ denotes the back-shift operator, $a$, $b$, and $d_{k}$ are nonnegative integers indicating the order of integration with respect to different (conjugate) unit roots, $l$ is the number of complex conjugate unit-root pairs, $\vartheta_{k} \in (0, \pi)$ denotes the location of a complex unit root, and $\psi_{n}(z) =1+ \sum_{s=1}^{\iota_{n}} a_{s,n}z^{s} \neq 0$ for all $|z|\leq 1$, with both $a_{s, n} \in \mathbb{R}$ and $\iota_{n} \in \mathbb{N} \cup \{0\}$ unknown.
In particular, we assume no prior knowledge about the unit roots in the model, so $a$, $b$, $l$, $d_{k}$ and $\vartheta_{k}$ are all unknown.
In model \eqref{Model1}, $\{\epsilon_{t,n}\}$ denotes a sequence of mean-zero random errors, $\{x_{t-l,j}^{(n)}\}$ and $\{\beta_{l,n}^{(j)}\}$, for $1\leq l\leq r_{j}^{(n)}, 1\leq j \leq p_n$, are observable exogenous variables and their respective unknown coefficients. 
Let $d=a+b+2\sum_{k=1}^{l}d_{k}$. The number of AR lags in \eqref{Model1}, $m_n=\iota_{n}+d$, is assumed to be smaller than $n$, whereas the number of exogenous predictors, $p_n^{*}=\sum_{j=1}^{p_n}r_{j}^{(n)},$ can be much greater than $n$. 
We adopt $y_{t, n}=0$ for $t \leq 0$ as the initial conditions, which are widely used in the literature for unit-root series \citep[e.g.][]{chan1988}.
Last but not least, we allow $\{x_{t,j}^{(n)}, 1\leq t \leq n\}$, for $1\leq j \leq p_n$, and $\{\epsilon_{t,n}\}$ to be conditionally heteroscedastic.

Numerous authors have investigated the model selection of model \eqref{Model1} for the special cases when $d=0$ or $p_{n}^{*} = 0$.
When $d = 0$, $y_{t,n}$ is stationary. In this case, under some strong sparsity conditions, the LASSO \citep{tib1996} and the adaptive LASSO \citep{zou2006} have been shown to achieve model selection consistency \citep{Han2020, medeiros2016}. 
In addition, \cite{ing2019p} proved that the orthogonal greedy algorithm (OGA), used in conjunction with a high-dimensional information criterion (HDIC), is rate-optimal adaptive to unknown sparsity patterns.
When $p_n^{\ast}=0$, model \eqref{Model1} reduces to a non-stationary AR model with unit roots. In this case, traditional information 
criteria such as AIC, BIC, and Fisher information criterion (FIC) can be employed to perform  model selection \citep{Ing2012, Tsay1984, Wei1992}. 
More recently, \cite{kock2016} applied the adaptive LASSO to the Dickey-Fuller regression of fixed AR order under the special case of a single unit root (i.e., $a=1$, $b=d_{1}=\ldots=d_{l} = 0$, and $\iota_n$ is a fixed positive integer). 

Nevertheless, applying those techniques to model \eqref{Model1} under full generality still remains challenging.
As pointed out earlier, the existence of unknown $\vartheta_{k}$ makes it difficult to transform $\{y_{t,n}\}$ into an asymptotically stationary time series.
Even worse, when applied to nonstationary time series, LASSO is known to perform poorly due to its internal standardization of unit-root variables 
\cite{Han2020, Lee2022}.
In fact, due to the near-perfect correlation of some (or all) lagged variables in model \eqref{Model1} when $d > 0$, the strong irrepresentable condition, which is almost necessary and sufficient for LASSO to achieve selection consistency in high-dimensional regression models \citep{Zhao2006}, is no longer valid.
This issue also undermines the effectiveness of other correlation-based feature selection methods, such as $L_2$-Boosting and OGA. 
Indeed, in Section \ref{sec::algorithm}, we prove that both LASSO and OGA can fail to achieve variable selection consistency in the presence of unit roots.
While AIC, BIC, and FIC are viable methods for selecting the AR order when $d > 0$ and $p_n^{*} = 0$, applying them when $p_n^{*}$ is large involves subset selection. Therefore they are not suitable for selecting exogenous variables, especially when $p_n^{*} \gg n$.

We address these difficulties by combining the strengths of the {\it least squares} method in unit-root AR models with {\it forward stepwise regression} (FSR, defined in Section \ref{sec::algorithm}) in high-dimensional regression models, and work directly with the observed nonstationary series.
Our procedure starts by rewriting \eqref{Model1} as
\begin{align} \label{Model2}
    y_{t} = \sum_{i=1}^{q_{n}}\alpha_{i} y_{t-i} + \sum_{j=1}^{p_{n}}\sum_{l=1}^{r_{j}^{(n)}}\beta_{l}^{(j)}x_{t-l,j} + \epsilon_{t},
\end{align}
where $q_n < n$ is a prescribed upper bound for $m_n$, $1-\sum_{i=1}^{q_n}\alpha_{i}z^{i}=(1-z)^{a}(1+z)^{b}\prod_{k=1}^{l}(1-2\cos \vartheta_{k} z +z^{2})^{d_{k}} \psi_{n}(z)$, and the dependence of $y_t$, $\alpha_i$, $\beta_l^{(j)}$, $x_{t-l,j}$, and $\epsilon_{t}$ on $n$ is suppressed for simplicity in notation.
Then, FSR is used to sequentially choose the exogenous predictors after $y_{t-1},\ldots, y_{t-q_n}$ are coerced into the model. By fitting an AR($q_n$) model by least squares in advance, this approach handles the nonstationarity of $\{y_t\}$ without recourse to any tests for (complex) unit roots, thereby facilitating the implementation of FSR without being encumbered by the highly correlated lagged dependent variables. 
Next, we use HDIC to guide the stopping rule of FSR, and use a backward elimination method also based on HDIC, which we call Trim, to remove redundant exogenous predictors that have been previously included by FSR. Finally, we introduce a data-driven thresholding (DDT) method to weed out irrelevant lagged dependent variables. Throughout the paper, the combined model selection procedure is called the FHTD algorithm.
Under a strong sparsity condition, which assumes that the number of relevant predictors in model \eqref{Model2} is smaller than $n$, we establish the sure screening property of FSR and the selection consistency of FHTD. Since complex unit roots, conditional heteroscedasticity, and high dimensionality are allowed simultaneously, this is one of the most comprehensive results to date on model selection consistency established for the ARX model.

The rest of the paper is organized as follows. We detail the FSR and FHTD algorithms in Section \ref{sec::algorithm}. 
Relevant theoretical properties of these methods are given in Section \ref{Sec::Theory}; see Theorems \ref{Theorem_3.2}--\ref{corr}.
The finite-sample performance of the proposed methods is illustrated using simulations and two U.S. monthly macroeconomic datasets in Sections \ref{Sec:simulation} and \ref{Sec::application}, respectively. Section \ref{Sec::conclusion} concludes.
We have moved the proofs and auxiliary results to the supplementary material. 
Nevertheless, it is noteworthy that, to tackle the nonstationary series, we derived a novel functional central limit theorem (FCLT) for linear processes driven by $\{\sum_{j=1}^{p_{n}}\sum_{l=1}^{r_{j}^{(n)}}\beta_{l}^{(j)}x_{t-l,j} + \epsilon_{t}\}$ and a uniform lower bound for the minimum eigenvalue of the sample covariance matrices associated with model \eqref{Model2}. 
These theoretical foundations, crucial for Theorems \ref{Theorem_3.2}--\ref{corr}, can be found in \hyperref[Sec::Appendix]{Appendix}. 

The following notation is used throughout the paper. For a matrix $\mathbf{A}$, $\lambda_{\min}(\mathbf{A})$, $\lambda_{\max}(\mathbf{A})$, $\Vert \mathbf{A} \Vert$, and $\mathbf{A}^{\top}$ denote its minimum eigenvalue, maximum eigenvalue, operator norm, and transpose, respectively. For a set $J$, $\sharp(J)$ is its cardinality. For two sequences of positive numbers, $\{a_n\}$ and $\{b_n\}$, $a_n \asymp b_n$ means $L<a_n/b_n<U$ for some $0<L\leq U<\infty$. For an event $E$, its complement and indicator function are denoted by $E^{c}$  and $\mathbb{I}_{E}$, respectively. For $k \in \{1, 2, \ldots\}$, $[k]=\{1,2,\ldots, k\}$. For $r \in \mathbb{R}$, $\lfloor r \rfloor$ is the largest integer $\leq r$. For two real numbers $a$ and $b$, $a \vee b = \max\{a,b\}$ and $a \wedge b = \min\{a,b\}$. For a vector $\mathbf{v}$, 
$\Vert \mathbf{v} \Vert$ denotes its Euclidean norm. For a random variable $X$,  $\Vert X \Vert_{q} = (\E|X|^{q})^{1/q}$. Generic absolute constants are denoted by $C$ whose value may vary at different places.
In what follows, we use Sx to refer to Sections in the supplementary material. 

\section{The FHTD algorithm} \label{sec::algorithm}
Let $\mathbf{y}_{n} = (y_{n},y_{n-1},\ldots,y_{\bar{r}_{n}+1})^{\top}$
and
$\bm{o}_{i} = (y_{n-i}, y_{n-i-1}, \ldots, y_{\Bar{r}_{n}-i+1})^{\top}$, where $i = 1, 2, \ldots, q_{n}$
and $\bar{r}_{n}=\{\max_{1\leq j\leq p}r_{j}^{(n)}\} \vee q_{n}$.
Define  
$\bm{x}_{l}^{(j)} = (x_{n-l,j}, x_{n-l-1,j}, \ldots, x_{\Bar{r}_{n}-l+1,j})^{\top}$, where $l=1,2,\ldots,r_{j}^{(n)}$
and 
$j = 1,2,\ldots, p_{n}$.
Then, it follows from \eqref{Model2} that
\begin{align*}
    \mathbf{y}_{n} = \mathbf{O}_{n}\bm{\alpha} + \mathbf{X}_{n}\bm{\beta} + \bm{\varepsilon}_{n} := \bm{\mu}_{n} + \bm{\varepsilon}_{n},
\end{align*}
where
$\mathbf{O}_{n} = (\bm{o}_{1}, \ldots, \bm{o}_{q_{n}})$,
$\mathbf{X}_{n} = (\bm{x}_{1}^{(1)}, \ldots, \bm{x}_{r_{1}^{(n)}}^{(1)}, \ldots, \bm{x}_{1}^{(p_{n})}, \ldots, \bm{x}_{r_{p_{n}}^{(n)}}^{(p_{n})})$, 
$\bm{\alpha} = (\alpha_{1},\ldots,\alpha_{q_{n}})^{\top}$, $\bm{\varepsilon}_{n} = (\epsilon_{n},\ldots,\epsilon_{\bar{r}_{n}+1})^{\top}$, and 
$\bm{\beta}=(\beta_{1}^{(1)},\ldots,\beta_{r_{1}^{(n)}}^{(1)},\ldots,\beta_{1}^{(p_n)},\ldots,
\beta_{r_{p_n}^{(n)}}^{(p_n)})^{\top}$. 
Note that $\bm{\mu}_{n}$ can be expressed as $(\mu_{n}, \ldots, \mu_{\bar{r}_{n}+1})^{\top}$, with
$\mu_{t} = \sum_{i=1}^{q_{n}} \alpha_{i }y_{t-i} +v_{t,n}$ and
\begin{eqnarray} \label{ingfeb31}
    v_{t,n} = \sum_{j=1}^{p_{n}}\sum_{l=1}^{r_{j}^{(n)}} \beta_{l}^{(j)}x_{t-l,j}.
\end{eqnarray}

We index the candidate variable $x_{t-l, j}$ by the tuple $(j, l)$. FSR is an iterative algorithm that greedily chooses variables from $\bar{J} :=\{ (j,l): j \in [p_n], l \in [r_{j}^{(n)}]\}$ after $y_{t-1}, \ldots, y_{t-q_n}$ are included in the regression model. Specifically, the  algorithm begins with  $\hat{J}_{0} = \emptyset$ and generates $\hat{J}_{m} \subset \bar{J} $ via $\hat{J}_{m} =\hat{J}_{m-1} \cup\{(\hat{j}_{m},\hat{l}_{m})\}$, where $m \geq 1$ and
\begin{align} \label{FSRselectrule}
    (\hat{j}_{m},\hat{l}_{m}) = \argmax_{(j,l) \in \bar{J} \setminus \hat{J}_{m-1}} \frac{n^{-1}|\mathbf{y}_{n}^{\top}(\mathbf{I}-\OPM_{[q_{n}]
    \oplus \hat{J}_{m-1}})\bm{x}_{l}^{(j)}|}{(n^{-1}\bm{x}_{l}^{(j) \top}(\I - \OPM_{[q_{n}] \oplus\hat{J}_{m-1}})\bm{x}_{l}^{(j)})^{1/2}},
\end{align}
where $\OPM_{Q \oplus J}$ is the orthogonal projection matrix associated with the linear space spanned by $\{\bm{o}_{l}: l \in Q \subseteq [q_{n}]\}\cup\{\bm{x}_{l}^{(j)}:(j,l) \in J \subseteq \bar{J} \}$. In the sequel, we also use $Q \oplus J$ to denote a candidate model consisting of predictor variables $\{y_{t-i}, i \in Q\}$ and $\{x_{t-l, j }, (j, l) \in J\}$.
When $m$ reaches a prescribed upper bound  $K_n \leq p_n^{*}$, the algorithm stops and outputs the index set $\hat{J}_{K_n}$.
By employing the lagged dependent variables $y_{t-1}, \ldots, y_{t-q_{n}}$ in the model, the effects of the unit roots are neutralized, and the algorithm is expected to include the set of relevant exogenous variables
\begin{eqnarray*}
    \mathcal{J}_{n} = \{ (j, l): \beta_{l}^{(j)} \neq 0,  l \in [r_{j}^{(n)}], j \in [p_{n}] \}.
\end{eqnarray*}

However, $[q_n]\oplus \hat{J}_{K_n}$ may contain some irrelevant variables, especially when $q_n$ or $K_n$ is large compared to $\sharp(\mathcal{Q}_{n})$ or $\sharp(\mathcal{J}_{n})$, where
    $\mathcal{Q}_{n} = \{q: \alpha_{q} \neq 0, q \in[q_{n}]\}$
is the set of relevant lagged dependent variables.
To alleviate this overfitting problem with FSR, we propose eliminating the irrelevant exogenous variables in $\hat{J}_{K_n}$ using the HDIC. 
Given a candidate model $Q \oplus J$, its HDIC value is given by
\begin{align} \label{HDICdef}
    \mbox{HDIC}(Q \oplus J) = n\log \hat{\sigma}^{2}_{Q \oplus J} + [\sharp(J)+\sharp(Q)]w_{n, p_n},
\end{align}
where $\hat{\sigma}_{Q\oplus J}^{2} = n^{-1}\mathbf{y}_{n}^{\top}(\I-\OPM_{Q\oplus J})\mathbf{y}_{n}$ and $w_{n, p_n}$, penalty for the model complexity $\sharp(J)+\sharp(Q)$, depends on the sample size $n$ as well as the number of candidate exogenous variables $p_{n}^{*}$. 

Our approach is to first find a ``promising'' subset $\hat{J}_{\hat{k}_n}$ of $\hat{J}_{K_n}$ that minimizes the HDIC values along the FSR path $\{\hat{J}_1,\ldots, \hat{J}_{K_n}\}$, where
\begin{align} \label{HDICstop}
    \hat{k}_{n} = \argmin_{1 \leq m \leq K_{n}} \mbox{HDIC}([q_{n}]\oplus\hat{J}_{m}).
\end{align}
We then refine $\hat{J}_{\hat{k}_n}$ by comparing the HDIC values of $[q_{n}]\oplus \hat{J}_{\hat{k}_n}$ and the leave-one-variable-out model $[q_{n}]\oplus(\hat{J}_{\hat{k}_n} \setminus \{(\hat{j}_{i},\hat{l}_{i})\}), 1 \leq i \leq \hat{k}_n$,
to judge whether the marginal contribution of $(\hat{j}_{i},\hat{l}_{i})$ is sufficient to warrant its inclusion in the final model. Then the refined set of variables is denoted by
\begin{align} \label{ing30}
    \hat{\mathcal{J}}_{n} = \{ (\hat{j}_{i}, \hat{l}_{i}): 1 \leq i \leq \hat{k}_n,
    \HDIC([q_{n}]\oplus(\hat{J}_{\hat{k}_n} \setminus \{(\hat{j}_{i},\hat{l}_{i})\})) >
    \HDIC([q_{n}]\oplus\hat{J}_{\hat{k}_n}) \},
\end{align}
and the method is called ``Trim.''

To weed out the potentially redundant AR variables, we employ a thresholding method
\begin{align} \label{ing31}
    \hat{\mathcal{Q}}_{n} = \{ 1 \leq q \leq q_{n}: |\hat{\alpha}_{q}(\hat{\mathcal{J}}_{n})| \geq \hat{H}_n \},
\end{align}
where $\hat{\alpha}_{q}(\hat{\mathcal{J}}_{n})$ are the least squares estimates defined as 
\begin{eqnarray*}
    (\hat{\alpha}_{1}(\hat{\mathcal{J}}_{n}), \ldots, \hat{\alpha}_{q_n}(\hat{\mathcal{J}}_{n}), \hat{\bm{\beta}}^{\top}(\hat{\mathcal{J}}_{n}))^{\top}
    =\left(\sum_{t=\bar{r}_n+1}^{n}\mathbf{w}_{t}(\hat{\mathcal{J}}_{n})\mathbf{w}^{\top}_{t}(\hat{\mathcal{J}}_{n})\right)^{-1}
    \sum_{t=\bar{r}_n+1}^{n}\mathbf{w}_{t}(\hat{\mathcal{J}}_{n})y_{t},
\end{eqnarray*}
in which, for $J \in \bar{J}$, $\mathbf{x}_{t}(J) = (x_{t-l,j}:(j,l)\in J)^{\top}$ and $\mathbf{w}_{t}(J) = (y_{t-1}, \ldots, y_{t-q_{n}},\mathbf{x}_{t}^{\top}(J))^{\top}$.
In \eqref{ing31}, $\hat{H}_n$ is a data-driven thresholding (DDT) value depending on $\hat{\mathcal{J}}_{n}$ and $q_n$; see Section \ref{consistency}.
Identifying $\hat{\mathcal{Q}}_n$ is crucial for accurate prediction because an overfitted model tends to cause a larger mean squared prediction error, especially in tackling nonstationary time series where the cost of overfitting is higher (see Example \ref{egPredictionError}). 
The final estimated model is $\hat{N}_{n} = \hat{\mathcal{Q}}_{n}\oplus\hat{\mathcal{J}}_{n}$. The above procedure, which combines FSR, HDIC, Trim, and DDT, is referred to as FHTD.

A number of methods in the literature are closely related to the FHTD.
The One Covariate at a Time Multiple Testing (OCMT) \cite{Chudik2018} employed a forward selection method similar to \eqref{FSRselectrule}, which controls the false positive rate and the false discovery rate in high-dimensional linear regression.
The selection rule \eqref{FSRselectrule} simplifies to the forward regression algorithm in \cite{wang2009} when $q_n = 0$, and it further reduces to the OGA \cite{ing2011} if the orthogonal projection matrix in the denominator is removed. 
However, the existing analyses for the greedy methods have all considered the stationary case and excluded the highly correlated unit-root covariates. 
In the context of independent observations, \eqref{HDICdef} and \eqref{ing30} have been employed in the OGA+HDIC+Trim method \cite{ing2011, ing2019p} to eliminate the redundant variables introduced by OGA; yet, the effectiveness of OGA+HDIC+Trim in identifying $\mathcal{J}_n$ and $\mathcal{Q}_n$ can be significantly compromised under nonstationarity; see Examples \ref{egSim2} and \ref{egSim3} of Section \ref{Sec:simulation}. 

The novelty of FHTD lies in its strategy of pre-including lagged dependent variables to neutralize the adverse effects of unit roots, followed by a carefully designed removal of redundant variables, which together yield provable model selection consistency. To clarify the rationale behind this design, we present two illustrative examples.
In the first example, we show that OGA, when applied directly to a mixed set of unit-root and stationary predictors, may fail to screen all relevant variables. In particular, once a lagged dependent variable is selected, the resulting residuals become stationary, causing subsequent iterations to overlook additional relevant lags. This phenomenon motivates the pre-inclusion of lagged dependent variables in FHTD.

\begin{example} \label{egOGA} 
Consider a special case of model \eqref{Model2},
\begin{align} \label{ingapr231}
    y_{t} = \alpha_{1} y_{t-1} + \alpha_{2} y_{t-2} + \sum_{j=1}^{p_n} \beta_{j} x_{t-1, j} + \epsilon_{t},
\end{align}
where $\alpha_{1} = 1 + a$, $\alpha_{2} = -a$, $|a|<1$, $\beta_{j}=0$, $1 \leq j \leq p_{n}$, and $\{(x_{t, 1}, \ldots, x_{t, p_n}, \epsilon_t)^{\top}\}$ is a sequence of independent normal random vectors with mean zero and identity covariance matrix.
It is easy to see that \eqref{ingapr231} is an AR(2) model whose characteristic polynomial $(1-az)(1-z)$ has a unit root of 1. 
In addition, all $x_{t-1, j}, 1\leq j \leq p_n$, are redundant.
Under this model, when OGA is directly applied to the set of candidate variables $\{y_{t-1}, y_{t-2}, x_{t-1,j}: 1 \leq j \leq p_{n}\}$, one of the relevant variables $\{y_{t-1}, y_{t-2}\}$ will \textit{not} be included in the OGA path.
 
To see this, let $F_{1,n} = (\mathbf{y}_{n}^{\top}\bm{o}_{1})^{2} / \Vert \bm{o}_{1} \Vert^{2}$, $F_{2,n} = (\mathbf{y}_{n}^{\top}\bm{o}_{2})^{2} / \Vert \bm{o}_{2} \Vert^{2}$, and $Q_{j,n} = (\mathbf{y}_{n}^{\top}\bm{x}^{(j)})^{2}$ $/ \Vert\bm{x}^{(j)}\Vert^{2}$ for $1 \leq j \leq p_n$,
where 
we define
$\mathbf{y}_{n} = (y_{n}, \ldots, y_{3})^{\top}$, $\bm{o}_{1} = (y_{n-1},\ldots, y_{2})^{\top}$, $\bm{o}_{2}=(y_{n-2},\ldots, y_{1})^{\top}$,
$\bm{x}^{(j)}=(x_{n-1, j}, \ldots, x_{2, j})^{\top}$, $1 \leq j \leq p_{n}$. 
Then
\begin{align} \label{ingapr232}
    \frac{F_{1,n}}{n^2} \Rightarrow (1 -a)^{-2} \int_{0}^{1}w^{2}(t)dt, \quad \frac{F_{2,n}}{n^2} \Rightarrow (1 -a)^{-2}\int_{0}^{1}w^{2}(t)dt,
\end{align}
where $w(t)$ is the standard Brownian motion and $\Rightarrow$ denotes convergence in law. 
Moreover, it is shown in Section \ref{sec06} of the supplement that
\begin{align} \label{ingapr235}
    \frac{1}{n}(F_{1,n}-F_{2,n}) \to \frac{1+2a}{1-a^2} \quad \mbox{in probability}.
\end{align}
By Bernstein's inequality, $\max_{1\leq j \leq p_n} Q_{j,n} = O_{p}(n \log p_n)$.
Hence, if $a > -0.5$, then \eqref{ingapr232}--\eqref{ingapr235} imply that with probability tending to 1, $y_{t-1}$ will be selected in the initial iteration of OGA, provided 
that $\log p_{n} = o(n)$. 

Assuming that $y_{t-1}$ is already included by OGA, define $\bm{\epsilon}=(\mathbf{I}-\bm{o}_{1}\bm{o}_{1}^{\top}/\|\bm{o}_{1}\|^{2})\mathbf{y}_{n}$, which is the residual vector obtained by regressing $y_t$ on $y_{t-1}$. It can be shown that $(\bm{o}_{2}^{\top}\bm{\epsilon})^2 / \Vert\bm{o}_{2}\Vert^{2} = O_{p}(1)$ and for some small $\underline{c} > 0$, $P(\max_{1\leq j \leq p_n}[(\bm{x}^{(j)})^{\top}\bm{\epsilon}]^2 / \Vert\bm{x}^{(j)}\Vert^2 >\underline{c} \log p_{n}) \rightarrow 1$. Hence, 
the probability of choosing $y_{t-2}$ in the second OGA iteration approaches 0 provided $\log p_n \to \infty$.
By a similar argument, $y_{t-2}$ will not be selected by OGA in the first $K_n$ iterations when $p_n \gg n \gg K_n$ with probability approaching 1.
Thus, while $y_{t-1}$ will likely be selected by OGA, it is very difficult for OGA to choose the other relevant lagged dependent variable in the presence of unit roots. 
If $a < -0.5$, $y_{t-2}$ will enter the model in the first iteration and $y_{t-1}$ will then be neglected by OGA due to the same argument.
\end{example}

The second example demonstrates that LASSO, a widely used method for high-dimensional linear regression, also fails to achieve model selection consistency in the presence of unit roots, regardless of the choice of penalty parameter. Taken together, these examples highlight the intrinsic difficulty of model selection in unit-root ARX models and motivate treating unit-root and stationary predictors differently in the selection procedure.

\begin{example} \label{egLASSO}
Consider the model $y_{t}= \beta_{1}^{*}y_{t-1} + \beta_{2}^{*}y_{t-2} + \beta_{3}^{*}x_{t-1} + \epsilon_{t}$, $t=1,2,\ldots,n$,
where $(\epsilon_{t}, x_{t})^{\top}$ are i.i.d. Gaussian with zero mean and an identity covariance matrix, and $\beta_{1}^{*} = \beta_{3}^{*} = 1$ and $\beta_{2}^{*}=0$. 
If we apply LASSO to estimate $(\beta_{1}^{*}, \beta_{2}^{*}, \beta_{3}^{*})$ with 
\begin{align*}
    \hat{\bm{\beta}}^{(\lambda_{n})} = (\hat{\beta}_{1}^{(\lambda_{n})}, \hat{\beta}_{2}^{(\lambda_{n})}, \hat{\beta}_{3}^{(\lambda_{n})})^{\top} \in \arg\min_{\{\beta_{j}\}_{j=1}^{3}}& \sum_{t=3}^{n}(y_{t}-\beta_{1}y_{t-1}-\beta_{2}y_{t-2}-\beta_{3}x_{t-1})^{2} \\
    & + \lambda_{n}\sum_{j=1}^{3}|\beta_{j}|,
\end{align*}
then LASSO will \textit{not} exhibit model selection consistency as in equations \eqref{ing20} and \eqref{ing21}.
This lack of consistency holds true whether the sequence $\{\lambda_{n}\}$ is chosen
such that (a) $\limsup_{n \to \infty}\lambda_{n}/n = \infty$, (b) $\liminf_{n \to \infty} \lambda_{n}/n = 0$, or (c) $\lambda_{n}\asymp n$.
In fact, one can show that for any sequence $\{\lambda_{n}\}$ satisfying (a), (b), or (c), we always have 
\begin{align} \label{sch731}
    \liminf_{n \rightarrow \infty} P(\hat{\beta}^{(\lambda_{n})}_{1} \neq 0, \hat{\beta}^{(\lambda_{n})}_{2}=0, \hat{\beta}^{(\lambda_{n})}_{3} \neq 0) \leq \frac{1}{2}.
\end{align}
The proof of \eqref{sch731} can be found in Section \ref{sec06} of the supplement. 
\end{example}

Finally, although it may appear feasible to establish model selection consistency for LASSO or OGA using residuals from a long autoregression, or by directly applying LASSO to jointly select lagged values of $y_t$ and stationary exogenous regressors $x_t$, followed by thresholding, our simulation results do not indicate satisfactory finite-sample performance for these approaches; see Section~\ref{Sec:simulation} for details.
In contrast, FHTD demonstrates more reliable finite-sample performance under unit-root dynamics, while retaining model selection consistency.

\section{Screening and selection consistency} \label{Sec::Theory}
In this section, we present the main results---sure-screening property of FSR and the 
model selection consistency of FHTD---in Sections \ref{sec3.3} and \ref{consistency}, respectively.
To this end, we introduce Assumptions (A1)--(A6) in Section \ref{Sec::modelassumption}.
Since Section \ref{Sec::modelassumption} introduces many technical notations, readers may choose to proceed directly to the main results and consult Section \ref{Sec::modelassumption} as needed.

\subsection{Model assumptions} \label{Sec::modelassumption}

Consider model \eqref{Model2}. Let $x_{t,j}$, $1\leq j \leq p_n$, and $\epsilon_t$ be $\mathcal{F}_{t}$-measurable random variables, where $\{\mathcal{F}_{t}\}$ is an increasing sequence of $\sigma$-fields representing available information up to time $t$. We impose the following assumptions.

\begin{description}
\item[\textbf{(A1)}]
$\{\epsilon_{t}, \mathcal{F}_{t}\}$ is a martingale difference sequence (m.d.s.) with $\E\epsilon_{t}^2 = \sigma^{2}$ and
\begin{align} \label{ing1}
    \epsilon_{t}^{2} - \sigma^{2} = \sum_{j=0}^{\infty} \theta_{j}^{\top}e_{t-j},
\end{align}
where $\theta_j$ are $l_0$-dimensional real vectors 
such that
\begin{eqnarray} \label{ingsep2}
    \sum_{j=0}^{\infty} \|\theta_{j}\| \leq C,
\end{eqnarray}
with $l_0$ being a fixed positive integer, and $\{e_{t}, \mathcal{F}_{t}\}$ is an $l_0$-dimensional m.d.s. with
\begin{eqnarray} \label{ingaug1}
    \sup_{t}\E \|e_{t}\|^{\eta} \leq C, \,\,\mbox{for some}\,\,\eta \geq 2.
    \end{eqnarray}
\end{description}

\begin{description}
\item[\textbf{(A2)}] For each $1 \leq s \leq p_{n}$, $\{x_{t,s}\}_{-\infty<t<\infty}$ is a covariance stationary time series with mean zero and admits a one-sided moving average representation,
\begin{align} \label{ing41}
    x_{t,s} = \sum_{k = 0}^{\infty} p_{k,s}\pi_{t-k,s},
\end{align}
where $p_{0,s} = 1$, $\{\pi_{t,s},\mathcal{F}_{t}\}$ is an m.d.s. and
\begin{eqnarray} \label{win1}
    \sum_{k= 0}^{\infty} \max_{1\leq s \leq p_{n}} \sqrt{k}\vert p_{k,s} \vert \leq C
\end{eqnarray}
for all $n$. Moreover, for $0\leq s_1 \leq s_2 \leq p_n$ and $s_1+s_2\geq 1$,
\begin{eqnarray} \label{ingsep1}
    \pi_{t, s_1}\pi_{t, s_2}-\sigma_{s_1, s_2}=\sum_{j=0}^{\infty} \theta_{j, s_1, s_2}^{\top}e_{t-j, s_1, s_2},
\end{eqnarray}
where $\pi_{t,0}=\epsilon_t$, $\sigma_{s_1, s_2}=\E(\pi_{t, s_1}\pi_{t, s_2})$, $\theta_{j, s_1, s_2}$ are $l_{s_1, s_2}$-dimensional real vectors, with $l_{s_1, s_2}$ being a fixed positive integer, such that
\begin{eqnarray} \label{ingsep3}
    \sum_{j=0}^{\infty} \|\theta_{j, s_1, s_2}\| \leq C,
\end{eqnarray}
and $\{e_{t, s_1, s_2}, \mathcal{F}_{t}\}$ is a $l_{s_1, s_2}$-dimensional m.d.s. satisfying for some $q_0 > 2$,
\begin{eqnarray} \label{ingaug2}
    \sup_{t}
    \E \|e_{t, s_1, s_2}\|^{q_0\eta} \leq C, \,\,\mbox{if} \,\,\min\{s_1, s_2\}>0,
\end{eqnarray}
\begin{eqnarray} \label{ingaug222}
    \sup_{t}
    \E \|e_{t, s_1, s_2}\|^{2q_0\eta/(1+q_0)} \leq C, \,\,\mbox{if} \,\,\min\{s_1, s_2\}=0,
\end{eqnarray}
where $\eta$ is defined in \eqref{ingaug1}. Note that $C$ does not depend on $s_{1}$ or $s_{2}$ in the above.
\end{description}

\begin{description}
\item[\textbf{(A3)}] There exists a positive definite sequence, $\{\gamma_{h}\}_{-\infty<h<\infty}$, of real numbers
such that
\begin{align} \label{ing7}
    \lim_{n \rightarrow \infty}\sum_{h=0}^{\infty} |\gamma_{h,n}-\gamma_{h}| = 0,
\end{align}
where $\gamma_{h,n} = \E(\delta_{t}\delta_{t+h})$
and $\delta_{t}=\delta_{t, n} = v_{t,n} + \epsilon_{t}$, with 
$v_{t,n}$ defined in \eqref{ingfeb31}.
\end{description}

\begin{description}
\item[\textbf{(A4)}] $\sum_{j=1}^{p_{n}}\sum_{l=1}^{r_{j}^{(n)}}|\beta_{l}^{(j)}| \leq C$ and $\sum_{j=1}^{\iota_n}j|a_{j}| \leq C $, where $a_{j}= a_{j,n}$ is defined after \eqref{Model1}.
\end{description}

\begin{description}
\item[\textbf{(A5)}]
$\max_{1\leq j\leq p_{n}} r_{j}^{(n)} = o(n^{1/2})$, $p_{n}^{*}\asymp n^{\nu}$, and $q_{n}=o(n^{1/2-\theta_{o}})$,
where 
$\nu \in [1, \eta/2]$ and $\theta_{o}=\nu(1+q_0)/(2\eta q_0)$.
\end{description}

Assumption (A1), implying 
\begin{align}
    \sup_{t}\E|\epsilon_t|^{2\eta} < C, \label{ingapr1} 
\end{align}
is satisfied by many conditionally heteroscedastic processes, including the stationary GJR-GARCH model with a finite $2\eta$-th moment.
Assumption (A2) assumes $\{x_{t, s}\}$ follows an MA($\infty$) process driven by the conditionally heteroscedastic innovations $\{\pi_{t,s}\}$, while also requiring 
\begin{align}
    \max_{1\leq s\leq p_n} \sup_{t}\E|x_{t,s}|^{2\eta q_0} < C. \label{ingapr2}    
\end{align}
Note that it allows $(\epsilon_t, \pi_{t, 1}\ldots,\pi_{t, p_n})^{\top}$ to be a multivariate GARCH process, with the diagonal VEC model
\cite{Bollerslev1988} being a particular instance.
Assumption (A3) is used to derive the FCLT for the multivariate linear process driven by $\{\delta_t\}$ (Theorem \ref{Theorem_FCLT}), and Assumption (A4), known as the weak sparsity condition, is frequently employed in the high-dimensional statistics literature.
Finally, by Assumption (A5), the covariate dimension, $p^{*}_n$, is allowed to grow at the same order as $n$, and can be much faster than $n$ if $\eta>2$.
It also permits that $q_{n}$, the prescribed upper bound of the number of AR variables, diverges at a rate slower than $n^{1/2}$. 
For a more detailed and comprehensive exploration of (A1)--(A5), readers are referred to Section \ref{s01} of the supplementary material.

Apart from (A1)--(A5), we also require Assumption (A6), which is stated below, on the covariance structure of $x_{t,j}$ and the stationary component of $y_{t}$, i.e., $z_{t}= [\psi^{-1}(B)\phi(B)]y_{t}$, where 
\begin{align*}
    \phi(z) = (1-z)^{a}(1+z)^{b}\prod_{k=1}^{l}(1-2\cos\vartheta_{k} z + z^{2})^{d_{k}}\psi(z),
\end{align*}
with $\psi(z)=\psi_n(z)$ defined after \eqref{Model1}. 
Since $\psi(z) \neq 0$ for all $|z| \leq 1$, by the second part of (A4) and Theorem 3.8.4 of \cite{brillinger1974},
$z_{t}$ can be expressed as $z_t=\sum_{j=0}^{t-1} b_{j} \delta_{t-j}$, with $b_0 = 1$, $\sum_{j=0}^{\infty} b_j z^{j} \neq 0$ for $|z| \leq 1$, and
$\sum_{j=0}^{\infty}|jb_{j}| \leq C$.
Define ${z}_{t,\infty} = \sum_{j=0}^{\infty}b_{j}\delta_{t-j}$, $\mathbf{z}_{t,\infty}^{\top}(k) = (z_{t-1,\infty},\ldots,z_{t-k,\infty})$, 
\begin{align*}
     \boldsymbol{\Gamma}_{n}(J) = \E \left\{ \begin{pmatrix} \mathbf{z}_{t,\infty}(q_{n}-d) \\ \mathbf{x}_{t}(J) \end{pmatrix}\begin{pmatrix}
    \mathbf{z}_{t,\infty}^{\top}(q_{n}-d), & \mathbf{x}_{t}^{\top}(J) \end{pmatrix} \right\}, \quad J \subseteq \bar{J},
\end{align*}
and $\mathbf{g}^{\top}_{J}(i,l)= (\E (\mathbf{z}^{\top}_{t,\infty}(q_{n}-d) x_{t-l,i}), \E (\mathbf{x}^{\top}_{t}(J)x_{t-l,i}))$ for $(i, l) \notin J$.
Now, (A6) is presented as follows:
\begin{description}
\item[\textbf{(A6)}]
\begin{eqnarray} \label{ing4}
    \max_{\sharp(J)\leq K_{n}} \lambda_{\min}^{-1}(\boldsymbol{\Gamma}_{n}(J)) \leq C,
\end{eqnarray}
and
\begin{align} \label{thm1-asmpt}
    \sum_{s=1}^{q_n-d}\max_{\sharp(J) \leq K_{n}, (i,l) \notin J}|a_{s, J}(i,l)|+ 
    \max_{\sharp(J) \leq K_{n}, (i,l) \notin J} \sum_{(i^*, l^*) \in J}|a_{(i^*, l^*)}(i,l)| \leq C,
\end{align}
where $(a_{1, J}(i, l), \ldots, a_{q_n-d, J}(i, l), (a_{(i^*, l^*)}(i, l):(i^*,l^*)\in J))^{\top} =
    \boldsymbol{\Gamma}_{n}^{-1}(J)\mathbf{g}_{J}(i,l).$
\end{description}
In Section \ref{s01} of the supplement, we provide examples where (A6) holds and highly correlated predictors are allowed in the model. 

\subsection{The sure screening property of FSR} \label{sec3.3}
In addition to (A1)--(A6), we require a \textit{strong sparsity} condition \textbf{(SS$_{{\rm X}}$)} on $\{\beta_{l}^{(j)}\}$ to ensure the sure screening property of FSR. 
\begin{description}
    \item[\textbf{(SS$_{{\rm X}}$)}] 
    \noindent \normalfont $s_0=\sharp(\mathcal{J}_n)$ and $\min_{\substack{(j, l) \in {\cal J}_n}} |{\beta_{l}^{(j)}}|$ obey
    \begin{eqnarray}
        \frac{s_0^{1/2} p_n^{*^{\bar{\theta}}}}{n^{1/2}}=o(\min_{\substack{(j, l) \in {\cal J}_n}} |{\beta_{l}^{(j)}}|), \label{SS_0}
    \end{eqnarray}
where
$\bar{\theta}=\max\{2/(q_0\eta), (q_0+1)/(2\eta q_0)\}$,
and $\eta$ and $q_0$
are defined in Assumptions (A1) and (A2).
\end{description}
A similar condition is used in \cite{medeiros2016} to derive the selection consistency of the adaptive LASSO for stationary time series. 
However, \textbf{(SS$_{\rm X}$)} is less stringent than the strong sparsity condition in \cite{medeiros2016}; see the discussion of Section \ref{s01}.
With this assumption, Theorem \ref{Theorem_3.2} below shows that FSR asymptotically screens all relevant variables. 

\begin{theorem} \label{Theorem_3.2}
Assume that {\rm (A1)--(A6)} and {\rm \textbf{(SS$_{{\rm X}}$)}} hold.
Then, for
\begin{eqnarray} \label{ingsep18}
    K_{n}\asymp (n/{p_{n}^{*}}^{2\bar{\theta}})^{\varsigma},
\end{eqnarray}
where $1/3<\varsigma<1/2$,
\begin{align} \label{ing15}
    \lim_{n \rightarrow \infty} P(\mathcal{J}_{n} \subseteq \hat{J}_{K_{n}}) = 1.
\end{align}
\end{theorem}


Theorem \ref{Theorem_3.2} requires a different set of techniques compared to the existing works on greedy-type methods \citep{Buhlmann2016, ing2011, ing2019p}. 
In particular, standard proofs typically rely on the convergence rates of the ``population'' OGA and its ``semi-population'' version (see Section 6 of \cite{Buhlmann2016}, Sections 2 and 3 of \cite{ing2011}, or Section 2 and Appendix A of \cite{ing2019p}). 
However, the population OGA can hardly be defined for nonstationary time series due to the varying covariances between the input variables and between the input and dependent variables.
To prove Theorem \ref{Theorem_3.2}, the key vehicle is a convergence rate for the mean squared error
\begin{eqnarray} \label{ingjune232}
    \hat{s}_m = n^{-1}\bm{\mu}_{n}^{\top}(\I-\OPM_{[q_{n}] \oplus \hat{J}_{m}})\bm{\mu}_{n}.
\end{eqnarray}
Our strategy is to first derive the convergence rate for 
\begin{eqnarray} \label{ingjune231}
\hat{a}_m=n^{-1}\bm{\mu}_{n}^{\top}(\I - \OPM_{[q_{n}]\oplus J_{m }})\bm{\mu}_{n},
\end{eqnarray}
where $J_m$ is the set of variables selected by the weak ``noiseless'' FSR, defined in Section \ref{theorems} in the supplementary material. 
Together with a probability bound for
\begin{align} \label{ing23june29}
    \max_{\sharp(J) \leq K_{n}} \lambda_{\min}^{-1}\left( n^{-1} \sum_{t=\bar{r}_{n}+1}^{n}
    \mathbf{w}_{t}(J)\mathbf{w}_{t}^{\top}(J) \right), 
\end{align}
developed in Theorem \ref{Theorem_3.1}, it leads to a convergence rate of \eqref{ingjune232}.
It is worth noting that in sharp contrast to conventional models where the sample covariance matrix of the explanatory variables can be accurately approximated by a non-random and positive definite matrix, the sample covariance matrix in \eqref{ing23june29}, $n^{-1} \sum_{t=\bar{r}_{n}+1}^{n}
\mathbf{w}_{t}(J)\mathbf{w}_{t}^{\top}(J)$, lacks a non-random limit due to the presence of highly correlated lagged dependent variables.
Our probability bound for \eqref{ing23june29} requires an intricate analysis based on an FCLT and moment bounds for linear processes driven by $\delta_{t}=\delta_{t, n} = v_{t,n}+ \epsilon_{t}$, where $v_{t,n}$ is defined in \eqref{ingfeb31}. Details are in the \hyperref[Sec::Appendix]{Appendix}.

\subsection{Selection Consistency} \label{consistency}
This section establishes the selection consistency of $\hat{N}_{n} = \hat{\mathcal{Q}}_{n}\oplus\hat{\mathcal{J}}_{n}$. 
To this end, we impose a sparsity condition slightly stronger than \textbf{(SS$_{{\rm X}}$)}.

\begin{description}
\item[\textbf{(SS)}] 
\noindent \normalfont
There exists  $d_n/\log n \to \infty$ such that 
\begin{eqnarray}
    \frac{s_0^{1/2} p_n^{*^{\bar{\theta}}}d_n^{1/2}}{n^{1/2}}=o(\min_{\substack{(j, l) \in {\cal J}_n}} |\beta_{l}^{(j)}|). \label{SS_1}
\end{eqnarray}
\end{description}
Note that the left-hand side in \eqref{SS_1} is larger than that of \eqref{SS_0} by a factor of about $(\log n)^{1/2}$. 
Further discussion of {\rm \textbf{(SS)}} is deferred to Section \ref{s01} of the supplement.
Based on {\rm \textbf{(SS)}}, among other conditions, Theorem \ref{Theorem_3.3} below ensures the consistency of Trim in selecting the exogenous variables.

\begin{theorem} \label{Theorem_3.3}
Assume that the assumptions of Theorem \ref{Theorem_3.2} hold with {\rm \textbf{(SS$_{{\rm X}}$)}} replaced by {\rm \textbf{(SS)}}. Let the $w_{n, p_n}$ in \eqref{HDICdef} satisfy
\begin{eqnarray} \label{penalty1}
    \frac{w_{n, p_n}}{{p_{n}^{*}}^{2\bar{\theta}}} \to \infty \,\,\mbox{ and }\,\,
    \frac{w_{n, p_n}}{{p_{n}^{*}}^{2\bar{\theta}} }=O((d_n/\log n)^{1-\delta}) \,\,\mbox{for any}\,\,0<\delta<1.
\end{eqnarray}
Then, $\hat{k}_n$ and $\hat{\mathcal{J}}_n$ defined in \eqref{HDICstop} and \eqref{ing30} satisfy
\begin{align} \label{ing32}
    \lim_{n \rightarrow \infty} P(\mathcal{J}_{n} \subseteq \hat{J}_{\hat{k}_{n}}) = 1,
\end{align}
\begin{align} \label{ing20}
    \lim_{n \rightarrow \infty} P( \hat{\mathcal{J}}_{n} = \mathcal{J}_{n} ) = 1.
\end{align}
\end{theorem}

\begin{remark}
    As an early stopping rule for FSR, $\hat{J}_{\hat{k}_n}$ not only preserves $\hat{J}_{K_n}$'s sure screening property \eqref{ing32}, but also substantially suppresses the impact of spurious variables greedily chosen by FSR, resulting in reliable performance of Trim. 
    We note that while in theory it is possible to perform Trim without applying the stopping rule \eqref{HDICstop}, the rule often enhances the finite-sample performance. 
\end{remark}

With the help of \eqref{ing20}, one can now establish the consistency of DDT in selecting the AR variables. 
We rely again on the following strong sparsity condition of the AR coefficients.

\begin{description}
\item[\textbf{(SS$_{{\rm A}}$)}]
\noindent \normalfont
\begin{eqnarray}
    \frac{\max\{q_n^{3/2}/\sqrt{n},\,\, [(s_0+q_n)^{1/2}\wedge (\underline{s}_0^{1/2}q_n^{1/\eta})]\}}{n^{1/2} }=o(\min_{q \in \mathcal{Q}_{n}} |\alpha_{q}| ),
    \label{SS_2}
\end{eqnarray}
where $\underline{s}_0=\sharp(\mathcal{D}_0) \,\,\mbox{and} \,\, \mathcal{D}_0=\{j: (j, l)\in \mathcal{J}_n\}$.
\end{description}
Compared with \textbf{(SS)} or \textbf{(SS$_{{\rm X}}$)}, the lower bound for the non-zero coefficients in  \textbf{(SS$_{{\rm A}}$)} is much smaller, enabling detection of weaker signals. Indeed, since the spurious exogenous variables chosen by FSR among the $p_{n}^{*}$ candidates have been (asymptotically) eliminated after the HDIC and Trim steps, $p_{n}^{*}$ is now replaced by the much smaller $q_{n}$ in the lower bound for $\min_{q \in \mathcal{Q}_{n}} |\alpha_{q}|$. 

\begin{theorem} \label{corr}
Assume that the assumptions of Theorem \ref{Theorem_3.3}, \eqref{penalty1}, and {\rm \textbf{(SS$_{{\rm A}}$)}} hold. Then, the DDT procedure, $\hat{\mathcal{Q}}_{n}$, defined in \eqref{ing31}, satisfies
\begin{align} \label{ing21}
    \lim_{n \rightarrow \infty} P( \hat{\mathcal{Q}}_{n} = \mathcal{Q}_{n}) = 1,
\end{align}
provided that the data-driven threshold satisfies 
\begin{eqnarray} \label{penalty2}
    \hat{H}_n= \frac{\max\{ q_n^{3/2}/\sqrt{n}, \,\, 
    [(q_{n} + \hat{s}_{0})^{1/2}\wedge (\underline{\hat{s}}^{1/2}_{0} q_n^{1/\eta})]\}}{n^{1/2}} \tilde{d}_n,
\end{eqnarray}
where $\tilde{d}_n$ diverges to $\infty$ at a sufficiently slow rate, $\hat{s}_0=\sharp(\hat{\mathcal{J}}_{n})$, and $\underline{\hat{s}}_{0}=\sharp(\{i: (i, l) \in \hat{\mathcal{J}}_{n}\})$.
\end{theorem}

\noindent
A few remarks are in order.
\begin{remark}
    Note that the unknown thresholding value in \textbf{(SS$_{{\rm A}}$)} is replaced by the feasible $\hat{H}_{n}$ in \eqref{penalty2}, hence the name data-driven thresholding (DDT).
\end{remark}
\begin{remark}
    Combining Theorems \ref{Theorem_3.3} and \ref{corr} yields that FHTD asymptotically captures exactly the relevant AR and exogenous covariates despite complex unit roots, conditional heteroscedasticity, and a large pool of candidate variables. 
\end{remark}
\begin{remark}
    If a fixed number of exogenous predictors are known to have unit roots, one may include them at the initial stage of the FSR along with the lagged dependent variables and threshold their coefficients at the final step. 
    This strategy is expected to preserve the model selection consistency of FHTD and can be used for predictive regression problems such as equity premium prediction \citep{Georgiev2018, Welch2007}.
    However, if there are many unit-root exogenous predictors, the problem is much more challenging because of the nonstationary predictors and the potential cointegration among them. 
    By a similar argument as Example~\ref{egOGA}, direct application of FSR and OGA to high-dimensional, possibly cointegrated predictors may yield undesirable variable selection. 
    The systematic treatment of this setting is beyond the scope of the present paper and is deferred for future research.    
\end{remark}

Recently, the variable selection consistency for a Twin Adaptive LASSO (TALasso) method for linear regression with mixed-root predictors is established in \cite{Lee2022}.
Our setting differs in several important respects.
Although \cite{Lee2022} considers both unit-root and cointegrated exogenous predictors, it excludes
autoregressive lag terms, complex unit roots, and conditionally heteroscedastic predictors and errors, and allows only a fixed number of predictors.
Moreover, the variable selection consistency in \cite{Lee2022} is built upon the sure-screening property of the adaptive LASSO, which fails to include all relevant variables in our simulation study in Section \ref{Sec:simulation} when nonstationary lagged dependent variables are
included among the candidate predictors.



It is also worth noting that \eqref{ing21}, achieving consistency for variable selection of the AR lags, is more desirable for prediction than consistent order selection which may still contain redundant AR variables. To the best of our knowledge, this type of consistency has not been reported elsewhere, even when $q_n$ is bounded, and $v_{t, n}$ (see \eqref{ingfeb31})
is dropped from model \eqref{Model2}.
The following example elucidates why achieving consistency in variable selection can offer greater advantages compared to order selection from a predictive standpoint.

\begin{example} \label{egPredictionError}
Consider the model
\begin{align} \label{ingaug001}
y_t=\sum_{j=1}^{k}\beta_jy_{t-j}+\epsilon_{t}, \quad t=1,\ldots, n,
\end{align}
where $k \geq 1$, $\beta_{1}=\cdots=\beta_{k-1}=0$, $\beta_{k}=1$, and $\{\epsilon_{t}\}$ are i.i.d. random variables with mean zero and a constant variance of $0<\sigma^2<\infty$. 
Clearly, model \eqref{ingaug001} is a nonstationary AR($k$) model containing $k$ unit roots.
If $k$ is known or can be consistently estimated by an order selection criterion such as BIC, then it is natural to predict $y_{n+1}$ using the least squares predictor, $\hat{y}_{n+1}(k)= \mathbf{y}_{n}^{\top}(k) \hat{\bm{\beta}}(k)$,
where
$\mathbf{y}_{t}(k) = (y_{t},\ldots,y_{t-k+1})^{\top}$ and $\hat{\bm{\beta}}(k)=(\sum_{t=k}^{n-1}\mathbf{y}_{t}(k) \mathbf{y}_{t}^{\top}(k))^{-1}\sum_{t=k}^{n-1}\mathbf{y}_{t}(k) y_{t+1}$.
The performance of $\hat{y}_{n+1}(k)$ can be evaluated using its mean squared prediction error (MSPE), defined as ${\rm MSPE}_{k}=\E(y_{n+1}-\hat{y}_{n+1}(k))^2
$.
Assume $\E|\epsilon_{1}|^{s}<\infty$ for some $s>4$, and a smoothness condition on $\epsilon_{t}$ described in Section 2 of \cite{ing2010}.
Then, by extending an argument used in
\cite{Ing2001}, \cite{ing2010}, and  \cite{Ing2014}, it can be shown that
\begin{eqnarray}
\label{ingaug002}
\lim_{n \to \infty} n ({\rm MSPE}_{k}-\sigma^{2}) = \sigma^2 {\rm plim}_{n \to \infty} \frac{\log \det(\sum_{t=k}^{n-1} \mathbf{y}_{t}(k) \mathbf{y}_{t}^{\top}(k))}{\log n} = 2k \sigma^2,
\end{eqnarray}
where the second equality is ensured by Theorem 5 of \cite{Wei1987}. 

Alternatively, if a method can consistently select the non-zero coefficients $\beta_{k}$ while excluding the redundant ones, as is offered by the FHTD, the least squares predictor,
\begin{eqnarray*}
\tilde{y}_{n+1}(k)= y_{n+1-k}\tilde{\beta}_k=
\frac{y_{n+1-k} \sum_{t=k}^{n-1}y_{t-k+1}y_{t+1}}{\sum_{t=k}^{n-1}y_{t-k+1}^2},
\end{eqnarray*}
would emerge as another appropriate predictor for
$y_{n+1}$, where 
$\tilde{\beta}_k$ is the least squares 
estimate of $\beta_k$
obtained from regressing $y_t$ on $y_{t-k}$.
By an argument similar to that used to prove
\eqref{ingaug002}, it can be shown that $\widetilde{{\rm MSPE}}_{k}=
\E(y_{n+1}-\tilde{y}_{n+1}(k))^2$ obeys
\begin{eqnarray}
\label{ingaug004}
\lim_{n \to \infty}
n(\widetilde{{\rm MSPE}}_{k}-\sigma^2)=
\sigma^2 {\rm plim}_{n \to \infty}
\frac{\log \det(\sum_{t=1}^{n-k} y_{t}^2)
}{\log n}=2 \sigma^2.
\end{eqnarray}
Equations \eqref{ingaug002} and \eqref{ingaug004} reveal that the least squares predictor constructed from a consistent order selection method could lead to significantly higher MSPE than the one derived from a consistent subset selection method, especially when the underlying unit-root model contains many irrelevant lagged dependent variables. 
\end{example}

In closing this subsection, we briefly discuss how to tune the FHTD algorithm.
First, the number of AR lags $q_n$ can be set to some large enough integer. In our simulation studies, $q_n = \lfloor 2n^{0.25} \rfloor$ is used.
Alternatively, one can use the AIC to select $q_n$ since AIC is known to be effective in selecting the order for AR models \cite{Tsay1984, Ing2012}.
Second, FHTD is not very sensitive to the number of iterations $K_n$ of FSR as long as it is large enough.
Limited experiences show that the relevant variables are picked up by FSR in the early iterations, so $K_n$ is often not crucial.
Nevertheless, since the theoretical value of $K_n$ depends on $\eta$ and $q_0$, which, in turn, are related to the moment assumptions on $\epsilon_t$ and $x_{t,j}$, one can select $\eta$ and $q_0$ either conservatively or according to the expected tail behavior of $\epsilon_t$ and $x_{t,j}$. 
This decision similarly applies to the HDIC.
Finally, after $\eta$ and $q_0$ are decided, one can still use a hold-out validation set to fine-tune the HDIC values, as is done in Section~\ref{Sec::application}. 
See \eqref{hdic_sim} for an example of such strategy, where an approximated penalty subject to further fine-tuning is used.

\section{Simulation studies} \label{Sec:simulation}

In this section, we examine the model selection performance of FHTD using synthetic data generated from model \eqref{Model1}, with coefficients, covariates, and error terms specified below. 
For the purpose of comparison, we employ LASSO, adaptive LASSO (ALasso), and OGA+HDIC+Trim (OGA-3), where the names in the parentheses are shorthands used throughout the rest of this paper. 
Since FHTD first coerces all candidate AR variables into the model, we modify ALasso and OGA-3 accordingly and consider the analogous methods, AR-ALasso and AR-OGA-3.
For AR-ALasso, the AR variables are not penalized in the first-stage LASSO and the resulting coefficients are used as the initial weights (weighted inversely) for the second-stage LASSO. 
Similarly, AR-OGA-3 forces the AR variables into the base model when implementing OGA.
Another potential method is to threshold the estimated coefficients of Lasso.
We found that such approach performed poorly in the settings considered; hence its simulation results are not reported. 

According to Theorem \ref{Theorem_3.3}, the penalty term, $w_{n, p_n}$, in HDIC can be taken to be $t_n p_n^{*^{2\bar{\theta}}}$, where $\bar{\theta}$ is defined in \textbf{(SS$_{{\rm X}}$)} and $\{t_n\}$ diverges to $\infty$ arbitrarily slowly. Here, we approximate $2\bar{\theta}$ using $1/\eta$ because the exogenous variables are often allowed to have finite higher-order moments. 
On the other hand, we set $\eta=2$ to include GARCH-type errors with relatively heavy tails. As a result, for the FSR- and OGA-based methods,
\begin{align} \label{hdic_sim}
    \HDIC(Q \oplus J) = n\log \hat{\sigma}_{Q \oplus J}^{2} + c p^{*^{1/\eta}}(\sharp(Q)+\sharp(J)), \quad \eta=2,
\end{align}
is used throughout all simulations, where $c>0$ is a tuning parameter. 
In view of Theorem \ref{corr} and since $\hat{H}_n$ in \eqref{penalty2} simplifies to $[(q_{n} + \hat{s}_{0})^{1/2}\wedge (\underline{\hat{s}}^{1/2}_{0} q_n^{1/\eta})]\tilde{d}_n / n^{1/2}$
if $q_n=o(n^{1/3})$, the threshold in DDT is set to
\begin{eqnarray} \label{ingapr12-2}
    \hat{H}_n = \frac{[(q_{n} + \hat{s}_{0})^{1/2}\wedge (\underline{\hat{s}}^{1/2}_{0} q_n^{1/2})]d}{n^{1/2}},
\end{eqnarray}
where $d$ is also subject to fine-tuning. 
In practice, one may use a hold-out validation set to determine $c$ and $d$. 
In the simulations, we set $c=d=0.5$ in all experiments for simplicity and only tune $c$ and $d$ in the real data analysis in Section \ref{Sec::application}. 
The number of iterations, $K_{n}$, of FSR and OGA is set to $40$. 
The tuning parameters for LASSO-type methods are selected using BIC as in \cite{medeiros2016}. 
Finally, 
$q_n$ and $r_{j}^{(n)}$ 
are set to $q_n=\lfloor 2n^{0.25} \rfloor$ and $r_{j}^{(n)}=r^{(n)}$ for all $1\leq j \leq p_n$, where $(n, p_n, r^{(n)})=(200, 100, 4), (400, 200, 5)$, and (800, 500, 6). Note that $p_n^{*}=p_n r^{(n)}> n$ in all cases. 

Let $\tilde{{\cal Q}}_i$ and $\tilde{{\cal J}}_i$ denote the sets of the AR and exogenous variables chosen by a model selection method in the $i$-th simulation. Then its performance is measured by the frequencies of selecting exactly the relevant variables (E) and including all relevant variables (SS) as well as the average numbers of true positives (TP) and false positives (FP), namely,
\begin{align*}
\begin{split}
    & {\rm E}=\sum_{i=1}^{1000} \mathbb{I}_{\{\tilde{{\cal Q}}_i={\cal Q}_n\}}\mathbb{I}_{\{\tilde{{\cal J}}_i={\cal J}_n\}}, \,\,
      {\rm SS}=\sum_{i=1}^{1000} \mathbb{I}_{\{{\cal Q}_n\subseteq \tilde{{\cal Q}}_i\}}\mathbb{I}_{\{{\cal J}_n\subseteq\tilde{{\cal J}}_i\}},\\
    & {\rm TP}=\frac{1}{1000}\sum_{i=1}^{1000} (\sharp{\{\tilde{{\cal Q}}_i \cap {\cal Q}_n\}} + \sharp{\{\tilde{{\cal J}}_i \cap {\cal J}_n\}}),
      {\rm FP}=\frac{1}{1000}\sum_{i=1}^{1000} (\sharp{\{\tilde{{\cal Q}}_i \cap {\cal Q}^{c}_n\}} + \sharp{\{\tilde{{\cal J}}_i \cap {\cal J}^{c}_n\}}),
\end{split}
\end{align*}
where ${\cal Q}^{c}=[q_n]\setminus{\cal Q}_n$ and ${\cal J}^{c}_n=\bar{J}\setminus{\cal J}_n$. 
All simulation results are based on 1,000 replicates.

\begin{example} \label{egSim2}
In this example, we generate $n$ observations from 
\begin{align} \label{example2-1}
    (1 - 0.45B^{4} - 0.45B^{5})(1 - B)y_{t} = \sum_{j=1}^{5}\beta_{1}^{(j)}x_{t-1,j} + \sum_{j=6}^{10}\beta_{2}^{(j)}x_{t-2,j} + \epsilon_{t},
\end{align}
where $\epsilon_{t}$ is independently drawn from a $t(6)$ distribution. The candidate covariates are generated according to the AR(1) model, $x_{t,j} = 0.8 x_{t-1, j} + 2w_{t} + v_{t,j}$, 
$j = 1, 2, \ldots, p_n$,
where $\{w_{t}\}$ and $\{v_{t,j}\}$ are independent standard Gaussian white noise processes and are independent of $\{\epsilon_t\}$.
The coefficients are given by $(\beta_{1}^{(1)},\beta_{1}^{(2)},\beta_{1}^{(3)}, \beta_{1}^{(4)},\beta_{1}^{(5)}) = (3, 3.75, 4.5, 5.25, 6)$, and 
$(\beta_{2}^{(6)},\beta_{2}^{(7)},\beta_{2}^{(8)}, \beta_{2}^{(9)}, \beta_{2}^{(10)}) = (6.75, 7.5, 8.25, 9, 9.25)$.
Since a unit-root is introduced in \eqref{example2-1}, $\{y_{t}\}$ is nonstationary and the model contains three relevant lagged dependent variables, $y_{t-1}, y_{t-4}$, $y_{t-6}$, and ten exogenous variables. In addition, the candidates $x_{t-l, j}$ are highly correlated because $\mathrm{Corr}(x_{t,i}, x_{t,j})=0.8$ for $i \neq j$.
\end{example}

Simulation results for Example \ref{egSim2} are summarized in Table \ref{table:1}.  
Clearly, the LASSO-type methods fail to identify the correct model. 
Their TP values are only slightly larger than 1, meaning on average they detect only one relevant variable. 
A closer look at the results reveals that $y_{t-1}$ is always included by these methods.
However, they include only another one or two variables at most, which are usually irrelevant, resulting in a low FP value. 
OGA-3 performs equally poorly in terms of TP values, and tends to select more irrelevant variables.
Although AR-OGA-3 has much higher TP values than OGA-3, its performance in variable screening and selection is still unsatisfactory.
This inferior performance of AR-OGA-3 is mainly ascribed to OGA's relatively poor selection path, which falls short of including all relevant exogenous variables after adding all candidate AR variables in the model. By contrast, FSR successfully includes all relevant exogenous variables. 
Based on the reliable screening capability of FSR, HDIC, Trim, and DDT further remove all redundant variables and identify the true ARX model over 90\% of the time when $n \geq 400$.

\begin{table}[t]
    \centering
    \caption{Values of E, SS, TP, and FP in Example \ref{egSim2}, 
    where E denotes selecting exactly the relevant variables and SS including all relevant variables, and TP and FP are the average numbers of true positives and false positives. Results are based on 1000 replications.} \label{table:1}
    \begin{tabular}{crrrrrr}
    \hline
    & LASSO & ALasso & OGA-3 & AR-ALasso & AR-OGA-3 & FHTD  \\
    \multicolumn{7}{l}{\footnotesize $(n, p_{n}^{*}, p_n, r^{(n)}, q_n) = (200, 400, 100, 4, 7)$} \\
    E  & 0    & 0    & 0    & 0    & 1     & 431 \\
    SS & 0    & 0    & 0    & 0    & 1     & 1000 \\
    TP & 1.02 & 1.02 & 1.16 & 1.12 & 6.67  & 13.00 \\
    FP & 0.73 & 0.39 & 3.36 & 0.50 & 12.49 & 0.98 \\ \hline
    \multicolumn{7}{l}{\footnotesize $(n, p_{n}^{*}, p_n, r^{(n)}, q_n) = (400, 1000, 200, 5, 8)$} \\
    E  & 0     & 0    & 0    & 0    & 78    & 919 \\
    SS & 0     & 0    & 0    & 0    & 78    & 1000 \\
    TP & 1.01  & 1.00 & 1.12 & 1.07 & 10.46 & 13.00 \\
    FP & 0.22  & 0.09 & 4.39 & 0.63 & 11.12 & 0.09 \\ \hline
    \multicolumn{7}{l}{\footnotesize $(n, p_{n}^{*}, p_n, r^{(n)}, q_n) = (800, 3000, 500, 6, 10)$} \\
    E  & 0     & 0    & 0    & 0    & 229   & 998 \\
    SS & 0     & 0    & 0    & 0    & 229   & 1000 \\
    TP & 1.03  & 1.00 & 1.32 & 1.06 & 11.87 & 13.00 \\
    FP & 0.13  & 0.00 & 5.62 & 0.66 & 9.29  & 0.00 \\ \hline
    \end{tabular}
\end{table}

\begin{example} \label{egSim3}
In this example, we generate data from
\begin{align} \label{ingapr14}
    (1-0.3B)(1-2\cos(0.1) B +B^{2})y_{t} = \sum_{j=1}^{5}\beta_{1}^{(j)}x_{t-1,j} + \sum_{j=6}^{10}\beta_{2}^{(j)}x_{t-2,j} + \epsilon_{t},
\end{align}
where $\{\epsilon_{t}\}$ is a GARCH(1,1) process satisfying
\begin{align*}
    \epsilon_{t} = \sigma_{t}Z_{t}, \quad
    \sigma_{t}^{2} = 5\times 10^{-2} + 0.05\epsilon_{t-1}^{2} + 0.9\sigma_{t-1}^{2},
\end{align*}
in which $\{Z_{t}\}$ is a sequence of i.i.d. standard Gaussian random variables. By Theorem 2.2 of \cite{LING2002109}, $\epsilon_{t}$ has finite sixth moment. Let $\mathbf{w}_{t} = \mathbf{A}\bm{\pi}_{t}$, where $\mathbf{A} = (a_{ij})_{1 \leq i,j \leq p_n}$, with $a_{ij}=0.6^{|i-j|}$ if $|i-j| \leq 7$ and $a_{ij} = 0$ otherwise, and $\{\bm{\pi}_{t}\}$, independent of $\{Z_{t}\}$, is a sequence of i.i.d. random vectors whose entries are independently drawn from a $t(13)$ distribution. 
We then generate $x_{t,j}$ by $(1 - 0.1B + 0.7B^{2})x_{t,j} = (1 + 0.7B)w_{t,j}$, $1 \leq j \leq p_n$,
where $w_{t,j}$ is the $j$-th component of $\mathbf{w}_{t}$. Note that $\{x_{t, j}\}$ is an ARMA(2,1) process. 
Moreover, the relevant coefficients are $(\beta_{1}^{(1)},\beta_{1}^{(2)},\beta_{1}^{(3)}, \beta_{1}^{(4)},\beta_{1}^{(5)}) = (0.82, -1.03, 1.92, -2.21, 2.42)$, and $(\beta_{2}^{(6)},\beta_{2}^{(7)},\beta_{2}^{(8)}, \beta_{2}^{(9)}, \beta_{2}^{(10)})= (-2.57, 3.28, -3.54, 3.72, -3.90)$.
\end{example}

In addition to conditionally heteroscedastic errors, in this example the AR component on the left-hand side of \eqref{ingapr14} contains complex unit roots; thus, $\{y_{t}\}$ cannot be made stationary through simple difference transforms. 
As shown in Table \ref{table:2}, this challenge hinders the performance of the OGA- and LASSO-type methods, all of which have zero SS and E values and low TP values even when $n=800$. In contrast, FHTD still fares well under the challenge. Specifically, it detects all relevant variables over 94\% of the time for $n \geq 200$.
In addition, its E value rapidly increases from 493 to over 840 when $n$ increases from 200 to 400 (or 800).

\begin{table}[t]
    \centering
    \caption{Values of E, SS, TP, and FP in Example \ref{egSim3}, where E, SS, TP and 
    FP are defined similarly as thos of Table~\ref{table:1}. Results are also based on 1000 replications.} \label{table:2}
    \begin{tabular}{crrrrrr}
    \hline
    & LASSO & ALasso & OGA-3 & AR-ALasso & AR-OGA-3 & FHTD  \\
    \multicolumn{7}{l}{\footnotesize $(n, p_{n}^{*}, p_n, r^{(n)}, q_n) = (200, 400, 100, 4, 7)$} \\
    E  & 0    & 0    & 0    & 0    & 0     & 493 \\
    SS & 0    & 0    & 0    & 0    & 1     & 943 \\
    TP & 1.45 & 1.24 & 1.33 & 1.19 & 5.40  & 12.93 \\
    FP & 3.04 & 2.22 & 2.09 & 1.73 & 6.33  & 0.80  \\ \hline
    \multicolumn{7}{l}{\footnotesize $(n, p_{n}^{*}, p_n, r^{(n)}, q_n) = (400, 1000, 200, 5, 8)$} \\
    E  & 0     & 0    & 0    & 0    & 0     & 845 \\
    SS & 0     & 0    & 0    & 0    & 0     & 999 \\
    TP & 1.21  & 1.10 & 1.83 & 1.05 & 5.63  & 13.00 \\
    FP & 1.93  & 1.58 & 3.19 & 1.50 & 6.40  & 0.24 \\ \hline
    \multicolumn{7}{l}{\footnotesize $(n, p_{n}^{*}, p_n, r^{(n)}, q_n) = (800, 3000, 500, 6, 10)$} \\
    E  & 0     & 0    & 0    & 0    & 0     & 850 \\
    SS & 0     & 0    & 0    & 0    & 0     & 1000 \\
    TP & 1.04  & 1.01 & 1.94 & 1.00 & 6.19  & 13.00 \\
    FP & 1.32  & 1.18 & 3.34 & 1.18 & 6.66  & 0.33 \\ \hline
    \end{tabular}
\end{table}

We also considered another challenging example, where the error term and all candidate exogenous variables are conditionally heteroscedastic in addition to two unit roots in the AR component. 
FHTD still substantially outperforms the other methods in this example. 
Details are deferred to Section \ref{supp::sim} of the supplement.

\section{Empirical applications} \label{Sec::application}
In this section, we apply the proposed FHTD to the U.S. monthly housing starts and unemployment series. 

\begin{figure}[t]
    \centering
    \begin{subfigure}{0.49\linewidth}
        \includegraphics[width=\linewidth]{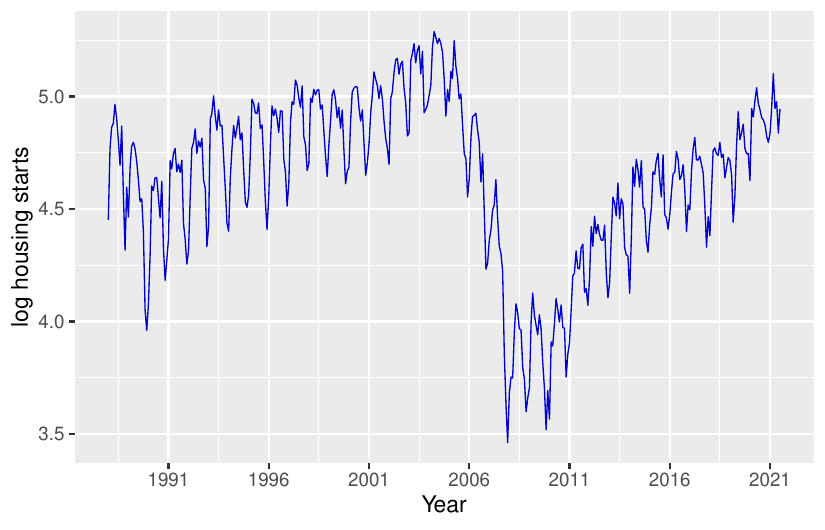}
        \caption{Logarithm of housing starts.}
        \label{fig:2}
    \end{subfigure}
    \begin{subfigure}{0.49\linewidth}
        \includegraphics[width=\linewidth]{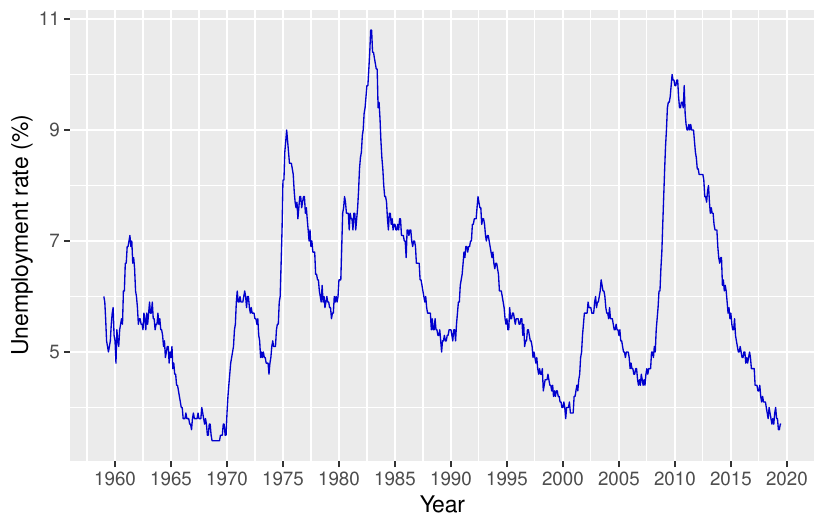}
        \caption{Unemployment rates, seasonally adjusted.}
        \label{fig:1}
    \end{subfigure}
    \caption{Time plots of U.S. monthly housing starts and unemployment series}
\end{figure}

\subsection{Housing starts in the U.S.}
In this application, we are interested in modeling the logarithm of U.S.~monthly housing starts. As depicted in Figure \ref{fig:2}, the series exhibits an apparent seasonal pattern along with a drastic level change around the subprime financial crisis of 2008. For covariates, we collect the monthly new private housing units authorized by building permits for each state\footnote{for instance, data for Illinois are retrieved from \url{https://fred.stlouisfed.org/series/ILBP1FH}}.
and the 30-year fixed rate mortgage averages from the Economic Data of St. Louis Federal Reserve\footnote{Freddie Mac, 30-Year Fixed Rate Mortgage Average in the United States [MORTGAGE30US], retrieved from FRED, Federal Reserve Bank of St. Louis; \url{https://fred.stlouisfed.org/series/MORTGAGE30US}, October 27, 2022}.
After removing series with missing values, we have 49 housing permits series $\{x_{t,j}, j=1,2,\ldots,49\}$ and the mortgage rate series $r_{t}$ from January 1988 through August 2022. We also remove the seasonality and unit root by taking $\Tilde{x}_{t,j} = (1-B^{12})(1-B)\log x_{t,j}$, $j=1,2,\ldots,49$, and $\Tilde{r}_{t} = r_{t} - r_{t-1}$. Consequently, we have 403 observations for each series.

We employ the following predictive model
\begin{align} \label{HS1}
    h_{t} = \sum_{l=1}^{18}\alpha_{l}h_{t-l} + \sum_{j=1}^{49}\sum_{k=1}^{18}\beta_{k}^{(j)}\Tilde{x}_{t-k,j} + \sum_{k=1}^{18}\beta_{k}^{(50)}\Tilde{r}_{t-k} + \epsilon_{t},
\end{align}
where $h_{t}$ denotes the logarithm of U.S. housing starts at month $t$. Note that there are 918 potential predictors. We also consider the model with a drift,
\begin{align} \label{HS2}
    h_{t} = \beta_{0} + \sum_{l=1}^{18}\alpha_{l}h_{t-l} + \sum_{j=1}^{49}\sum_{k=1}^{18}\beta_{k}^{(j)}\Tilde{x}_{t-k,j} + \sum_{k=1}^{18}\beta_{k}^{(50)}\Tilde{r}_{t-k} +  
    \epsilon_{t}.
\end{align}
In implementing FHTD, we estimate \eqref{HS2} via the following procedure. Subtract first from each variable (including the dependent variable) its own sample average, and then apply FHTD to the transformed data. Afterwards, estimate \eqref{HS2} by OLS with the selected variables and an intercept. 
For FHTD and AR-OGA-3, we use HDIC in \eqref{hdic_sim}, $\hat{H}_n$ in \eqref{ingapr12-2}, and choose $c$ and $d$ therein over a grid of values between 0.1 and 0.7 via a hold-out validation set consisting of the last $20\%$ of the training data in each window. The BIC is used to select the penalty parameters for LASSO-type methods.

\begin{figure}
    \centering
    \includegraphics[width=\linewidth]{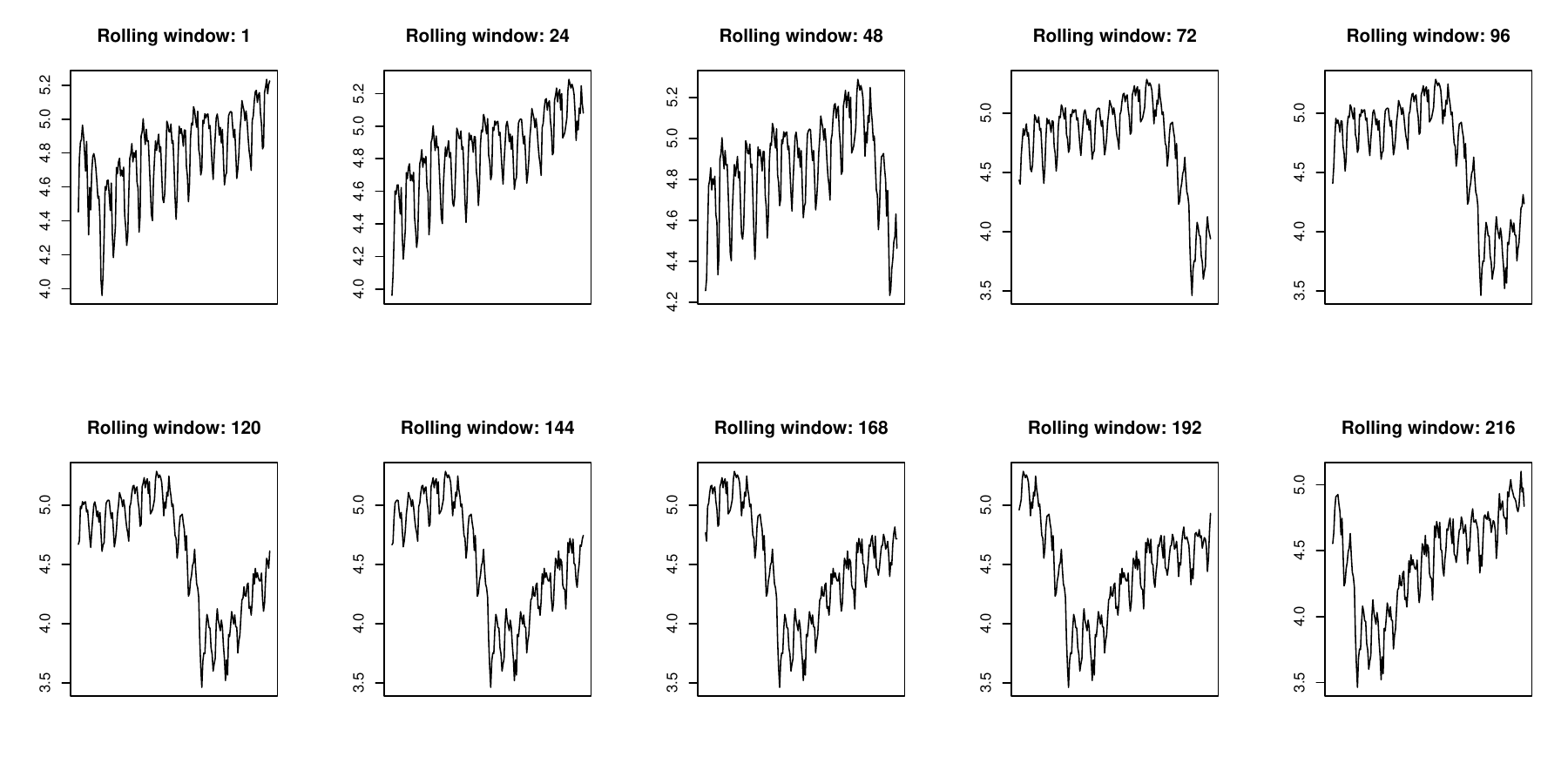}
    \caption{Time plots of logarithm of monthly U.S. Housing Starts, $h_{t}$, of selected windows}
    \label{fig:HS2}
\end{figure}

We perform rolling-window one-step-ahead prediction using FHTD as well as the other methods described in Section \ref{Sec:simulation}. We reserve the last 18 years of data as the test set, resulting in $W = 216$ windows. Each window contains 169 observations as training data. Figure \ref{fig:HS2} plots some selected windows. As shown in the figure, the methods are challenged to forecast the sharp dip around 2008 and the following recovery. Since the true model is unknown, the performance of the methods is measured by the root mean squared prediction error (RMSE) and the median absolute prediction error (MAE), where $\mathrm{RMSE} = \{W^{-1}\sum_{w=1}^{W}( h_{T-w+1} - \hat{h}_{T-w+1} )^{2}\}^{1/2}$
and MAE is the median of $\{|h_{T-w+1}-\hat{h}_{T-w+1}|:w=1,2\ldots,W\}$, in which $T$ is the time index for the last data point and $\hat{h}_{T-w+1}$ is the predicted value of $h_{T-w+1}$. 
We also conduct the Diebold-Mariano (DM) test \cite{Diebold2002} for testing the equality of prediction accuracy with power $= 1$ in the loss function.

The prediction results are recorded in Table \ref{table:HS1}. 
The LASSO-type methods are highly sensitive to the specification of the intercept. They performed poorly when the intercept is omitted. 
In view of Figure \ref{fig:HS2}, fitting the drift term to the upward trend in the first few windows may help alleviate the unit-root property in the data in finite sample, and without the drift, LASSO-type methods are unable to adapt to the unit-root behavior in the data. 
On the contrary, FHTD remains stable whether or not an intercept is included, and its prediction errors are substantially lower than the other methods. 
The DM tests also support the FHTD's prediction superiority over the other methods with $p$-values all less than 5\%, except for the AR-ALasso with intercept.

\begin{table}[t]
\centering
\caption{Out-of-sample RMSEs and MAEs of competing methods applied to \eqref{HS1} and \eqref{HS2}. The $p$-values from the two-sided Diebold-Mariano test against FHTD are also reported.}
\label{table:HS1}
\begin{tabular}{ccccccc}
\hline
    & FHTD    & OGA-3 & AR-OGA-3 & LASSO  & ALasso & AR-ALasso \\ 
Model $\eqref{HS1}$ \\
RMSE\footnotesize{($\times 10$)}  & 1.08 & 1.34 & 1.32 & 2.14 & 2.13 & 1.43 \\ 
MAE\footnotesize{($\times 10$)}   & 0.74 & 0.81 & 0.88 & 1.21 & 1.16 & 0.97 \\ 
$p$-values                        & -    & 0.00 & 0.00 & 0.00 & 0.00 & 0.00 \\
\hline
Model $\eqref{HS2}$ \\
RMSE\footnotesize{($\times 10$)}  & 1.11 & 1.31 & 1.29 & 1.24 & 1.21 & 1.14 \\ 
MAE\footnotesize{($\times 10$)}   & 0.70 & 0.82 & 0.88 & 0.86 & 0.88 & 0.81 \\ 
$p$-values                        & -    & 0.00 & 0.00 & 0.02 & 0.03 & 0.14 \\
\hline
\end{tabular}
\end{table}

\subsection{U.S. unemployment rate}

Next, we consider the U.S. monthly unemployment rate $\{u_{t}\}$, shown in Figure~\ref{fig:1}.
In some empirical studies, the unemployment rate is considered difference-stationary. 
Nevertheless, \cite{bierens2001} has found some evidence that the fluctuations in $\{u_{t}\}$ may be due to complex unit roots. 
\cite{Montgomery1998} have also noted the possibility of complex unit roots. Regardless of such complications, we can directly apply FHTD to select a model for $\{u_t\}$ and to construct predictions. 
The data used are from the FRED-MD dataset\footnote{available on \url{https://research.stlouisfed.org/econ/mccracken/fred-databases/}}, which contains 128 U.S. monthly macroeconomic variables from January 1959 to July 2019. 



We use data from January 1973 to June 2019 and again consider rolling-window one-step-ahead predictions. After discarding the series with missing values during the time span, there remain 124 macroeconomic time series that can be used to forecast $u_t$. Following \cite{McC2016}, we transform some series by taking logs, differencing, or both. 
Denote these series by $\{x_{t,j}\}, j=1, \ldots, 124$. Then we apply FHTD to the following model, 
\begin{align} \label{5-1}
    u_{t} = \sum_{i=1}^{6}\alpha_{i}u_{t-i} + \sum_{j=1}^{124}\sum_{l=1}^{6}
   \beta_{l}^{(j)}x_{t-l, j}+ \epsilon_{t},
\end{align}
which contains 750 candidate predictors. The last two years of data are reserved as test samples, 
resulting in a window size of 310 observations. 

The results are reported in Table \ref{table:4}. 
In both performance measures, FHTD outperforms all other competing methods. 
The DM tests also favor FHTD over AR-OGA-3, LASSO, ALasso, and AR-ALasso with at least 10\% significance level.
Note that for LASSO, ALasso, and AR-ALasso, we only report their performance when an intercept is included, since, as observed in the previous application, these methods performed poorly when the intercept is omitted. 
In this particular application, the RMSEs and MAEs of LASSO and ALasso without the intercept can be more than 14 times as large as their counterparts for the AR-AIC.
The prediction accuracy of OGA-3 is indistinguishable from that of FHTD according to the DM test, which might be a consequence of the high correlation between their error series. 
In examining the selected variables using full data, FHTD identifies there variables as relevant: (number of) all employees in goods-producing industries, average weekly hours in the manufacturing industry, and new housing permits. The first two variables are related to different segments of the labor market, and the third variable is conceivably associated with business cycles.
Therefore the selected variables are indeed aligned with economic intuitions.

The two empirical applications in this section show that FHTD performs well with general unit-root time series and is stable across specifications, indicating that FHTD is a viable method in practice.

\begin{table}[t]
\centering
\caption{Out-of-sample RMSEs and MAEs of competing methods applied to \eqref{5-1} for U.S. monthly unemployment rate series. The $p$-values from the (two-sided) Diebold-Mariano test against FHTD are also reported.}
\label{table:4}
\begin{tabular}{ccccccc}
\hline
    & FHTD    & OGA-3 & AR-OGA-3 & LASSO  & ALasso & AR-ALasso  \\
RMSE\footnotesize{($\times 10$)} & 1.35 & 1.39 & 1.48 & 1.53 & 1.40 & 1.39 \\
MAE\footnotesize{($\times 10$)}  & 0.88 & 0.94 & 0.93 & 1.12 & 1.04 & 0.99 \\
$p$-value                        & -    & 0.35 & 0.08 & 0.00 & 0.00 & 0.08 \\
\hline
\end{tabular}
\end{table}

\section{Concluding remarks} \label{Sec::conclusion}
This paper proposed the FHTD algorithm for variable selection in high-dimensional nonstationary ARX models with conditionally heteroscedastic covariates and errors. 
Under strong sparsity conditions, we established its selection consistency, valid even when the lagged dependent variables are highly correlated and sample covariance matrices lack deterministic limits. 
Finally, we point out some potential research directions. 
First, extending from \textbf{(SS)} and \textbf{(SS$_{{\rm X}}$)} to weak sparsity assumptions (where coefficients are mostly non-zero but only a few are significant) may prioritize optimal forecasting over variable selection consistency. 
Addressing this issue remains challenging, particularly in the presence of complex unit roots. 
Another intriguing avenue is the model selection for cointegrated data, common in economics and environmental studies, for high-dimensional data analysis. 
In addition, the structural break detection problem can be cast into a model selection problem in the ARX framework (see, for example, \cite{Chan2014}). 
Investigating such problem in the nonstationary setting is also an important direction for future research.


\begin{appendix}
\section*{Appendix: Key technical tools} \label{Sec::Appendix} 
As noted earlier, the analysis of FHTD relies on a number of key theoretical tools which are of independent interests. 
In particular, Theorems \ref{Theorem_FCLT}--\ref{Theorem_3.1} presented in this section not only are indispensable to our analysis, but may also be useful in future research. 
Their proofs can be found in Section \ref{sec6} of the supplementary material.

Let
\begin{align} \label{ingapr10}
    \mathbb{B}_{n}(t_{1},t_{2},\ldots,t_{2l+2})  = \frac{1}{\sqrt{n}} &\left( \sum_{k=1}^{\lfloor nt_{1} \rfloor}\delta_{k}, \sum_{k=1}^{\lfloor nt_{2} \rfloor}(-1)^{k}\delta_{k}, \sum_{k=1}^{\lfloor nt_{3} \rfloor }
    \sqrt{2}\sin(k\vartheta_{1})\delta_{k}, \sum_{k=1}^{\lfloor nt_{4} \rfloor } \sqrt{2}\cos(k\vartheta_{1})\delta_{k}, \right. \notag \\
    & \left. \ldots, \sum_{k=1}^{\lfloor nt_{2l+2} \rfloor }\sqrt{2}\cos(k\vartheta_{l})\delta_{k}  \right).
\end{align}
Note that $\mathbb{B}_{n}$ is a random element in $D^{2l+2}$, where $D$ is the Skorohod space $D=D[0,1]$ (see \cite{billingsley1999}).
Our first theoretical apparatus is a novel FCLT for the multivariate linear processes \eqref{ingapr10}  under a set of mild conditions.

\begin{theorem} \label{Theorem_FCLT}
Assume that {\rm (A1)--(A4)} hold with $\eta$ in \eqref{ingaug1}, \eqref{ingaug2}, and \eqref{ingaug222} replaced by some $\eta_{1} > 1$. 
In addition, assume
\begin{eqnarray} \label{a7'}
    \max_{1\leq j\leq p_{n}} r_{j}^{(n)} = o(n^{\kappa}),
\end{eqnarray}
where $\kappa = \min\{1/2, 1-\eta_{1}^{-1}\}$.
Then
\begin{align} \label{LemmaC1-0}
    \mathbf{V}^{-1/2}\mathbb{B}_{n} \Rightarrow \mathbb{W},
\end{align}
where $\Rightarrow$ denotes convergence in law, $\mathbb{W}$ is a $(2l+2)$-dimensional standard Brownian motion, and $\mathbf{V} = \mathrm{diag}(v_{1}^{2}, v_{2}^{2}, \ldots, v_{2l+2}^{2})$ is a $(2l+2)$-dimensional diagonal matrix with 
\begin{align*}
    v_{1}^{2} = \sum_{h=-\infty}^{\infty} \gamma_{h}, \,\, v_{2}^{2} = \sum_{h=-\infty}^{\infty} (-1)^{h}\gamma_{h}, \,\,
    v_{2k+1}^{2} = v_{2k+2}^{2} = \sum_{h=-\infty}^{\infty} \cos(h \vartheta_{k})\gamma_{h},
\end{align*}
$k = 1,2,\ldots, l$. 
\end{theorem}
Note that $ v_{i}^{2} >0,i=1, \ldots, 2l+2$, are ensured by (A3). 
If \eqref{Model2} reduces to an AR($p$) model with a fixed $p$ and a GARCH error $\epsilon_{t}$, then $v_{m}^{2}=\sigma^{2}$ for all $m$, and the weak limit in Theorem~\ref{Theorem_FCLT} coincides with Theorem 3.3 of \cite{ling1998}.  
Furthermore, when the error $\epsilon_{t}$ reduces to an m.d.s. with constant conditional variance, Theorem \ref{Theorem_FCLT} reduces to Theorem 2.2 of \cite{chan1988}.
Here, due to the presence of $v_{t, n}=\sum_{j=1}^{p_{n}}\sum_{l=1}^{r_{j}^{(n)}}\beta_{l}^{(j)}x_{t-l,j}$ in $\delta_t$, the FCLT is quite different from the classical ones, where the linear processes are driven by $\{\epsilon_{t}\}$ only.

\begin{theorem} \label{LemmaS1}
Let
\begin{align*}
    X_{i} = \sum_{l=-\infty}^{i} a_{i,l}\tilde{x}_{l},\;\; Y_{j} = \sum_{k=-\infty}^{j} b_{j,k}\tilde{y}_{k},
\end{align*}
where
$\sum_{l=-\infty}^{i}a^2_{i,l}<\infty$
and $\sum_{k=-\infty}^{j}b^2_{j,k}<\infty$, for all $1\leq i,j \leq n$,
and $\{(\tilde{x}_{t},\tilde{y}_{t}),\mathcal{F}_{t}\}$ is an m.d.s. Suppose also
\begin{align} \label{ingapr11}
    \tilde{x}_{t}\tilde{y}_{t} - \E(\tilde{x}_{t}\tilde{y}_{t}) = \sum_{j=0}^{\infty} \bm{\theta}_{j}^{\top} \tilde{\mathbf{z}}_{t-j},
\end{align}
where $\{\tilde{\mathbf{z}}_{t},\mathcal{F}_{t}\}$ is a multivariate m.d.s. and $\sum_{j=0}^{\infty} \Vert \bm{\theta}_{j} \Vert \leq C$, and
\begin{align} \label{LemmaS1-0-1}
    \sup_{t}\{ \E|\tilde{x}_{t}|^{p}+ \E|\tilde{y}_{t}|^{q} + \E\Vert \tilde{\mathbf{z}}_{t}\Vert^{2r} \} \leq C,
\end{align}
where $1/p + 1/q = 1/(2r)$ for some $r \geq 1$.
Define
$Q_{n}=\sum_{i,j=1}^{n}q_{i,j}X_{i}Y_{j}$, where
$q_{i,j}$ are real numbers.
Then,
\begin{align*}
    \E|Q_{n}-\E Q_{n}|^{2r} \leq C_{r}\left\{ \E\left( \sum_{i,j=1}^{n}q_{i,j}X_{i}^{*}Y_{j}^{*} \right)^{2} \right\}^{r},
\end{align*}
where $C_{r}$ is a positive constant depending only on $r$,
\begin{align*}
    X_{i}^{*} = \sum_{l=-\infty}^{i} a_{i,l}x_{l}^{*}, \quad Y_{j}^{*} = \sum_{k=-\infty}^{j} b_{j,k}y_{k}^{*},
\end{align*}
and $\{x_{t}^{*}\}$ and $\{y_{t}^{*}\}$ are independent processes that are identically distributed with $\{\tilde{x}_{t}\}$ and $\{\tilde{y}_{t}\}$, respectively.
\end{theorem} 

Theorem \ref{LemmaS1} provides a novel moment bound for quadratic forms of linear processes driven by conditional heteroscedastic m.d.s. 
The First Moment Bound Theorem of \cite{findley1993} also provides similar bounds. 
However, Theorem \ref{LemmaS1} has some essential improvements. 
First, their theorem assumes that
the $\tilde{x}_{t}$ and $\tilde{y}_{t}$ in \eqref{ingapr11}
obey $\E(\tilde{x}_{t}\tilde{y}_{t}|\mathcal{F}_{t-1})=\E(\tilde{x}_{t}\tilde{y}_{t})$ a.s. Hence conditional heteroscedasticity is precluded.
By contrast, as discussed in Section \ref{s01} in the supplementary material, many conditionally heteroscedastic processes satisfy \eqref{ingapr11}. 
Second, \eqref{LemmaS1-0-1} allows $\tilde{x}_{t}$ and $\tilde{y}_{t}$ to be integrable up to different orders, which helps deal with quadratic forms involving $\{\epsilon_{t}\}$ and $\{x_{t,j}\}$ in \eqref{Model2} since they may exhibit different tail behaviors. 
Theorem \ref{Theorem_FCLT} and Theorem \ref{LemmaS1} are crucial for the following uniform lower bound for the minimum eigenvalues of the sample covariance matrices.

\begin{theorem} \label{Theorem_3.1}
Assume {\rm (A1)--(A5)} and \eqref{ing4}. Then, for
\begin{eqnarray} \label{ingjan1}
    K_{n} = o(n^{1/2}/{p_{n}^{*}}^{\bar{\theta}}),
\end{eqnarray}
where $\bar{\theta}$ is defined in {\rm \textbf{(SS$_{{\rm X}}$)}},
\begin{align} \label{ing3}
    \max_{\sharp(J) \leq K_{n}} \lambda_{\min}^{-1}\left( n^{-1} \sum_{t=\bar{r}_{n}+1}^{n}\mathbf{s}_{t}(J) \mathbf{s}_{t}^{\top}(J) \right) = O_{p}(1),
\end{align}
where $\mathbf{s}_{t}(J)$ is defined in Section \ref{sec6} of the supplementary material. Moreover, 
\begin{align} \label{ww'eigenvalue}
    \max_{\sharp(J) \leq K_{n}} \lambda_{\min}^{-1}\left( n^{-1} \sum_{t=\bar{r}_{n}+1}^{n}
    \mathbf{w}_{t}(J)\mathbf{w}_{t}^{\top}(J) \right) = O_{p}(1).
\end{align}
\end{theorem}
Note that $\mathbf{s}_t(J)$, obtained via a nonsingular linear transformation of $\mathbf{w}_t(J)$, is introduced to stabilize the regressor vector by decomposing and rescaling unit-root components so that all regressors have comparable stochastic magnitudes.
Nevertheless, in contrast to the stationary case, the associated sample covariance matrix
\begin{align*}
    n^{-1}\sum_{t=\bar r_n+1}^n \mathbf{s}_t(J)\mathbf{s}^\top_t(J)
\end{align*}
fails to admit a deterministic limit. Moreover, since $\mathbf{w}_t(J)$ comprises highly correlated autoregressive variables, it is not known \it{a priori} whether the sample covariance matrix arising directly from least-squares fitting,
\begin{align*}
    n^{-1}\sum_{t=\bar r_n+1}^n \mathbf{w}_t(J)\mathbf{w}^\top_t(J)
\end{align*}
is asymptotically ill-conditioned or even singular. With the aid of \eqref{ing3}, this issue is resolved through \eqref{ww'eigenvalue}, which further indicates that least-squares-based model selection criteria are still able to discriminate among candidate predictors despite the presence of $q_n$ highly correlated AR variables. Related results for finite-order nonstationary AR models with conditionally homoscedastic errors can be found in \cite{lai1982} (Equation 3.10) and \cite{chan1988} (Equation 3.5.1).
\end{appendix}




\begin{supplement}
\stitle{}
\sdescription{}
The supplemental material contains five sections. Section \ref{s01} provides comments on Assumptions (A1)–(A6) and discusses the sparsity conditions \textbf{(SS$_{{\rm X}}$)} and {\rm \textbf{(SS)}}. Section \ref{sec6} presents proofs for the main results, Theorems \ref{Theorem_3.2}--\ref{corr} of the paper, as well as the proofs of Theorems \ref{Theorem_FCLT}--\ref{Theorem_3.1}. Further details related to Section \ref{sec6} can be found in Section \ref{Appendix_C2}. Section \ref{sec06} contains the proofs regarding Examples \ref{egOGA} and \ref{egLASSO}. Finally, Section \ref{supp::sim} offers additional simulation results.  

\setcounter{page}{1}
\setcounter{section}{0}
\setcounter{equation}{0}
\def\theequation{S\arabic{section}.\arabic{equation}}
\def\thetable{S\arabic{section}.\arabic{table}}
\def\thesection{S\arabic{section}}
\setcounter{table}{0}
\setcounter{figure}{0}
\renewcommand\thetheorem{ST\arabic{section}.\arabic{theorem}}
\renewcommand\thelemma{SL\arabic{section}.\arabic{lemma}}

\section{Comments on Assumptions (A1)--(A6), \textbf{(SS$_{{\rm X}}$)}, and \textbf{(SS)}} \label{s01}

In this section, we provide some comments on Assumptions (A1)--(A6). 
Assumption (A1) is fulfilled by many conditionally heteroscedastic processes. For example, consider a stationary GJR-GARCH$(p_0^{\prime}, q_0^{\prime})$ model \citep{GJR-GARCH},
\begin{equation}\label{GJR.1}
\begin{split}
    \epsilon_{t}=\nu_t z_t,\,\,
    \nu_t^2=\E(\epsilon_t^2|{\cal F}_{t-1})=\varphi_{0,0}+\sum_{i=1}^{p^{\prime}_{0}}\varphi_{0,i}\epsilon_{t-i}^2+\sum_{j=1}^{q^{\prime}_{0}}\psi_{0,j}\nu_{t-j}^2
     +\sum_{k=1}^{p_{0}^\prime}\zeta_{0,k}\epsilon_{t-k}^{2} \mathbb{I}_{\{\epsilon_{t-k}<0\}},
\end{split}
\end{equation}
where $z_{t}$ are independent and identically distributed (i.i.d.) symmetric random variables with zero mean, unit variance, and finite 2$\eta$-th moment, $\eta \geq 2$, 
$\mathcal{F}_t=\sigma(z_s, s\leq t)$, $p^{\prime}_{0}$ and $q^{\prime}_{0}$ are positive integers, $\varphi_{0,0}>0$, and $\varphi_{0,i}$, $\psi_{0,j}$, and $\zeta_{0,k}$ are non-negative constants that guarantee $\E|\epsilon_{1}|^{2\eta}<\infty$. 
Following \cite{huang2022}, it can be shown that (A1) is fulfilled by \eqref{GJR.1} with $l_0=2$, $\mathbf{e}_t=(w_{1,t},w_{2,t})^{\top}$, and $\bm{\theta}_s=(\tilde{b}_s, \tilde{c}_s)^{\top}$, where $w_{1,t}=\epsilon_t^{2}-\nu_t^2$, $w_{2,t}=\epsilon_t^{2}I_{\{\epsilon_t<0\}}-\frac{1}{2}\epsilon_t^2$, and $\tilde{b}_j$ and $\tilde{c}_j$, respectively, satisfy
\begin{eqnarray*}
    \sum_{j=0}^{\infty}\tilde{b}_jz^{j}=\frac{1-\sum_{j=1}^{q_0^{'}}\psi_{0,j}z^{j}}
    {1-\sum_{i=1}^{\max\{p_{0}^\prime, q_{0}^\prime\}}(\varphi_{0,i}+\psi_{0,i}+\frac{\zeta_{0,i}}{2})z^{j}}
\end{eqnarray*}
and
\begin{eqnarray*}
    \sum_{j=0}^{\infty}\tilde{c}_jz^{j}=\frac{\sum_{j=1}^{p_0^{'}}\zeta_{0,j}z^{j}}
    {1-\sum_{i=1}^{\max\{p_{0}^\prime, q_{0}^\prime\}}(\varphi_{0,i}+\psi_{0,i}+\frac{\zeta_{0,i}}{2})z^{j}},
\end{eqnarray*}
in which $\psi_{0,j}=0$ if $j>q_{0}^\prime$, $\varphi_{0,i}=\zeta_{0,i}=0$ if $i>p_{0}^\prime$, and $\sum_{i=1}^{p^{\prime}_{0}}\varphi_{0,i}+\sum_{j=1}^{q^{\prime}_{0}}\psi_{0,j}+\sum_{k=1}^{p^{\prime}_{0}}\frac{\zeta_{0,k}}{2}<1$. 
When $\zeta_{0,k}=0$ for all $k$, \eqref{GJR.1} reduces to the well-known stationary GARCH ($p_0^{\prime}, q_0^{\prime}$) process; the same argument ensures that (A1) holds with $l_0=1$, $e_t=w_{1,t}$, and $\theta_s=\tilde{b}_s$.

Assumption (A2) requires that $\{x_{t, s}\}$ is an MA($\infty$) process driven by the conditionally heteroscedastic innovations $\{\pi_{t,s}\}$. This type of assumption is broadly adopted in time series analysis. 
In fact, (A2) allows $(\epsilon_t, \pi_{t, 1}\ldots,\pi_{t, p_n})^{\top}$ to be a multivariate GARCH process. 
By the same argument used in the previous paragraph, it can be shown that the diagonal VEC model of \cite{Bollerslev1988} is a special case of \eqref{ingsep1}.
Furthermore, in the notable special case where $\{\pi_{t, s}\}$ is a sequence of
independent and identically distributed random variables with $\E(\pi_{t, s})=0$ and $\E(\pi_{t, s_1}\pi_{t, s_2})=\sigma_{s_1, s_2}$, \eqref{ingsep1} remains valid, with $e_{t, s_1, s_2}=\pi_{t, s_1}\pi_{t, s_2}-\sigma_{s_1, s_2}$, $\theta_{0, s_1, s_2}=1$, and $\theta_{j, s_1, s_2}=0$ for $j>0$.
Moment conditions \eqref{ingaug2} and \eqref{ingaug222} are more stringent than \eqref{ingaug1}. These stronger moment assumptions ensure a reliable screening performance of FSR when the number of exogenous covariates is larger than the sample size, as allowed by Assumption (A5).

Assumption (A3) is used to derive Theorem \ref{Theorem_FCLT}, leading to a uniform lower bound for the minimum eigenvalues of the sample covariance matrices of dimensions less than or equal to $q_n+K_n$; see the proof for Theorem \ref{Theorem_3.1}. 
Assumption (A4), referred to as the weak sparsity condition, is commonly used in the literature on high-dimensional data analysis. 
It follows from \eqref{ing41}, \eqref{win1}, and (A4) that 
\begin{eqnarray} \label{ing42}
    \sup_{n \geq 1}\sum_{h=-\infty}^{\infty}|\gamma_{h,n}| \leq C, 
\end{eqnarray}
which, together with \eqref{ing7}, leads to
\begin{eqnarray} \label{ing43}
    \sum_{h=-\infty}^{\infty}|\gamma_{h}| \leq C. 
\end{eqnarray}

When the moment conditions are controlled, (A5) is more flexible than the assumptions on model dimensions in \cite{medeiros2016}, where $\{y_t\}$ is assumed to be stationary, corresponding to the case of  $a=b=d_1=\cdots=d_l=0$. To see this, note that (A1) and (A2) imply
\eqref{ingapr1} and \eqref{ingapr2} respectively. Moreover, (A1), together with (A2), yields
\begin{eqnarray} \label{ingapr3}
    \sup_{t}\E|y_t|^{2\eta}<C,
\end{eqnarray}
provided $a=b=d_1=\cdots=d_l=0$. By 
\eqref{ingapr1}, \eqref{ingapr2}, \eqref{ingapr3}, and H\"{o}lder's inequality,
\begin{eqnarray*}
    \sup_{t}\E|y_{t-i}\epsilon_t|^{\eta}<C \,\,\mbox{and}\,\,\sup_{t}\E|x_{t-l,j}\epsilon_t|^{2\eta q_0/(q_0+1)}<C,
\end{eqnarray*}
for all $1\leq i\leq q_n$, $1\leq l\leq r_j^{(n)}$, and $1\leq j\leq p_n$. Therefore, the $m$ in Assumption DGP(4) of \cite{medeiros2016} obeys
\begin{eqnarray} \label{ingapr6}
    m=\min\{\eta, 2\eta q_0/(q_0+1)\}=\eta.
\end{eqnarray}
Equation \eqref{ingapr6} 
and the discussion after Assumption (REG) of \cite{medeiros2016} lead to a restriction on the number of candidate variables such that
\begin{eqnarray} \label{ingapr4}
    q_n+p_n^{*}=o(n^{\alpha \eta(\eta-2)/(2\eta+4b)}),
\end{eqnarray}
where $0<\alpha<1$ and $b>0$ are positive numbers defined therein. Equation \eqref{ingapr4} requires $p_n^{*}$ to be much smaller than $n$ unless $\eta>4$. In contrast, (A5) allows $p_n^{*}>n$ even if $\eta=2$.

We also make a few comments on (A6). For $D \subset \{1, \ldots, p_n\}$, let
\begin{align} \label{ingfeb29}
\begin{split}
    & \bm{\pi}_{t}(D)=(\pi_{t,s}: s \in D)^{\top},\,\, \bm{\mu}_{t}(D) = (\epsilon_t, \bm{\pi}^{\top}_{t}(D))^{\top},\\
    & \mathbf{\Sigma}_{n}(D) = \E (\bm{\mu}_{t}(D)\bm{\mu}^{\top}_{t}(D)).
\end{split}
\end{align}
Then, it can be shown that \eqref{ing4} holds if $\{x_{t, j}\}$ admits an infinite-order AR representation with absolutely summable coefficients and
\begin{eqnarray} \label{ing400}
    \max_{\sharp(D)\leq K_{n}+\underline{s}_0} \lambda_{\min}^{-1}(\mathbf{\Sigma}_{n}(D)) \leq C,
\end{eqnarray}
where
$\underline{s}_0$ is defined in \textbf{(SS$_{{\rm A}}$)}.
When the AR components are deleted from model \eqref{Model2}, \eqref{thm1-asmpt} reduces to (3.2) of \cite{ing2011}, which is closely related to the ``exact recovery condition'' introduced by \cite{Tropp2004} in the analysis of the orthogonal matching pursuit and plays a role similar to the ``restricted eigenvalue assumption'' introduced by \cite{Bickel2009} in the study of LASSO. Condition \eqref{thm1-asmpt} is a natural generalization of (3.2) of \cite{ing2011} when the (asymptotically) stationary AR component, $\mathbf{z}_{t}(q_n-d)=(z_{t-1},\ldots, z_{t-q_n+d})^{\top}$, is taken into account.

We now present an example that illustrates the validity of \eqref{thm1-asmpt} even when model \eqref{Model2} includes highly correlated 
lagged dependent variables and highly correlated exogenous variables.
Assume in model \eqref{Model2} that $r_{j}^{(n)}=1$ for all $1\leq j \leq p_n$ and $\{(x_{t-1, 1}, \ldots, x_{t-1, p_n}, \epsilon_t)^{\top}\}$ is a sequence of white noise vectors obeying $\E(\epsilon_t^2)=\E(x_{t, j}^2)=1$ for all $1\leq j \leq p_n$ and $0\leq \E(x_{t, i}x_{t, j})=\E(x_{t,l}\epsilon_t)=\lambda<1$ for all $1\leq i \neq j \leq p_n$ and $1\leq l \leq p_n$.
In this model specification, not only are
$y_{t-j}, 1\leq j \leq q_n$, highly correlated, but also $x_{t-1, j}, 1\leq j \leq p_n$, especially when $\lambda$ is close to 1.
Define $\mathbf{G}(q_n-d)=[\mathbf{G}_{ij}]_{1\leq i, j\leq q_n-d}=\E^{-1} (\mathbf{z}_{t, \infty}(q_n-d)\mathbf{z}_{t, \infty}^{\top}(q_n-d))$ and $c_{J}^2=\lambda^2\sharp(J)/(1-\lambda+\sharp(J)\lambda)$. Since $0<c_{J}^2<\lambda$ and $0< \mathbf{G}_{11}\leq1$, it holds that
\begin{eqnarray} \label{ingfeb5}
    \max_{\sharp(J) \leq K_{n}, (i,1) \notin J} \sum_{(i^*, 1) \in J}|a_{i^*, 1}(i,1)|\leq
    \max_{\sharp(J) \leq K_{n}}\frac{(1-\lambda \mathbf{G}_{11})\lambda\sharp(J)}{(1-c_{J}^2\mathbf{G}_{11})(1-\lambda+\sharp(J)\lambda)} \leq 1.
\end{eqnarray}
Moreover, one has
\begin{align} \label{ingfeb6}
\begin{split}
    \sum_{s=1}^{q_n-d}\max_{\sharp(J) \leq K_{n}, (i,1) \notin J}|a_{s, J}(i,1)| \leq \sum_{s=1}^{q_n-d}\max_{\sharp(J) \leq K_{n}}
    \frac{\lambda-c_{J}^2}{1-c_{J}^2\mathbf{G}_{11}} |\mathbf{G}_{s1}| \leq \sum_{s=1}^{q_n-d} |\mathbf{G}_{s1}|.
\end{split}
\end{align}
Define $a^2=1+\lambda (\sum_{j=1}^{p_n}\beta_{1}^{(j)})^2+(1-\lambda)\sum_{j=1}^{p_n}(\beta_{1}^{(j)})^2$, $b=\lambda \sum_{j=1}^{p_n}\beta_{1}^{(j)}$, and $h^2=(a^2+(a^4-4b^2)^{1/2})/2$, noting that $|b|<a^2/2$ and 
\begin{eqnarray} \label{ingfeb11}
    h^2> \max\{1, \lambda (\sum_{j=1}^{p_n}\beta_{1}^{(j)})^2+(1-\lambda)\sum_{j=1}^{p_n}(\beta_{1}^{(j)})^2\}.
\end{eqnarray}
Then, it can be shown that $\{z_{t, \infty}\}$ admits an infinite-order AR representation,
\begin{eqnarray} \label{ingfeb9}
    z_{t, \infty}+\sum_{j=1}^{\infty}\phi_j z_{t-j,\infty}=\eta_t,
\end{eqnarray}
where $1+\sum_{j=1}^{\infty} \phi_j z^{j} \neq 0$, for $|z|\leq 1$, $\sum_{j=0}^{\infty}|\phi_j| \leq C$, and $\{\eta_t\}$ is a white noise sequence with variance $h^2$. By a modified Cholesky decomposition (e.g., \cite{Ing2016}), \eqref{ingfeb11}, \eqref{ingfeb9}, and Baxter's inequality \citep{Baxter1962}, one gets $\sum_{s=1}^{q_n-d} |\mathbf{G}_{s1}| \leq Ch^{-2}\sum_{j=0}^{\infty}|\phi_j|\leq C$, which, together with \eqref{ingfeb5} and \eqref{ingfeb6}, leads to \eqref{thm1-asmpt}.

Finally, we provide a brief discussion of the strong sparsity conditions \textbf{(SS$_{{\rm X}}$)} and {\rm \textbf{(SS)}} in Sections 3.2 and 3.3.
As mentioned earlier, a condition similar to  \textbf{(SS$_{{\rm X}}$)} has been utilized by \cite{medeiros2016} to establish the selection consistency of the adaptive LASSO when $\{y_t\}$ is stationary. Specifically, they assume
\begin{eqnarray}
    \frac{\lambda s_0^{1/2}}{n^{1-\xi/2}\phi_{\min}}=o(\min_{\substack{(j, l) \in {\cal J}_n}} |{\beta_{l}^{(j)}}|), \label{ingapr5}
\end{eqnarray}
where $0<\xi<1$ is some constant defined in their Assumption (WEIGHTS) and $2\phi_{\min}$ is a lower bound for the minimum eigenvalue of the covariance matrix of the random vector formed by all relevant predictors. Assuming that $\phi_{\min}$ is bounded away from 0 and choosing $\lambda$ to be the value suggested after Assumption (REG) of \cite{medeiros2016}, \eqref{ingapr5} becomes
\begin{eqnarray}
    \frac{s_0^{1/2}p_n^{*^{1/m}}n^{\xi/m}}{n^{1/2}}=o(\min_{\substack{(j, l) \in {\cal J}_n}} |{\beta_{l}^{(j)}}|). \label{ingapr7}
\end{eqnarray}
In view of \eqref{ingapr6} and the definitions of $\bar{\theta}$ and $\xi$, we conclude that \eqref{ingapr7} is more stringent than \eqref{SS_0} in 
\textbf{(SS$_{{\rm X}}$)}.
While the left-hand side of
\eqref{SS_1} in {\rm \textbf{(SS)}} is larger than that of \eqref{SS_0} by a factor about $(\log n)^{1/2}$, it is still smaller than that of \eqref{ingapr7}.

\setcounter{equation}{0}
\section{Main proofs}\label{sec6}

In this section, we present the proofs of the main results in the paper, namely Theorems \ref{Theorem_3.2}--\ref{corr}.
The proofs, shown in Section \ref{theorems}, are proceeded by the proofs for Theorems \ref{Theorem_FCLT}--\ref{Theorem_3.1}.
Some further details regarding the proofs are collected in Section \ref{sec_further}.

\subsection{Proofs of Theorems \ref{Theorem_FCLT}--\ref{Theorem_3.1} and discussions} \label{sec3.1}

\begin{proof}[{\bf Proof of Theorem \ref{Theorem_FCLT}}]
Let $a_t^{(m)}$, $t \geq 1$ and $1\leq m \leq 2l+2$, be defined by $a_{t}^{(1)} = 1, a_{t}^{(2)} = (-1)^{t}, a_{t}^{(3)} = \sqrt{2}\sin(t\vartheta_{1}), a_{t}^{(4)} = \sqrt{2}\cos(t\vartheta_{1}),\ldots, a_{t}^{(2l+1)} = \sqrt{2}\sin(t\vartheta_{l})$, and $a_{t}^{(2l+2)} = \sqrt{2}\cos(t\vartheta_{l})$.
Note first that each element of $\mathbb{B}_{n}$ is of the form
\begin{align} \label{LemmaC1-1}
    \frac{1}{\sqrt{n}}\sum_{t=1}^{\lfloor nu \rfloor}a_{t}\delta_{t},
\end{align}
where $u \in [0,1]$ and $\{a_{t}\}$ is one of $\{a_t^{(m)}\}, 1\leq m\leq 2l+1$.
In Section \ref{sec_further}, we show that \eqref{LemmaC1-1} has the following decomposition
\begin{align} \label{LemmaC1-2}
    \frac{1}{\sqrt{n}}\sum_{t=1}^{\lfloor nu \rfloor}a_{t}\delta_{t} = R_{n}(u) + \sum_{t=1}^{\lfloor nu \rfloor-1}
    \left( X_{t,n} +
    \frac{a_{t}}{\sqrt{n}}\epsilon_{t} \right) + \frac{a_{\lfloor nu \rfloor}}{\sqrt{n}}\epsilon_{\lfloor nu \rfloor},
\end{align}
where
\begin{align*}
    X_{t,n} = \frac{1}{\sqrt{n}}\sum_{j=1}^{p_{n}}
    \left[ \sum_{s=t+1}^{n}a_{s}\left( \sum_{l=1}^{(s-t)\wedge r_{j}} \beta_{l}^{(j)}p_{s-l-t,j} \right) \right] \pi_{t,j},
\end{align*}
and $R_{n}(u)$ is a remainder term whose explicit form is given in Section \ref{sec_further}.
By Assumption (A1),
\begin{align} \label{ingsep5}
    \sup_{t}\mathbb{E}|\epsilon_t|^{2\eta_1}<C,    
\end{align}
yielding
\begin{eqnarray} \label{ing44}
    \sup_{u \in [0,1]}\frac{|a_{\lfloor nu\rfloor}\epsilon_{\lfloor nu\rfloor}|}{\sqrt{n}} = o_{p}(1)
\end{eqnarray}
and
\begin{eqnarray} \label{ing45}
    \mathbb{E}\big(\frac{a^2_{\lfloor nu\rfloor}\epsilon^2_{\lfloor nu\rfloor}}{n}\big) = o(1),\,\,u \in [0,1].
\end{eqnarray}

By Assumption (A2), 
\begin{align} \label{ingsep4}
    \sup_t\max_{1\leq j \leq p_n}\E|\pi_{t, j}|^{2\eta_1} \leq C.
\end{align}
Using \eqref{ingsep4}, after some algebraic manipulations, we show in Section \ref{sec_further} that
\begin{align}
    \sup_{u \in [0,1]}|R_{n}(u)| =& o_{p}(1), \label{LemmaC1-3} \\
    \E R_{n}^{2}(u) =& o(1), \; u \in [0,1]. \label{LemmaC1-4}
\end{align}
By \eqref{LemmaC1-2}, \eqref{ing44}, and \eqref{LemmaC1-3}, \eqref{LemmaC1-0} in the main paper holds if
\begin{align} \label{LemmaC1-5}
    \left( v_{1}^{-1}\sum_{t=1}^{\lfloor n u_{1} \rfloor-1} \left( X_{t,n}^{(1)} + \frac{a_{t}^{(1)}}{\sqrt{n}}\epsilon_{t} \right),
    \ldots, v_{2l+2}^{-1}\sum_{t=1}^{\lfloor n u_{2l+2} \rfloor-1} \left( X_{t,n}^{(2l+2)} + \frac{a_{t}^{(2l+2)}}{\sqrt{n}}\epsilon_{t} \right) \right) \Rightarrow \mathbb{W},
\end{align}
where $X_{t,n}^{(m)}$ is $X_{t,n}$ with $a_{s}=a_{s}^{(m)}$, and $u_{j} \in [0,1]$ for $1 \leq j \leq 2l+2$. Define
\begin{align*}
    \xi_{t,n}^{(m)} = v_{m}^{-1}(X_{t,n}^{(m)} + \frac{a_{t}^{(m)}}{\sqrt{n}}\epsilon_{t}).
\end{align*}
Note that $\{(\xi_{t,n}^{(1)}, \ldots, \xi_{t,n}^{(2l+2)})^{\top}, \mathcal{F}_{t}\}$ is an m.d.s. By Theorem 3.3 of \cite{helland1982}, \eqref{LemmaC1-5} follows if
\begin{align}
    \sum_{t=1}^{\lfloor nu \rfloor} \E( {\xi_{t,n}^{(m)}}^{2} | \mathcal{F}_{t-1} ) \stackrel{p}{\rightarrow} u, \; 1 \leq m \leq 2l+2, \label{LemmaC1-6} \\
    \sum_{t=1}^{\lfloor nu \rfloor} \E( {|\xi_{t,n}^{(m)}}|^{2+c} | \mathcal{F}_{t-1} ) \stackrel{p}{\rightarrow} 0, \; 1 \leq m \leq 2l+2, \label{LemmaC1-7}
\end{align}
for some $c>0$, and
\begin{align}
    \sum_{t=1}^{\lfloor nu \rfloor}\E(\xi_{t,n}^{(k)}\xi_{t,n}^{(m)}|\mathcal{F}_{t-1}) \stackrel{p}{\rightarrow} 0, \; 1 \leq k \neq m \leq 2l+2. \label{LemmaC1-8}
\end{align}

In the following, we only prove \eqref{LemmaC1-6} because \eqref{LemmaC1-8} can be proved similarly and \eqref{LemmaC1-7} follows directly from \eqref{ingsep13} in Section \ref{sec_further}.
Let $\zeta_{t,n}^{(m)}=v_{m}\xi_{t,n}^{(m)}$. Then,
\begin{align}
\begin{split} \label{LemmaC1-9}
v_{m}^{2} \sum_{t=1}^{\lfloor nu \rfloor} \E( {\xi_{t,n}^{(m)}}^{2} | \mathcal{F}_{t-1} ) =& \sum_{t=1}^{\lfloor nu \rfloor} \E({\zeta_{t,n}^{(m)}}^{2})
- \sum_{t=1}^{\lfloor nu \rfloor} \big\{{\zeta_{t,n}^{(m)}}^{2}-\E({\zeta_{t,n}^{(m)}}^{2}|\mathcal{F}_{t-1})\big\}\\
& + \sum_{t=1}^{\lfloor nu \rfloor}\big\{{\zeta_{t,n}^{(m)}}^{2}-\E({\zeta_{t,n}^{(m)}}^{2}) \big\}.
\end{split}
\end{align}
We show
in Section \ref{sec_further} that
\begin{align} \label{LemmaC1-10}
    \E\big\{ \sum_{t=1}^{\lfloor nu \rfloor} \zeta_{t,n}^{(m)} \big\}^{2} \leq C,
\end{align}
which, together with
\eqref{LemmaC1-2}, \eqref{ing45}, \eqref{LemmaC1-4}, Assumption (A3), \eqref{ing42}, and \eqref{ing43}, gives, for $1\leq m\leq2l+2$,
\begin{eqnarray}
\label{LemmaC1-11}
  \lim_{n \rightarrow \infty}  \sum_{t=1}^{\lfloor nu \rfloor} \E({\zeta_{t,n}^{(m)}}^{2}) =
  \lim_{n \rightarrow \infty} \E\left( \frac{1}{\sqrt{n}}\sum_{t=1}^{\lfloor nu \rfloor} a_{t}^{(m)}\delta_{t} \right)^{2} =uv_m^2.
\end{eqnarray}
Moreover, by making use of \eqref{ingsep5}, \eqref{ingsep4}, and Assumptions (A1) and (A2), we prove in Section \ref{sec_further} that
\begin{align} \label{LemmaC1-13.5}
 \big|\sum_{t=1}^{\lfloor nu \rfloor} \big\{{\zeta_{t,n}^{(m)}}^{2}-\E({\zeta_{t,n}^{(m)}}^{2}|\mathcal{F}_{t-1})\big\}\big|+
\big|\sum_{t=1}^{\lfloor nu \rfloor}\big\{{\zeta_{t,n}^{(m)}}^{2}-\E({\zeta_{t,n}^{(m)}}^{2}) \big\}\big|
     =o_{p}(1).
\end{align}
Combining \eqref{LemmaC1-13.5} with \eqref{LemmaC1-11} and \eqref{LemmaC1-9}, we have \eqref{LemmaC1-6}. Thus, the proof is complete.
\end{proof}

\begin{proof}[\bf Proof of Theorem \ref{LemmaS1}]
    In the following, we denote by $C_{r}$ any generic absolute constant depending only on $r$ whose value
may vary at different places. By some straightforward algebra, it is not difficult to see
\begin{align*}
    \E|Q_{n}-\E Q_{n}|^{2r} = \E\left\vert \sum_{l=-\infty}^{n}\sum_{k=-\infty}^{n} c_{k,l} \tilde{x}_{l}\tilde{y}_{k} - \E Q_{n} \right\vert^{2r},
\end{align*}
where $c_{k,l} = \sum_{i= l \vee 1}^{n}\sum_{j=k \vee 1}^{n} q_{i,j}a_{i,l}b_{j,k}$. Then observe that
\begin{align*}
    \E\left\vert \sum_{l=-\infty}^{n}\sum_{k=-\infty}^{n} c_{k,l} \tilde{x}_{l}\tilde{y}_{k} - \E Q_{n} \right\vert^{2r} \leq&
    C_{r} \left\{ \E\left| \sum_{l=-\infty}^{n}c_{l,l}(\tilde{x}_{l}\tilde{y}_{l}-\E(\tilde{x}_{l}\tilde{y}_{l})) \right|^{2r} \right. \\
    &+ \E\left| \sum_{k=-\infty}^{n}\left( \sum_{l=-\infty}^{k-1}c_{k,l}\tilde{x}_{l} \right) \tilde{y}_{k} \right|^{2r} \\
    &+ \left. \E\left| \sum_{l=-\infty}^{n}\left( \sum_{k=-\infty}^{l-1}c_{k,l}\tilde{y}_{k} \right)\tilde{x}_{l} \right|^{2r} \right\} \\
    :=& C_{r}\{{\rm (I)+(II)+(III)}\}.
\end{align*}
By Burkholder's inequality, Minkowski's inequality, Cauchy-Schwart inequality, and the fact that $\sup_{t}\E \Vert \tilde{\mathbf{z}}_{t} \Vert^{2r} \leq C$, we have
\begin{align}
    {\rm (I)} = \E\left| \sum_{j=-\infty}^{n}\left(\sum_{l=j}^{n}c_{l,l} \bm{\theta}_{l-j}^{\top} \right) \tilde{\mathbf{z}}_{j} \right|^{2r} 
    \leq& C_{r} \left\{ \sum_{j=-\infty}^{n} \left\Vert \left( \sum_{l=j}^{n} c_{l,l} \bm{\theta}_{l-j}^{\top} \right) \tilde{\mathbf{z}}_{j} \right\Vert^{2}_{2r} \right\}^{r}    \notag \\
    \leq& C_{r}\left\{ \sum_{j=-\infty}^{n} \left\Vert \sum_{l=j}^{n}c_{l,l} \bm{\theta}_{l-j} \right\Vert^{2} \right\}^{r}.
\label{LemmaS1-1}
\end{align}
Note that
\begin{align*}
    \sum_{j=-\infty}^{n} \left\Vert \sum_{l=j}^{n}c_{l,l} \bm{\theta}_{l-j} \right\Vert^{2} = \sum_{l=-\infty}^{n}
    \sum_{k=-\infty}^{n} c_{l,l}c_{k,k} \sum_{j=0}^{\infty} \bm{\theta}_{j}^{\top}\bm{\theta}_{j+|l-k|} \leq C
    \left( \sum_{j=0}^{\infty} \Vert \bm{\theta}_{j} \Vert \right)^{2}\left( \sum_{l=-\infty}^{n}c_{ll}^{2} \right). 
\end{align*}
This, combined with \eqref{LemmaS1-1} and $\sum_{j=0}^{\infty} \Vert \bm{\theta}_j \Vert \leq C$, yields
\begin{align}
    {\rm (I)} \leq C_{r} \left\{ \sum_{l=-\infty}^{n}c_{l,l}^{2} \right\}^{r}. \label{LemmaS1-3}
\end{align}
Next, by Burkholder's inequality, Minkowski's inequality, H\"{o}lder's inequality, and \eqref{LemmaS1-0-1} in the main paper, we have
\begin{align}
    {\rm (II)} \leq& \left\{ C_{r} \left\Vert \sum_{k=-\infty}^{n} \left( \sum_{l=-\infty}^{k-1} c_{k,l}\tilde{x}_{l} \right)^{2} \tilde{y}_{k}^{2} \right\Vert_{r} \right\}^{r} \notag \\
    \leq& C_{r} \left\{ \sum_{k=-\infty}^{n} \left\Vert \sum_{l=-\infty}^{k-1}c_{k,l}\tilde{x}_{l} \right\Vert_{p}^{2} \Vert \tilde{y}_{k} \Vert_{q}^{2} \right\}^{r} \notag \\
    \leq& C_{r}\left\{ \sum_{k=-\infty}^{n} \sum_{l=-\infty}^{k-1}c_{k,l}^{2} \right\}^{r} \label{LemmaS1-4}.
\end{align}
By a similar argument, it can be shown that
\begin{align*}
    {\rm (III)} \leq C_{r} \left\{ \sum_{l=-\infty}^{n}\sum_{k=-\infty}^{l-1}c_{k,l}^{2} \right\}^{r}.
\end{align*}
This, in conjunction with \eqref{LemmaS1-3} and \eqref{LemmaS1-4}, yields the desired conclusion.
\end{proof}

Theorems \ref{Theorem_FCLT} and \ref{LemmaS1} can be used to bound from below the minimum eigenvalues of the sample covariance matrices of the candidate models; see Theorem \ref{Theorem_3.1}.
To state this result, we need to introduce some notations.

Recall $\phi(B)$ in Section 3.1.
Inspired by \cite{chan1988}, we define
\begin{align} \label{ingfeb4}
\begin{split}
    u_{t}(j) &= [(1-B)^{-j}\phi(B)]y_{t}, \\
    v_{t}(j) &= [(1+B)^{-j}\phi(B)]y_{t}, \\
  g_{t}(k,j) &= [(1-2\cos\vartheta_{k}B+B^{2})^{-j}\phi(B)]y_{t},
\end{split}
\end{align}
where $k=1,\ldots,l$.
For $k=1,\ldots,l$, it can be shown that
\begin{align*}
    g_{t}(k,1) =& \frac{1}{\sin \vartheta_{k}} \sum_{s=1}^{t}\sin[(t-s+1)\vartheta_{k}]\delta_{s} \\
    =& \sum_{s=0}^{t-1} \frac{\sin[(s+1)\vartheta_{k}]}{\sin\vartheta_{k}} \delta_{t-s} := \sum_{s=0}^{t-1} \kappa_{s}(k, 1) \delta_{t-s},
\end{align*}
where $|\kappa_{s}(k, 1)| \leq C$ for all $s \geq 0$. By induction it follows that
\begin{align*}
    g_{t}(k,j) = \sum_{s=0}^{t-1}\kappa_{s}(k,j)\delta_{t-s},
\end{align*}
where
\begin{eqnarray} \label{ingsep16}
    |\kappa_{s}(k,j)| \leq C (s+1)^{j-1},
\end{eqnarray}
for all $s \geq 0$, $1\leq k \leq l$, and $1 \leq j \leq d_{k}$. Similarly,
\begin{eqnarray*}
    u_{t}(j_1)=\sum_{s=0}^{t-1}\iota_{s}(j_1)\delta_{t-s}, \,\, v_{t}(j_2)=\sum_{s=0}^{t-1}\vartheta_{s}(j_2)\delta_{t-s},
\end{eqnarray*}
where
\begin{eqnarray} \label{ingsep17}
    |\iota_{s}(j_1)| \leq C (s+1)^{j_{1}-1}, \,\,|\vartheta_{s}(j)|  \leq C (s+1)^{j_{2}-1},
\end{eqnarray}
for all $s \geq 0$, $1 \leq j_1 \leq a$, and $1 \leq j_2 \leq b$.

Let $\mathbf{Q}_{n}$ be defined  implicitly by
\begin{align*}
    \mathbf{Q}_{n}\mathbf{w}_{t}(J) =(\mathbf{u}_{t}^{\top},\mathbf{v}_{t}^{\top},\mathbf{g}_{t}^{\top}(1),
    \ldots,\mathbf{g}_{t}^{\top}(l),\mathbf{z}_{t}^{\top}(q_{n}-d),\mathbf{x}_{t}^{\top}(J))^{\top},
\end{align*}
where
\begin{align*}
    \mathbf{u}_{t} &= (u_{t-1}(a),\ldots,u_{t-1}(1))^{\top}, \\
    \mathbf{v}_{t} &= (v_{t-1}(b),\ldots,v_{t-1}(1))^{\top},\\
    \mathbf{g}_{t}(k) &= (g_{t-1}(k,1),g_{t-2}(k,1),\ldots,g_{t-1}(k,d_{k}),g_{t-2}(k,d_{k}))^{\top}, \,\,1 \leq k \leq l,
\end{align*}
and recall that $\mathbf{w}_{t}(J) = (y_{t-1}, \ldots, y_{t-q_{n}},\mathbf{x}_{t}^{\top}(J))^{\top}$, $\mathbf{x}_{t}(J)  = (x_{t-l,j}:(j,l)\in J)^{\top} $, and $\mathbf{z}_{t}(k)=(z_{t-1},\ldots, z_{t-k})^{\top}$. 
It is not difficult to show that $\mathbf{Q}_{n}$ is nonsingular and satisfies for all $J \subset \{(j,l): 1 \leq j \leq p_{n}, 1 \leq l \leq r_{j}^{(n)}\}$,
\begin{align} \label{Q_eigenmax}
    \lambda_{\max}(\mathbf{Q}_{n}^{\top}\mathbf{Q}_{n}) \leq C.
\end{align}
It is well known that each element in $\mathbf{Q}_{n}\mathbf{w}_{t}(J)$ has certain stochastic order of magnitude \cite{chan1988, Tsay1984, ing2010}.
Thus, we may consider a normalized version,
\begin{align*}
    \mathbf{s}_{t}(J) =(\tilde{y}_{t,1},\ldots, \tilde{y}_{t,d}, \mathbf{z}_{t}^{\top}(q_{n}-d),\mathbf{x}_{t}^{\top}(J))^{\top}=G_{n}\mathbf{Q}_{n}\mathbf{w}_{t}(J),
\end{align*}
of $Q_{n}\mathbf{w}_{t}(J)$, where 
\begin{align*}
    G_{n} = \mathrm{diag}(G_{n,u},G_{n,v},G_{n,g}(1),\ldots,G_{n,g}(l),\I_{q_{n}+\sharp(J)-d}) \in \mathbb{R}^{(q_{n}+\sharp(J))\times(q_{n}+\sharp(J))},
\end{align*}
with
\begin{align*}
\begin{split}
    & G_{n,u}= \mathrm{diag}(n^{-a+1/2},\ldots,n^{-1/2}), \,\,G_{n,v}= \mathrm{diag}(n^{-b+1/2},\ldots,n^{-1/2}), \\
    &G_{n,g}(k) = \mathrm{diag}(\underbrace{n^{-1/2},n^{-1/2},n^{-3/2},n^{-3/2},\ldots,n^{-d_{k}+1/2},n^{-d_{k}+1/2}}_{2d_{k}}), \; k=1,\ldots,l.
\end{split}
\end{align*}

The following moment bounds imply useful concentration inequalities for quadratic forms involving the covariates and the lagged dependent variables that lend a helping hand throughout this paper.
These bounds can be verified using Theorem \ref{LemmaS1} and the definitions above.

\begin{lemma} \label{LemmaA3}
Assume that {\rm (A1), (A2)}, and {\rm (A4)} hold. Then,
\begin{align}
    \max_{\substack{1 \leq j_{1},j_{2} \leq p_{n} \\ 1 \leq l_{1} \leq r_{j_{1}}^{(n)}, \; 1 \leq l_{2} \leq r_{j_{2}}^{(n)}}}
    & \mathbb{E}\left\vert n^{-1/2}\sum_{t=\bar{r}_{n}+1}^{n}
    \left\{x_{t-l_{1},j_{1}}x_{t-l_{2},j_{2}} - \E(x_{t-l_{1},j_{1}}x_{t-l_{2},j_{2}}) \right\}\right\vert ^{\eta q_0}=
    O\left( 1  \right), \label{LemmaA3-1} \\
    \max_{1\leq i,j \leq q_{n}-d} &
    \mathbb{E} \left\vert n^{-1/2}\sum_{t=\bar{r}_{n}+1}^{n}\left\{
    z_{t-i}z_{t-j} - \E(z_{t-i,\infty}z_{t-j,\infty}) \right\}\right\vert^{\eta} =
    O\left(  1  \right), \label{LemmaA3-2} \\
    \max_{\substack{1 \leq j \leq p_{n}, 1 \leq l \leq r_{j}^{(n)} \\ 1 \leq k \leq q_{n}-d}} &
    \mathbb{E} \left\vert n^{-1/2}
    \sum_{t=\bar{r}_{n}+1}^{n}\left\{x_{t-l,j}z_{t-k} - \E(x_{t-l,j}z_{t-k,\infty})\right\} \right\vert^{\frac{2 \eta q_{0}}{q_{0}+1}} =
    O\left( 1  \right), \label{LemmaA3-3} \\
    \max_{\substack{1 \leq j \leq p_{n}, 1 \leq l \leq r_{j}^{(n)} \\ 1 \leq i \leq d }}
    &  \mathbb{E}
    \left\vert n^{-1/2}\sum_{t=\bar{r}_{n}+1}^{n} x_{t-l,j}
    \tilde{y}_{t, i} \right\vert^{\frac{2 \eta q_{0}}{q_{0}+1}} = O\left( 1  \right), \label{LemmaA3-4} \\
    \max_{\substack{ 1 \leq k \leq q_{n}-d\\1 \leq i \leq d}} &
    \mathbb{E}
    \left\vert n^{-1/2}\sum_{t=\bar{r}_{n}+1}^{n} z_{t-k}\tilde{y}_{t, i} \right\vert^{\eta}
    = O\left( 1  \right). \label{LemmaA3-5}
\end{align}
\end{lemma}

Now we can prove Theorem \ref{Theorem_3.1}.

\begin{proof}[\bf Proof of Theorem \ref{Theorem_3.1}]
Define $\tilde{\mathbf{y}}_{t}=(\tilde{y}_{t,1},\ldots, \tilde{y}_{t,d})^{\top}$. 
By Theorem \ref{Theorem_FCLT}, it can be shown that 
\begin{align} \label{ing5}
   \mathbf{P}_n\equiv n^{-1}\sum_{t=\bar{r}_{n}+1}^{n} \tilde{\mathbf{y}}_{t}\tilde{\mathbf{y}}^{\top}_{t} \Rightarrow \begin{pmatrix}
    v_{1}^{2}F & & &\\
    & v_{2}^{2}\tilde{F} & & & \\
    & & v_{3}^{2}H_{1} & &\\
    & & & \ddots & \\
    & & & & v_{2l+1}^{2}H_{l}
    \end{pmatrix},
\end{align}
where $F, \tilde{F}, H_{1}, \ldots, H_{l}$ are independent and almost surely nonsingular random matrices defined in Theorem 3.5.1 of \cite{chan1988}. 
Since $v_{i}^2>0$, it follows from \eqref{ing5} and the continuous mapping theorem that
\begin{align} \label{Thm3.1-1}
    \lambda_{\min}^{-1}\left( \mathbf{P}_n \right) = O_{p}(1).
\end{align}
Let
\begin{align*} 
    \mathbf{S}_{n}(J) =& \begin{pmatrix}  \mathbf{P}_n & \mathbf{0} \\
    \mathbf{0} & \mathbf{\Gamma}_{n}(J)
    \end{pmatrix}.
\end{align*}
Then,
\begin{align} \label{Thm3.1-2}
    &\min_{\sharp(J)\leq \bar{K}_{n}}\lambda_{\min}\left( n^{-1}\sum_{t=\bar{r}_{n}+1}^{n}\mathbf{s}_{t}(J)\mathbf{s}_{t}^{\top}(J) \right) \notag \\
    \geq& \min_{\sharp(J)\leq \bar{K}_{n}}\lambda_{\min}\left( \mathbf{S}_{n}(J) \right) - \max_{\sharp(J) \leq \bar{K}_{n}}
    \left\Vert n^{-1}\sum_{t=\bar{r}_{n}+1}^{n}\mathbf{s}_{t}(J)\mathbf{s}_{t}^{\top}(J) - \mathbf{S}_{n}(J) \right\Vert.
\end{align}
Note that if
\begin{align}
    \max_{\sharp(J) \leq \bar{K}_{n}}
      \left\Vert n^{-1}\sum_{t=\bar{r}_{n}+1}^{n}\mathbf{s}_{t}(J)\mathbf{s}_{t}^{\top}(J) - \mathbf{S}_{n}(J) \right\Vert =& O_{p}\left(
       \frac{\bar{K}_{n} p_n^{*^{\frac{2}{\eta q_0}}}}{n^{1/2}}+\frac{q_n^{1/2}\bar{K}_{n}^{1/2}p_n^{*^{\frac{q_0+1}{2\eta q_0}}}}{n^{1/2}}+\frac{q_n}{n^{1/2}} \right) \notag \\
       =& o_{p}(1), \label{ingjan2}
\end{align}
then \eqref{ing3} in the main paper follows from \eqref{Thm3.1-1}--\eqref{ingjan2} and (A6).

To show \eqref{ingjan2}, note first that by \eqref{LemmaA3-1}, 
\begin{align} \label{ingjan3}
\begin{split}
    & \max_{\sharp(J) \leq \bar{K}_{n}} \sqrt{\sum_{(l_2, j_2)\in J}
      \sum_{(l_1, j_1) \in J} \big(n^{-1}\sum_{t=\bar{r}_{n}+1}^{n}
      \left\{x_{t-l_{1},j_{1}}x_{t-l_{2},j_{2}} - \E(x_{t-l_{1},j_{1}}x_{t-l_{2},j_{2}}) \right\}\big)^{2}
      }\\
    &\leq \bar{K}_{n}
     \max_{\substack{1 \leq j_{1},j_{2} \leq p_{n} \\ 1 \leq l_{1} \leq r_{j_{1}}^{(n)}, \; 1 \leq l_{2} \leq r_{j_{2}}^{(n)}}}
     \left\vert n^{-1}\sum_{t=\bar{r}_{n}+1}^{n}
     \left\{x_{t-l_{1},j_{1}}x_{t-l_{2},j_{2}} - \E(x_{t-l_{1},j_{1}}x_{t-l_{2},j_{2}}) \right\}\right\vert \\
    & = O_{p}\big(\frac{\bar{K}_{n} p_n^{*^{\frac{2}{\eta q_0}}}}{n^{1/2}}\big).
\end{split}
\end{align}
In addition, \eqref{LemmaA3-2} and \eqref{LemmaA3-3} ensure
\begin{align} \label{ingjan4}
    \sqrt{\sum_{i=1}^{q_n-d} \sum_{j=1}^{q_n-d} \big(n^{-1}\sum_{t=\bar{r}_{n}+1}^{n}
      \left\{z_{t-i}z_{t-j} - \E(z_{t-i, \infty}z_{t-j,\infty}) \right\}\big)^{2}
      }  = O_{p}\big(\frac{q_n}{n^{1/2}}\big),
\end{align}
and
\begin{align} \label{ingjan5}
\begin{split}
    & \sqrt{\sum_{k=1}^{q_n-d} \max_{\sharp(J) \leq \bar{K}_{n}} \sum_{(j, l) \in J} \big(n^{-1}\sum_{t=\bar{r}_{n}+1}^{n}
    \left\{x_{t-l,j}z_{t-k} - \E(x_{t-l,j}z_{t-k, \infty}) \right\}\big)^{2}} \\
    & = O_{p}\big(\frac{q^{1/2}_n \bar{K}_{n}^{1/2}p_n^{*^{\frac{q_0+1}{2 \eta q_0}}}}{n^{1/2}}\big),
\end{split}
\end{align}
respectively. Similarly, \eqref{LemmaA3-4} and \eqref{LemmaA3-5} imply
\begin{align} \label{ingjan6}
\begin{split}
    \sqrt{\sum_{i=1}^{d} \max_{\sharp(J) \leq \bar{K}_{n} } \sum_{(j, l) \in J} \big(n^{-1}\sum_{t=\bar{r}_{n}+1}^{n}
    x_{t-l,j}\tilde{y}_{t,i} \big)^{2}}
    = O_{p}\big(\frac{\bar{K}_{n}^{1/2}p_n^{*^{\frac{q_0+1}{2 \eta q_0}}}}{n^{1/2}}\big)
\end{split}
\end{align}
and
\begin{align} \label{ingjan7}
\begin{split}
    \sqrt{\sum_{i=1}^{d} \sum_{k=1}^{q_n-d} \big(n^{-1}\sum_{t=\bar{r}_{n}+1}^{n}
    z_{t-k}\tilde{y}_{t,i} \big)^{2}}
    = O_{p}\big(\frac{q_n^{1/2}}{n^{1/2}}\big).
\end{split}
\end{align}
Combining \eqref{ingjan3}--\eqref{ingjan7} yields the first equality of \eqref{ingjan2}. The second equality of \eqref{ingjan2} is ensured by (A5) and that $\bar{K}_{n} = o(n^{1/2}/{p_{n}^{*}}^{\bar{\theta}})$.

Finally, by noticing
\begin{align*}
    &\lambda_{\min}\left( n^{-1} \sum_{t=\bar{r}_{n}+1}^{n} \mathbf{w}_{t}(J)\mathbf{w}_{t}^{\top}(J) \right) \\
    \geq& \lambda_{\min}\left( n^{-1}\sum_{t=\bar{r}_{n}+1}^{n}\mathbf{s}_{t}(J)\mathbf{s}_{t}^{\top}(J) \right)
    \lambda_{\max}^{-1}( \mathbf{Q}_{n}^{\top} \mathbf{Q}_{n} )\lambda_{\min}\left( \mathbf{G}_{n}^{-2} \right),
\end{align*}
\eqref{ww'eigenvalue} in the main paper follows from \eqref{ing3}, $\lambda_{\min}(\mathbf{G}_{n}^{-2}) = 1$, and \eqref{Q_eigenmax}.
\end{proof}

As the above proof shows, the delicacy of Theorem \ref{Theorem_3.1} lies in the fact that $ \mathbf{P}_{n} := n^{-1}\sum_{t=\bar{r}_{n}+1}^{n} \tilde{\mathbf{y}}_{t}\tilde{\mathbf{y}}^{\top}_{t}$ does not converge in probability to a deterministic limit; hence the analysis of its minimum eigenvalue is much more involved. 
The problem is resolved through the novel FCLT developed in Theorem \ref{Theorem_FCLT} that ensures $\mathbf{P}_n$'s weak limit exists and is almost surely positive definite. Consequently, as long as the size of a candidate model is equal to (or less than) $K_{n}+q_n$, Theorem \ref{Theorem_3.1} guarantees that the corresponding sample covariance matrix is well-behaved.

\subsection{Proofs of Theorems \ref{Theorem_3.2}--\ref{corr}} \label{theorems}

We first introduce the weak noiseless FSR, a device used in proving a convergence for the FSR.
Define
\begin{align*}
    \psi_{J,(i,l)} = \frac{n^{-1} \bm{\mu}_{n}^{\top}(\I - \OPM_{[q_{n}]\oplus J})\bm{x}_{l}^{(i)} }{ (n^{-1} {\bm{x}_{l}^{(i)}}^{\top}
    (\I - \OPM_{[q_{n}]\oplus J})\bm{x}_{l}^{(i)} )^{1/2} } \quad \mbox{and } \hat{\psi}_{J,(i,l)} = \frac{n^{-1} \mathbf{y}_{n}^{\top}
    (\I - \OPM_{[q_{n}]\oplus J})\bm{x}_{l}^{(i)} }{ (n^{-1} {\bm{x}_{l}^{(i)}}^{\top} (\I - \OPM_{[q_{n}]\oplus J})\bm{x}_{l}^{(i)} )^{1/2} }.
\end{align*}
The main distinction between 
$\psi_{J,(i,l)}$ and $\hat{\psi}_{J,(i,l)}$ is that the $\mathbf{y}_{n}$ in the latter is substituted with its noiseless counterpart $\bm{\mu}_{n}$ in the former.
Recall that $\hat{J}_{0}=\emptyset$. FSR chooses
\begin{eqnarray*}
(\hat{j}_m, \hat{l}_m)= \arg\max_{(j,l) \in \bar{J} \setminus \hat{J}_{m-1}} \hat{\psi}_{\hat{J}_{m-1},(i,l)}, \,\,m\geq 1,
\end{eqnarray*}
at the $m$-th iteration, and then updates $\hat{J}_{m-1}$ by $\hat{J}_m=\hat{J}_{m-1} \cup \{(\hat{j}_m, \hat{l}_m)\}$.
In contrast, the weak noiseless FSR selects $(j_{m},l_{m})$ satisfying
\begin{align} \label{selectionrule}
    |\psi_{J_{m-1},(j_m,l_m)}| \geq \xi \max_{(j,l) \in \bar{J} \setminus {J}_{m-1}} |\psi_{J_{m-1},(j,l)}|, \quad m \geq 1,
\end{align}
where $J_0 = \emptyset$ and $0 < \xi \leq 1$ is a constant. 
It subsequently updates $J_{m-1}$ by $J_{m}=J_{m-1} \cup \{(j_{m},l_{m})\}$. 

The performance of the weak noiseless FSR is evaluated by the ``noiseless" mean squared error $\hat{a}_m$ defined in \eqref{ingjune231}.
In \eqref{L1} of Section \ref{Appendix_C2}, we derive a convergence rate of $\hat{a}_m$ as $m$ increases.
The convergence of $\hat{a}_m$, along with Theorem \ref{Theorem_3.1}, enables us to establish
in \eqref{ing12} the convergence rate of  $\hat{a}_m$'s semi-noiseless counterpart,
$\hat{s}_m$, defined in \eqref{ingjune232}. As we will see later, \eqref{ing12} serves as the key vehicle for us to develop the sure-screening property of $\hat{J}_{m}$.

\begin{proof}[\bf Proof of Theorem \ref{Theorem_3.2}]
By \eqref{SS_0}, there exists $l_n \to \infty$ such that
\begin{eqnarray*}
    \frac{l_n s_0 p_n^{*^{2\bar{\theta}}}}{n\min_{\substack{(j, l) \in {\cal J}_n}} {\beta_{l}^{(j)}}^2}=o(1).
\end{eqnarray*}
Define
\begin{align*}
    \mathcal{A}_{n}(K_n) = \left\{ \max_{ \substack{\sharp(J) \leq K_n - 1 \\ (i,l) \notin J} } | \psi_{J,(i,l)} - \hat{\psi}_{J,(i,l)}| \leq 
    \frac{l_n^{1/2} p_{n}^{*^{(q_0+1)/(2\eta q_0)}}}{n^{1/2}}  \right\}
\end{align*}
and
\begin{align*}
    \mathcal{B}_{n}(K_n) = \left\{ \min_{0 \leq m \leq K_n - 1 } \max_{ (j,l) \notin \hat{J}_{m} } |\psi_{\hat{J}_{m},(j,l)}| > \tilde{\xi}
    \frac{l_n^{1/2}p_{n}^{*^{(q_0+1)/(2\eta q_0)}}}{n^{1/2}}   \right\},
\end{align*}
where $\tilde{\xi}>2$ is some constant. On $\mathcal{A}_{n}(K_n) \cap \mathcal{B}_{n}(K_n)$, 
it holds that for all $1 \leq m \leq K_n$, 
\begin{align} \label{ingaug4}
\begin{split}
    & |\psi_{\hat{J}_{m-1},(\hat{j}_{m},\hat{l}_{m})}| \geq -| \hat{\psi}_{\hat{J}_{m-1},(\hat{j}_{m},\hat{l}_{m})}
      - \psi_{\hat{J}_{m-1},(\hat{j}_{m},\hat{l}_{m})}| + |\hat{\psi}_{\hat{J}_{m-1},(\hat{j}_{m},\hat{l}_{m})}|  \\
    &\geq - \max_{\substack{ \sharp(J)\leq m-1 \\ (j,l) \notin J}}|\hat{\psi}_{J,(j,l)} - \psi_{J,(j,l)}| + \max_{(j,l) \notin \hat{J}_{m-1}} |\hat{\psi}_{\hat{J}_{m-1},(j,l)}|  \\
    &\geq -2  l_n^{1/2} p_{n}^{*^{(q_0+1)/(2\eta q_0)}}n^{-1/2} + \max_{(j,l) \notin \hat{J}_{m-1}} |\psi_{\hat{J}_{m-1},(j,l)}| \\
    &\geq \xi \max_{(j,l) \notin \hat{J}_{m-1}} |\psi_{\hat{J}_{m-1},(j,l)}|,
\end{split}
\end{align}
where $0<\xi=1-2/\tilde{\xi}<1$. 
By \eqref{ingaug4}
and \eqref{L1}, we show in Section \ref{Appendix_C2} that for all $1\leq m \leq K_n$,  
\begin{align} \label{ingaug3}
    \hat{s}_{m} \mathbb{I}_{\mathcal{A}_{n}(K_n) \cap \mathcal{B}_{n}(K_n)}
    \leq C_{n} {\rm exp}(-m\xi^2D_n/s_0),
\end{align}
where
\begin{align}
\begin{split} \label{ing16}
    & C_n=n^{-1}\sum_{t=\bar{r}_n+1}^{n}v_{t,n}^2, \\
    & D_{n} = \frac{\min_{1\leq \sharp(J)\leq K_n}\lambda_{\min}(n^{-1}\sum_{t=\bar{r}_n+1}^{n}\mathbf{w}_t({\cal J}_n\cup J)
      \mathbf{w}^{\top}_t({\cal J}_n\cup J))} {\max_{(j, l) \in {\cal J}_n}n^{-1}\|\bm{x}_{l}^{(j)}\|^2},
\end{split}
\end{align}
recalling that $v_{t, n}$ is defined in \eqref{ingfeb31}. We also show in Section \ref{Appendix_C2} that for all $1\leq m \leq K_n$,
\begin{eqnarray} \label{ing11}
    \hat{s}_{m} \mathbb{I}_{\mathcal{B}^{c}_{n}(K_n)}
    \leq \frac{s_0 l_n \tilde{\xi}^2p_n^{*^{(q_0+1)/(\eta q_0)}}}{n D_n},
\end{eqnarray}
and 
\begin{align} \label{Thm3.2-AppB}
    \lim_{n \rightarrow\infty} P(\mathcal{A}_{n}^{c}(K_{n})) = 0.
\end{align}
By (A4) and \eqref{SS_0},
\begin{eqnarray} \label{ing19}
    s_0=o((n/p_n^{*^{2\bar{\theta}}})^{1/3}),
\end{eqnarray}
which, together with \eqref{ingsep18}, yields $s_0u_n\log n=o(K_n)$ for some $u_n \to \infty$. 
It follows from \eqref{ww'eigenvalue}, (A2), (A4), and \eqref{ing19} that
\begin{eqnarray} \label{ingaug5}
    C_n=O_{p}(1)\,\,\mbox{and}\,\, D^{-1}_n=O_{p}(1).
\end{eqnarray}
According to \eqref{ingaug3}--\eqref{Thm3.2-AppB} and \eqref{ingaug5},
\begin{eqnarray} \label{ing12}
    \max_{1\leq m \leq K_n} \frac{\hat{s}_{m}}
    {{\rm exp}(-m\xi^2D_n/s_0)+ (s_0 l_n p_n^{*^{(q_0+1)/(\eta q_0)}}/n)}=O_p(1).
\end{eqnarray}

Let $\tilde{m}_{n}=s_0 u_n \log n$. The second equation of \eqref{ingaug5} implies for any $\bar{M}>0$,
\begin{eqnarray*}
    {\rm exp}(-\tilde{m}_n\xi^2D_n/s_0)=O_{p}(n^{-\bar{M}}),
\end{eqnarray*}
and hence \eqref{ing12} leads to
\begin{eqnarray} \label{ing14}
    \hat{s}_{\tilde{m}_{n}}=
    O_{p}(s_0 l_n p_n^{*^{(q_0+1)/(\eta q_0)}}/n).
\end{eqnarray}
On the set $\{\mathcal{J}_{n} \not\subseteq \hat{J}_{\tilde{m}_{n}}\}$, one has
\begin{align} \label{ing13}
\begin{split}
   & \hat{s}_{\tilde{m}_{n}}
   \geq \lambda_{\min}\left( n^{-1}\sum_{t=\bar{r}_{n}+1}^{n}\mathbf{w}_{t}({\cal J}_n\cup \hat{J}_{\tilde{m}_n})
    \mathbf{w}_{t}^{\top}({\cal J}_n\cup \hat{J}_{\tilde{m}_n})\right)
    \min_{(j,l)\in \mathcal{J}_{n}} |\beta_{l}^{(j)}|^{2}\\
   &\geq \min_{\sharp(J) \leq K_{n}} \lambda_{\min}\left( n^{-1}\sum_{t=\bar{r}_{n}+1}^{n}\mathbf{w}_{t}(J)\mathbf{w}_{t}^{\top}(J) \right)
    \min_{(j,l)\in \mathcal{J}_{n}} |\beta_{l}^{(j)}|^{2}.
\end{split}
\end{align}
Combining \eqref{ing13}, \eqref{ing14}, \eqref{ww'eigenvalue}, and \eqref{SS_1} leads to
\begin{align} \label{ing18}
   P(\mathcal{J}_{n} \not\subseteq \hat{J}_{K_{n}}) \leq   P(\mathcal{J}_{n} \not\subseteq \hat{J}_{\tilde{m}_{n}})
    \leq P\left(O_{p}(s_0 l_n p_n^{*^{(q_0+1)/(\eta q_0)}}/n) \geq \min_{(j,l)\in \mathcal{J}_{n}} |\beta_{l}^{(j)}|^{2} \right) = o(1).
\end{align}
Thus, the desired conclusion \eqref{ing15} follows.    
\end{proof}

\begin{proof}[\bf Proof of Theorem \ref{Theorem_3.3}]
Let $\tilde{k}_{n} = \min\{ 1 \leq k \leq K_{n}: \mathcal{J}_{n} \subseteq \hat{J}_{k}\}$ if $\mathcal{J}_{n} \subseteq \hat{J}_{K_{n}}$ and $K_{n}$ if $\mathcal{J}_{n} \not\subseteq \hat{J}_{K_{n}}$.
We start by showing that
\begin{align} \label{Thm3.3_1}
    \lim_{n \rightarrow \infty} P(\hat{k}_{n}=\tilde{k}_{n}) = 1.
\end{align}
By Theorem \ref{Theorem_3.2}, \eqref{ing32} is an immediate consequence of \eqref{Thm3.3_1}. In the rest of the proof, we suppress the dependence on $n$ and write $\hat{k}$ and $\tilde{k}$ instead of $\hat{k}_{n}$ and $\tilde{k}_{n}$. Let $\tilde{m}_n^*=s_0 \log n \min\{u_n, (d_n/\log n)^{\delta}\}$, where $u_n$ is defined after \eqref{ing19} and $\delta$ is defined in \eqref{penalty1}. By an argument similar to that used to prove \eqref{ing18}, it holds that
\begin{eqnarray} \label{ingfeb15}
    \lim_{n \rightarrow \infty} P(\mathcal{D}_{n}) = 1,
\end{eqnarray}
where $\mathcal{D}_{n} = \{\mathcal{J}_{n} \subseteq \hat{J}_{\tilde{m}^*_{n}}\} = \{\tilde{k}_{n} \leq \tilde{m}^*_{n}\}$.
Therefore, \eqref{Thm3.3_1} is ensured by
\begin{eqnarray} \label{ing25}
    P(\hat{k}<\tilde{k}, \mathcal{D}_{n} ) = o(1)
\end{eqnarray}
and
\begin{eqnarray} \label{ing26}
    P(\tilde{k}<\hat{k}, \mathcal{D}_{n}) = o(1).
\end{eqnarray}

By the definition of HDIC,
\begin{align} \label{Thm4.2-1}
    \hat{\sigma}_{M_{\tilde{k}-1}}^{2} - \hat{\sigma}_{M_{\tilde{k}}}^{2} \leq&
    \hat{\sigma}_{M_{\tilde{k}}}^{2}\left\{ \exp\left(\frac{(\tilde{k}- \hat{k})w_{n, p_n}}{n}\right)-1\right\} \,\,\,\,\mbox{on}\,\, \{\hat{k}<\tilde{k}\},
\end{align}
where $M_{k}=[q_{n}]\oplus\hat{J}_{k}$. Straightforward calculations give
\begin{align} \label{Thm4.2-2}
\begin{split}
    & \hat{\sigma}_{M_{\tilde{k}-1}}^{2} - \hat{\sigma}_{M_{\tilde{k}}}^{2}  \\
    &= n^{-1}(\beta_{\hat{l}_{\tilde{k}}}^{(\hat{j}_{\tilde{k}})}\bm{x}_{\hat{l}_{\tilde{k}}}^{(\hat{j}_{\tilde{k}})} +
    \bm{\varepsilon}_{n})^{\top}(\OPM_{M_{\tilde{k}}}-\OPM_{M_{\tilde{k}-1}})
    (\beta_{\hat{l}_{\tilde{k}}}^{(\hat{j}_{\tilde{k}})}\bm{x}_{\hat{l}_{\tilde{k}}}^{(\hat{j}_{\tilde{k}})} + \bm{\varepsilon}_{n}) \\
    &= {\beta_{\hat{l}_{\tilde{k}}}^{(\hat{j}_{\tilde{k}})}}^{2}\hat{A}_{n} + 2 \beta_{\hat{l}_{\tilde{k}}}^{(\hat{j}_{\tilde{k}})} \hat{B}_{n} + \hat{A}_{n}^{-1}\hat{B}_{n}^{2}
    \,\,\,\,\mbox{on} \,\, \mathcal{D}_{n},
\end{split}
\end{align}
in which
\begin{align*}
    \hat{A}_{n} =& n^{-1} {\bm{x}_{\hat{l}_{\tilde{k}}}^{(\hat{j}_{\tilde{k}})}}^{\top} (\OPM_{M_{\tilde{k}}}-\OPM_{M_{\tilde{k}-1}})\bm{x}_{\hat{l}_{\tilde{k}}}^{(\hat{j}_{\tilde{k}})},  \\
    \hat{B}_{n} =& n^{-1} {\bm{x}_{\hat{l}_{\tilde{k}}}^{(\hat{j}_{\tilde{k}})}}^{\top}(\OPM_{M_{\tilde{k}}}-\OPM_{M_{\tilde{k}-1}})\bm{\varepsilon}_{n}.
\end{align*}
In view of \eqref{Thm4.2-1}, \eqref{Thm4.2-2}, and
\begin{eqnarray} \label{ingfeb12}
\frac{w_{n, p_n}\tilde{m}^*_{n}}{n} =o(\min_{\substack{(j, l) \in {\cal J}_n}} (\beta_{l}^{(j)})^2)=o(1)
\end{eqnarray}
(which is ensured by \eqref{SS_1}, the second part of \eqref{penalty1}, and the definition of $\tilde{m}^*_{n}$), we have for all large $n$,
\begin{align} \label{Thm4.2-3}
    {\beta_{\hat{l}_{\tilde{k}}}^{(\hat{j}_{\tilde{k}})}}^{2}\hat{A}_{n} + 2 \beta_{\hat{l}_{\tilde{k}}}^{(\hat{j}_{\tilde{k}})} \hat{B}_{n} + \hat{A}_{n}^{-1}\hat{B}_{n}^{2}
    \leq (\hat{C}_{n}+\sigma^{2}) \lambda_1 \tilde{m}^*_{n}w_{n, p_n}/n\,\,\,\,\mbox{on} \,\, \mathcal{D}_{n} \cap \{\hat{k}<\tilde{k}\},
\end{align}
where $\hat{C}_{n} = \hat{\sigma}_{M_{\tilde{k}}}^{2}-\sigma^{2}$ and $\lambda_1>1$ is some constant.
In addition, by making use of Theorem \ref{Theorem_3.1} and Lemma \ref{LemmaA3}, we show in Section \ref{Appendix_C2} that
\begin{align}
    \hat{A}_{n}^{-1} =& O_{p}(1), \label{Thm4.2-4} \\
    |\hat{B}_{n}| =& O_{p}\left( \left(\frac{p_{n}^{*^{\frac{q_0+1}{2\eta q_0}}}}{n^{1/2}} \right) \right), \label{Thm4.2-5} \\
    |\hat{C}_{n}| =& O_{p}\left( \frac{n^{1/2}+q_{n}+\tilde{m}_{n}^{*}p_{n}^{*^{ \frac{q_0+1}{\eta q_0}}}}{n} \right)=o_{p}(1). \label{Thm4.2-6}
\end{align}
As a result, \eqref{ing25} follows from \eqref{ingfeb12}--\eqref{Thm4.2-6}.

On the other hand,
\begin{align} \label{Thm4.2-8}
    \hat{\sigma}^{2}_{M_{\tilde{k}}} - \hat{\sigma}^{2}_{M_{\hat{k}}} \geq \hat{\sigma}^{2}_{M_{\tilde{k}}}\{ 1-\exp(n^{-1}w_{n, p_n}(\tilde{k}-\hat{k}) )\}
    \,\,\mbox{on} \,\, \{\tilde{k}<\hat{k}\}.
\end{align}
Let $F_{\hat{k},\tilde{k}} = (\bm{x}_{l}^{(j)}:(j,l)\in \hat{J}_{\hat{k}}\cap\hat{J}_{\tilde{k}}^{c})$. Then on $\{\tilde{k}<\hat{k}\} \cap \mathcal{D}_{n}$,
\begin{align} \label{Thm4.2-9}
\begin{split}
    & \hat{\sigma}_{M_{\tilde{k}}}^{2} - \hat{\sigma}_{M_{\hat{k}}}^{2} = n^{-1}\bm{\varepsilon}_{n}^{\top}(\OPM_{M_{\hat{k}}}-\OPM_{M_{\tilde{k}}})\bm{\varepsilon}_{n}  \\
    &\leq 2 \left\Vert\left\{ n^{-1}F_{\hat{k},\tilde{k}}^{\top}(\I-\OPM_{M_{\tilde{k}}})F_{\hat{k},\tilde{k}} \right\}^{-1} \right\Vert \left\{ \left\Vert n^{-1}
    F_{\hat{k},\tilde{k}}^{\top}\bm{\varepsilon}_{n} \right\Vert^{2} + \left\Vert n^{-1}F_{\hat{k},\tilde{k}}^{\top} \OPM_{M_{\tilde{k}}}\bm{\varepsilon}_{n}
    \right\Vert^{2} \right\} \\
    &\leq 2 (\hat{k}-\tilde{k})(\hat{a}_{n}+\hat{b}_{n}),
\end{split}
\end{align}
where
\begin{align*}
    \hat{a}_{n} =& \lambda_{\min}^{-1}\left( n^{-1}\sum_{t=\bar{r}_{n}+1}^{n} \mathbf{w}_{t}(\hat{J}_{\hat{k}})\mathbf{w}_{t}^{\top}(\hat{J}_{\hat{k}}) \right) \max_{\substack{1 \leq j \leq p_{n} \\ 1 \leq l \leq r_{j}^{(n)}}}\left\vert n^{-1}\sum_{t=\bar{r}_{n}+1}^{n} x_{t-l,j}\epsilon_{t} \right\vert^{2}, \\
    \hat{b}_{n} =& \lambda_{\min}^{-1}\left( n^{-1}\sum_{t=\bar{r}_{n}+1}^{n} \mathbf{w}_{t}(\hat{J}_{\hat{k}})
    \mathbf{w}_{t}^{\top}(\hat{J}_{\hat{k}}) \right) \max_{\substack{\sharp(J) \leq \tilde{m}_n^{*} \\ (j,l) \notin J}} \left\vert n^{-1}
    \sum_{t=\bar{r}_{n}+1}^{n}\hat{x}_{t-l,j;J}\epsilon_{t} \right\vert^{2},
\end{align*}
with
\begin{align*}
\begin{split}
    \hat{x}_{t-l,i;J} &:= \mathbf{w}_{t}^{\top}(J) \left(\sum_{t=\bar{r}_{n}+1}^{n}\mathbf{w}_{t}(J)\mathbf{w}_{t}^{\top}(J)\right)^{-1}
    \sum_{t=\bar{r}_{n}+1}^{n}\mathbf{w}_{t}(J)x_{t-l,i}  \\
    & = \mathbf{s}_{t}^{\top}(J) \left(\sum_{t=\bar{r}_{n}+1}^{n}\mathbf{s}_{t}(J)\mathbf{s}_{t}^{\top}(J)\right)^{-1}\sum_{t=\bar{r}_{n}+1}^{n}\mathbf{s}_{t}(J)x_{t-l,i}.
\end{split}
\end{align*}
Combining \eqref{Thm4.2-8} and \eqref{Thm4.2-9}, we have
\begin{align} \label{Thm4.2-10}
    2(\hat{k}-\tilde{k})(\hat{a}_{n}+\hat{b}_{n}) \geq \hat{\sigma}_{M_{\tilde{k}}}^{2}\{ 1 - \exp(n^{-1}w_{n, p_n}(\tilde{k}-\hat{k})) \}
    \,\, \mbox{on} \,\, \{\tilde{k}<\hat{k}\}\cap {\cal D}_n.
\end{align}
With the help of the first part of \eqref{penalty1} and Theorem \ref{Theorem_3.1}, we also show in Section \ref{Appendix_C2} that for any $\delta>0$,
\begin{align} \label{Thm4.2-11}
    P\{ (\hat{k}-\tilde{k})(\hat{a}_{n}+\hat{b}_{n}) \geq& \delta [1-\exp(n^{-1}w_{n, p_n}(\tilde{k}-\hat{k}))], \tilde{k}<\hat{k} \} = o(1).
\end{align}
As a consequence of \eqref{Thm4.2-8}--\eqref{Thm4.2-11} and \eqref{Thm4.2-6}, \eqref{ing26} follows. Thus, the proof of \eqref{Thm3.3_1} is complete.
Moreover, by an argument similar to that used to prove \eqref{Thm3.3_1}, it can be shown that \eqref{ing20} holds true.
The details are omitted here for brevity.
\end{proof}

\begin{proof}[\bf Proof of Theorem \ref{corr}]
By Theorem \ref{Theorem_3.3},
\begin{eqnarray} \label{ingfeb17}
    \lim_{n \to \infty}P(\hat{s}_0=s_0)=\lim_{n \to \infty} P(\hat{\underline{s}}_0=\underline{s}_0)=1.
\end{eqnarray}
In view of \eqref{ingfeb17}, Theorem \ref{Theorem_3.3}, and \textbf{(SS$_{{\rm A}}$)}, it suffices for Theorem \ref{corr} to show that
\begin{eqnarray} \label{ingfeb18}
    \lim_{n \to \infty} P(\max_{1\leq i \leq q_n}|\hat{\alpha}_i(\mathcal{J}_n)-\alpha_i| \geq H_n )=0,
\end{eqnarray}
where $H_n$ is $\hat{H}_n$ with $\hat{s}_0$ and $\hat{\underline{s}}_0$ replaced by $s_0$ and $\underline{s}_0$, respectively.

Straightforward calculations yield
\begin{eqnarray} \label{ingfeb20}
    \max_{1\leq i \leq q_n}|\hat{\alpha}_i(\mathcal{J}_n)-\alpha_i| \leq C(J_{1,n}+ J_{2,n}+ J_{3,n}),
\end{eqnarray}
where
\begin{align*}
\begin{split}
    &J_{1, n}=\|(n^{-1}\sum_{t=\bar{r}_n+1}^{n}\mathbf{s}_{t}(\mathcal{J}_n)\mathbf{s}^{\top}_{t}(\mathcal{J}_n))^{-1}-\mathbf{S}_{n}^{-1}(\mathcal{J}_n))\|
     \|n^{-1}\sum_{t=\bar{r}_n+1}^{n}\mathbf{s}_{t}(\mathcal{J}_n)\epsilon_t\|,\\
    &J_{2, n}=n^{-1/2}\|(n^{-1}\sum_{t=\bar{r}_n+1}^{n}\tilde{\mathbf{y}}_{t}\tilde{\mathbf{y}}^{\top}_{t})^{-1}
     n^{-1}\sum_{t=\bar{r}_n+1}^{n}\tilde{\mathbf{y}}_{t}\epsilon_t\|,\\
    &J_{3, n}=\max_{1\leq i\leq q_n-d}
     |\bm{\nu}_{i}^{\top}n^{-1}\sum_{t=\bar{r}_n+1}^{n}(\mathbf{z}_{t}^{\top}(q_n-d), \mathbf{x}^{\top}_{t}(\mathcal{J}_n))^{\top}\epsilon_t|,
\end{split}
\end{align*}
where $\bm{\nu}_{i}$ is the $i$-th column (row) of $\mathbf{\Gamma}^{-1}(\mathcal{J}_n)$.

By Lemma \ref{LemmaA3}, Theorem \ref{Theorem_3.1}, \eqref{ing19}, and \eqref{ingjan13} and \eqref{ingfeb19} in Section \ref{Appendix_C2}, it can be shown that
\begin{eqnarray} \label{ingfeb25}
    J_{1, n}=O_{p}\left( \frac{(s_0+q_n)^{3/2}}{n}\right) = O_{p}\left(\frac{n^{1/2}+q_n^{3/2}}{n} \right)\,\, \mbox{and}\,\,J_{2, n}=O_{p}(n^{-1}).
\end{eqnarray}
Since \eqref{ing4} ensures
\begin{eqnarray} \label{ingfeb27}
    \max_{1\leq i\leq q_n-d}\|\bm{\nu}_{i}\|\leq C,
\end{eqnarray}
it follows from \eqref{ingjan13} that
\begin{eqnarray} \label{ingfeb26}
    J_{3, n}\leq C \|n^{-1}\sum_{t=\bar{r}_n+1}^{n}(\mathbf{z}_{t}(q_n-d), \mathbf{x}^{\top}_{t}(\mathcal{J}_n))^{\top}\epsilon_t\|=
    O_{p}\left(\frac{(s_0+q_n)^{1/2}}{n^{1/2}}\right).
\end{eqnarray}
In addition, we write
\begin{eqnarray*}
    \bm{\nu}_{i}^{\top}(\mathbf{z}^{\top}_{t}(q_n-d), \mathbf{x}^{\top}_{t}(\mathcal{J}_n))^{\top}=
    \sum_{m=0}^{\infty}(\bm{c}_{m}^{(i)})^{\top}\bm{\mu}_{t-1-m}(\mathcal{D}_0),
\end{eqnarray*}
where
$\mathcal{D}_0$ and
$\bm{\mu}_{t}(\cdot)$ are defined in \textbf{(SS$_{\rm A}$)} 
and \eqref{ingfeb29}, respectively, and $\{\bm{c}_{m}^{(i)}\}$ is a sequence of $(\underline{s}_0+1)$-dimensional vectors
depending on $\bm{\nu}_{i}$, $\{p_{m, j}\}, 1\leq j \leq p_n$, $\{b_m\}$, and $\beta_{j}^{(l)}, 1\leq j \leq r_{j}^{(n)}, 1\leq j \leq p_n$.
By \eqref{ing4} and \eqref{ingfeb27},
\begin{eqnarray*}
    \max_{1\leq i \leq q_n-d}\sum_{m=0}^{\infty}\|(c_{m,1}^{(i)}, \ldots, c_{m, \underline{s}_0+1}^{(i)})^{\top}\|^2 \equiv
    \max_{1\leq i \leq q_n-d}\sum_{m=0}^{\infty}\|\bm{c}_m^{(i)}\|^2
    \leq C,
\end{eqnarray*}
which, together with Theorem \ref{LemmaS1}, gives,
\begin{align*}
\begin{split}
    &\max_{1\leq i \leq q_n-d}
     \E\left|n^{-1/2}\sum_{t=\bar{r}_n+1}^{n} [\sum_{m=0}^{\infty}(\bm{c}_{m}^{(i)})^{\top}\bm{\mu}_{t-1-m}(\mathcal{D}_0)]\epsilon_t
     \right|^{\eta}\\
    &\leq C \max_{1\leq i \leq q_n-d}\left\{ \sum_{j=1}^{\underline{s}_0+1} (\sum_{m=0}^{\infty}{c_{m,j}^{(i)}}^2)^{1/2} \right\}^{\eta} \leq
     C(\max_{1\leq i \leq q_n-d}\sum_{m=0}^{\infty}\|\bm{c}_m^{(i)}\|^2)^{\eta/2}(\underline{s}_0+1)^{\eta/2}\\
    &\leq C\underline{s}_0^{\eta/2},
\end{split}
\end{align*}
yielding
\begin{eqnarray} \label{ingfeb30}
    J_{3, n}\leq
    \max_{1\leq i \leq q_n-d} \left|n^{-1}\sum_{t=\bar{r}_n+1}^{n}
    [\sum_{m=0}^{\infty}(\bm{c}_{m}^{(i)})^{\top}\bm{\mu}_{t-1-m}(\mathcal{D}_0)]\epsilon_t
    \right|
    =O_{p}\left(\frac{q_{n}^{1/\eta}\underline{s}_0^{1/2}}{n^{1/2}}\right).
\end{eqnarray}
Consequently,
\eqref{ingfeb18} follows from \eqref{ingfeb20}, \eqref{ingfeb25}, \eqref{ingfeb26}, and \eqref{ingfeb30}. Thus, the proof of Theorem \ref{corr} is complete.
\end{proof}

\subsection{Proofs of \eqref{LemmaC1-2}, \eqref{LemmaC1-3}, \eqref{LemmaC1-4}, \eqref{LemmaC1-10}, and \eqref{LemmaC1-13.5}} \label{sec_further}

\begin{proof}[\bf Proof of \eqref{LemmaC1-2}] 
Let
\begin{align*}
    v_{t,n}^{*} =& \sum_{j=1}^{p_{n}}\sum_{l=1}^{r_{j}} \beta_{l}^{(j)}x_{t-l,j}^{*}, \\
    x_{t,j}^{*} =& \sum_{s=1}^{t} p_{t-s,j}\pi_{s,j}.
\end{align*}
Then,
\begin{align*}
    \frac{1}{\sqrt{n}}\sum_{t=1}^{\lfloor nu \rfloor} a_{t}\delta_{t} =& H_{n}(u) + \frac{1}{\sqrt{n}}\sum_{t=1}^{\lfloor nu \rfloor} a_{t}v_{t,n}^{*} + \frac{1}{\sqrt{n}}\sum_{t=1}^{\lfloor nu \rfloor} a_{t}\epsilon_{t},
\end{align*}
where $H_{n}(u)=n^{-1/2}\sum_{t=1}^{\lfloor nu \rfloor} a_{t}(v_{t,n}-v_{t,n}^{*})$. By
changing the order of summation, we can write
\begin{align*}
    \frac{1}{\sqrt{n}}\sum_{t=1}^{\lfloor nu \rfloor} a_{t}v_{t,n}^{*} =& \sum_{s=1}^{\lfloor nu \rfloor-1} \sum_{j=1}^{p_{n}}\frac{1}{\sqrt{n}}\left[ \sum_{t=s+1}^{\lfloor nu \rfloor}a_{t} \left( \sum_{l=1}^{(t-s) \wedge r_{j}} \beta_{l}^{(j)}p_{t-l-s,j} \right)\right]\pi_{s,j} \\
    =& \sum_{s=1}^{\lfloor nu \rfloor-1} X_{s,n} - \sum_{s=1}^{\lfloor nu \rfloor-1} \sum_{j=1}^{p_{n}}\frac{1}{\sqrt{n}}\left[ \sum_{t=\lfloor nu \rfloor+1}^{n}a_{t} \left( \sum_{l=1}^{(t-s) \wedge r_{j}} \beta_{l}^{(j)}p_{t-l-s,j} \right)\right]\pi_{s,j} \\
    :=& \sum_{s=1}^{\lfloor nu \rfloor-1} X_{s,n} - G_{n}(u).
\end{align*}
Hence \eqref{LemmaC1-2} is proved with $R_{n}(u) = H_{n}(u) - G_{n}(u)$.
\end{proof}

\begin{proof}[\bf Proofs of \eqref{LemmaC1-3} and \eqref{LemmaC1-4}] 
Since $R_{n}(u)=H_{n}(u)-G_{n}(u)$, it is sufficient for
\eqref{LemmaC1-3}--\eqref{LemmaC1-4} to show
\begin{align}
    \sup_{0 \leq u \leq 1} |H_{n}(u)| =& o_{p}(1), \label{Supp-S3.2-1} \\
    \E H_{n}^{2}(u) =& o(1), 0 \leq u \leq 1, \label{Supp-S3.2-2}\\
    \E\left\{ \sup_{0 \leq u \leq 1} |G_{n}(u)|^{2\eta_{1}} \right\} =& o(1). \label{Supp-S3.2-3}
\end{align}
For \eqref{Supp-S3.2-1}, with $p_{s,j} = 0$ for $s<0$ notice that
\begin{align*}
    \sup_{u \in [0,1]}|H_{n}(u)| \leq& \sum_{j=1}^{p_{n}}\sum_{l=1}^{r_{j}}|\beta_{l}^{(j)}|\max_{1 \leq k \leq n} \left\vert \sum_{s=-\infty}^{0}\left( \frac{1}{\sqrt{n}} \sum_{t=1}^{k}a_{t}p_{t-l-s,j} \right)\pi_{s,j} \right\vert \\
    \leq& \sqrt{2} \sum_{j=1}^{p_{n}}\sum_{l=1}^{r_{j}}|\beta_{l}^{(j)}| \sum_{s=-\infty}^{0} \frac{1}{\sqrt{n}}\sum_{t=1}^{n}|p_{t-l-s,j}||\pi_{s,j}|.
\end{align*}
This and \eqref{ingsep4} imply
\begin{align} \label{Supp-S3.2-4}
    \E \sup_{u \in [0,1]}|H_{n}(u)| \leq& C \sum_{j=1}^{p_{n}}\sum_{l=1}^{r_{j}}|\beta_{l}^{(j)}| \frac{1}{\sqrt{n}} \sum_{s=-\infty}^{0}\sum_{t=1}^{n}|p_{t-l-s,j}|.
\end{align}
By \eqref{a7'} in the main paper, we can choose a sequence of positive integers $\{C_{n}\}$
such that $C_{n} \rightarrow \infty$ with $C_{n}=o(n^{1/2})$ and $(\max_{1\leq j \leq
p_{n}} r_{j}^{(n)})/C_{n} \rightarrow 0$. Then for each $1 \leq j \leq p_{n}$ and $1
\leq l \leq r_{j}^{(n)}$, it can be shown that
\begin{align*}
    \frac{1}{\sqrt{n}}\sum_{s=-\infty}^{0}\sum_{t=1}^{n}|p_{t-l-s,j}| =& \frac{1}{\sqrt{n}}\sum_{s=-l}^{\infty}\sum_{t=1}^{n}|p_{t+s,j}| \\
    \leq& \sqrt{n} \sum_{s=n-l}^{\infty}|p_{s,j}| + \frac{1}{\sqrt{n}}\sum_{s=1-l}^{C_{n}-l}|p_{s,j}|(s+l) + \frac{1}{\sqrt{n}}\sum_{s=C_{n}-l+1}^{n-l-1}|p_{s,j}|(s+l) \\
    \leq& C\left\{ \max_{1 \leq j \leq p_{n}}\sum_{s=n-\bar{r}}^{\infty}\sqrt{s}|p_{s,j}| + \frac{C_{n}}{\sqrt{n}}\max_{1 \leq j\leq p_{n}}\sum_{s=0}^{\infty}|p_{s,j}| \right. \\
    &+ \left. \sum_{s=C_{n}-l+1}^{\infty} \max_{1 \leq j\leq p_{n}} \sqrt{s}|p_{s,j}| \right\} \\
    =& o(1),
\end{align*}
where $\bar{r} = \max_{1 \leq j \leq p_{n}} r_{j}^{(n)}$ and the last equality uses \eqref{win1} in the main paper.
This, Assumption (A4), and
\eqref{Supp-S3.2-4} prove \eqref{Supp-S3.2-1}.

Next we prove \eqref{Supp-S3.2-2}. Assume, without loss of generality, that $u>0$. By Minkowski's
inequality and \eqref{ingsep4},
\begin{align}
    \E H_{n}^{2}(u) =& \E\left( \sum_{j=1}^{p_{n}}\sum_{l=1}^{r_{j}}\beta_{l}^{(j)} \sum_{s=-\infty}^{0}\left( \frac{1}{\sqrt{n}}\sum_{t=1}^{\lfloor nu \rfloor} a_{t} p_{t-l-s,j} \right)\pi_{s,j} \right)^{2} \notag \\
    \leq&\left\{ \sum_{j=1}^{p_{n}}\sum_{l=1}^{r_{j}}|\beta_{l}^{(j)}| \frac{C}{\sqrt{n}} \left[ \sum_{s=-\infty}^{0}\left( \sum_{t=1}^{\lfloor \sqrt{nu} \rfloor} a_{t}p_{t-l-s,j}\right)^{2} + \sum_{s=-\infty}^{0}\left( \sum_{t=\lfloor \sqrt{nu} \rfloor+1}^{\lfloor nu \rfloor} a_{t}p_{t-l-s,j}\right)^{2} \right]^{1/2} \right\}^{2} \notag \\
    :=& \left\{ \sum_{j=1}^{p_{n}}\sum_{l=1}^{r_{j}}|\beta_{l}^{(j)}| \frac{C}{\sqrt{n}} \left[ K_{1,n,j,l}(u) +  K_{2,n,j,l}(u) \right]^{1/2} \right\}^{2}. \label{Supp-S3.2-5}
\end{align}
Observe that if $\{h_{t}:t=1,2,\ldots\}$ is a sequence of nonnegative numbers, then
$\sum_{s=0}^{\infty}(\sum_{i=1}^{n} h_{s+i})^{2}$ $ \leq n(\sum_{s=0}^{\infty}
h_{s})^{2}$. By applying this inequality to $K_{1,n,j,l}(u)$ and $K_{2,n,j,l}(u)$, we
have
\begin{align} \label{Supp-S3.2-6}
    K_{1,n,j,l}(u) \leq C\lfloor \sqrt{nu} \rfloor\left( \max_{1 \leq j \leq p_{n}} \sum_{s=0}^{\infty} |p_{s,j}| \right)^{2} \leq C\sqrt{n}
\end{align}
and
\begin{align}
    K_{2,n,j,l}(u) =& \sum_{s=0}^{\infty}\left( \sum_{t=1}^{\lfloor nu \rfloor-\lfloor \sqrt{nu} \rfloor} a_{t+\lfloor \sqrt{nu} \rfloor}p_{s+t+\lfloor \sqrt{nu} \rfloor-l,j} \right)^{2} \notag \\
    \leq& C n\left( \max_{1 \leq j \leq p_{n}} \sum_{s=\lfloor \sqrt{nu} \rfloor-\bar{r}}^{\infty} |p_{s,j}| \right)^{2} = o(n). \label{Supp-S3.2-7}
\end{align}
\eqref{Supp-S3.2-5}--\eqref{Supp-S3.2-7} and Assumption (A4) together prove
\eqref{Supp-S3.2-2}.

Finally, we prove \eqref{Supp-S3.2-3}. Notice that
\begin{align*}
   | G_{n}(u)| \leq
   \sum_{s=1}^{\lfloor nu \rfloor-1} \sum_{j=1}^{p_{n}}\sum_{l=1}^{r_{j}}|\beta_{l}^{(j)}|
   \left|\left( \frac{1}{\sqrt{n}} \sum_{t=(l+s)\vee(\lfloor nu \rfloor+1)}^{n}a_{t} p_{t-l-s,j} \right) \pi_{s,j}\right|.
\end{align*}
Therefore, by Minkowski's inequality,
\begin{align}
    \E\sup_{u \in [0,1]}|G_{n}(u)|^{2\eta_{1}} \leq& \left\{ \sum_{j=1}^{p_{n}}\sum_{l=1}^{r_{j}}|\beta_{l}^{(j)}| \left\Vert \max_{1 \leq k \leq n-1} \left\vert \sum_{s=1}^{k}\left( \frac{1}{\sqrt{n}} \sum_{t=(l+s)\vee(k+2)}^{n} a_{t}p_{t-l-s,j} \right)\pi_{s,j} \right\vert \right\Vert_{2\eta_{1}} \right\}^{2\eta_{1}} \notag \\
    :=& \left\{ \sum_{j=1}^{p_{n}}\sum_{l=1}^{r_{j}}|\beta_{l}^{(j)}| f_{n}(j,l) \right\}^{2\eta_{1}}. \label{Supp-S3.2-8}
\end{align}
For $2\leq l\leq r_j$ and $1\leq j\leq p_n$,
we obtain from \eqref{ingsep4}, Burkholder's inequality, and Minkowski's inequality that
\begin{align*}
    (f_{n}(j,l))^{2\eta_{1}} =& n^{-\eta_{1}} \E \max_{1 \leq k \leq n-1} \left\vert \sum_{s=1}^{k}\left( \sum_{t=(l+s)\vee(k+2)}^{n}a_{t}p_{t-l-s,j} \right) \pi_{s,j} \right\vert^{2\eta_{1}} \\
    \leq& n^{-\eta_{1}}\sum_{k=1}^{n-1}\E\left\vert \sum_{s=1}^{k}\left( \sum_{t=(l+s)\vee(k+2)}^{n}a_{t}p_{t-l-s,j} \right) \pi_{s,j} \right\vert^{2\eta_{1}} \\
    \leq& Cn^{-\eta_{1}}\sum_{k=1}^{n-1}\left\{ \sum_{s=1}^{k} \left( \sum_{t=(l+s)\vee(k+2)}^{n}a_{t}p_{t-l-s,j} \right)^{2} \right\}^{\eta_{1}} \\
    \leq& C n^{-\eta_{1}} \sum_{k=1}^{l-1}\left\{ \sum_{s=1}^{l-1}\left(\max_{1\leq j \leq p_{n}}\sum_{t=l+s}^{n}|p_{t-l-s,j}|\right)^{2} \right\}^{\eta_{1}} \\
    &+ Cn^{-\eta_{1}}\sum_{k=l}^{n-1}\left\{ \sum_{s=1}^{k-l+1} \frac{1}{k+2-l-s}\left( \sum_{t=k+2}^{n}\max_{1\leq j \leq p_{n}} (t-l-s)^{1/2} |p_{t-l-s,j}| \right)^{2} \right\}^{\eta_{1}} \\
    &+ C n^{-\eta_{1}}\sum_{k=l}^{n-1}\left\{ \sum_{s=k-l+2}^{k}\left(\max_{1\leq j \leq p_{n}} \sum_{t=l+s}^{n}|p_{t-l-s,j}| \right)^{2} \right\}^{\eta_{1}} \\
    \leq& C \frac{\bar{r}^{\eta_{1}+1}+ n \log^{\eta_{1}}n + n \bar{r}^{\eta_{1}}}{n^{\eta_{1}}}.
\end{align*}
 Hence, by \eqref{a7'} in the main paper,
\begin{align}
\label{ingsep6}
    \max_{1\leq j \leq p_{n}, 2 \leq l \leq r_{j}} f_{n}(j,l) = o(1).
\end{align}
Similarly, it can be shown that
\begin{align}
\label{ingsep7}
    \max_{1\leq j \leq p_{n}} f_{n}(j,1) \leq C \frac{n (\log n)^{\eta_1}}{n^{\eta_1}}=o(1).
\end{align}
Consequently, \eqref{Supp-S3.2-3} is ensured by \eqref{Supp-S3.2-8}--\eqref{ingsep7} and Assumption (A4).
\end{proof}

\begin{proof}[\bf Proof of \eqref{LemmaC1-10}]
By Minkowski's inequality, \eqref{ingsep4}, \eqref{win1} in the main paper, and (A4),
\begin{align}
\begin{split}
\label{ingsep8}
\E\left(\sum_{t=1}^{\lfloor nu \rfloor} X_{t,n}^{(m)} \right)^{2} =&
\sum_{t=1}^{\lfloor nu \rfloor} \E \left( X_{t,n}^{(m)} \right)^{2} \\
   \leq& \frac{C}{n}
    \sum_{t=1}^{\lfloor nu\rfloor}\left(
    \sum_{j=1}^{p_n}\left| \sum_{s=t+1}^{n}a_{s}^{(m)}\sum_{l=1}^{(s-t)\wedge r_{j}} \beta_{l}^{(j)}p_{s-l-t,j} \right| \right)^{2} \\
    \leq& \frac{C}{n}
    \sum_{t=1}^{\lfloor nu\rfloor}\left(
    \sum_{j=1}^{p_n}\sum_{l=1}^{r_{j}\wedge (n-t)} |\beta_{l}^{(j)}| \sum_{s=t+l}^{n}|p_{s-l-t,j}|  \right)^{2} \leq C.
\end{split}
\end{align}
In addition, Assumption (A1) and \eqref{ingsep5} ensure
\begin{eqnarray*}
\E(\sum_{t=1}^{\lfloor nu \rfloor} \frac{a_{t}^{(m)}\epsilon_{t}}{\sqrt{n}})^2\leq C,
\end{eqnarray*}
which, together with \eqref{ingsep8}, yields \eqref{LemmaC1-10}.
\end{proof}

\begin{proof}[\bf Proof of \eqref{LemmaC1-13.5}] 
Equation \eqref{LemmaC1-13.5} follows from
\begin{eqnarray}
\label{ingsep9}
 \sum_{t=1}^{\lfloor nu \rfloor} \big\{{\zeta_{t,n}^{(m)}}^{2}-\E({\zeta_{t,n}^{(m)}}^{2}|\mathcal{F}_{t-1})\big\}=o_{p}(1)
\end{eqnarray}
and
\begin{eqnarray}
\label{ingsep10}
\sum_{t=1}^{\lfloor nu \rfloor}\big\{{\zeta_{t,n}^{(m)}}^{2}-\E({\zeta_{t,n}^{(m)}}^{2}) \big\}=o_{p}(1).
\end{eqnarray}
It suffices for \eqref{ingsep9} to show that
\begin{eqnarray}
\label{ingsep11}
\E\big| \sum_{t=1}^{\lfloor nu \rfloor} \big\{{\zeta_{t,n}^{(m)}}^{2}-\E({\zeta_{t,n}^{(m)}}^{2}|\mathcal{F}_{t-1})\big\}\big|^{\eta_1}=o(1),
\end{eqnarray}
where we may assume, without loss of generality, that $1<\eta_1<2$.
By Burkholder's inequality, Jensen's inequality for conditional expectations,
and $1<\eta_1<2$, it holds that
\begin{eqnarray}
\label{ingsep12}
\E\big| \sum_{t=1}^{\lfloor nu \rfloor} \big\{{\zeta_{t,n}^{(m)}}^{2}-\E({\zeta_{t,n}^{(m)}}^{2}|\mathcal{F}_{t-1})\big\}\big|^{\eta_1}
\leq C \sum_{t=1}^{\lfloor nu \rfloor} \E|\zeta_{t,n}^{(m)}|^{2\eta_1}.
\end{eqnarray}
Moreover, by \eqref{ingsep5}, \eqref{ingsep4}, and an argument used in \eqref{ingsep8},
\begin{align}
\label{ingsep13}
\begin{split}
\sum_{t=1}^{\lfloor nu \rfloor} \E|\zeta_{t,n}^{(m)}|^{2\eta_1}
\leq& C \big\{
\sum_{t=1}^{\lfloor nu \rfloor} \E\left\vert \frac{a_t^{(m)}\epsilon_t}{n^{1/2}}\right\vert^{2\eta_1}+
\sum_{t=1}^{\lfloor nu \rfloor} \E| X_{t,n}^{(m)}|^{2\eta_1} \big\} \\
\leq& C
\left\{n^{-\eta_1+1}+n^{-\eta_1} \sum_{t=1}^{\lfloor nu \rfloor}
\left(
\sum_{j=1}^{p_n}\sum_{l=1}^{r_{j}\wedge (n-t)} |\beta_{l}^{(j)}| \sum_{s=t+l}^{n}|p_{s-l-t,j}| \right)^{2\eta_1}
\right\}\\
=& o(1).
\end{split}
\end{align}
Consequently, \eqref{ingsep11} (and hence \eqref{ingsep9}) follows from \eqref{ingsep12} and \eqref{ingsep13}.

Since
\begin{align*}
\sum_{t=1}^{\lfloor nu \rfloor}\big\{{\zeta_{t,n}^{(m)}}^{2}-\E({\zeta_{t,n}^{(m)}}^{2}) \big\} =&
\sum_{t=1}^{\lfloor nu \rfloor}\big\{{X_{t,n}^{(m)}}^{2}-\E({X_{t,n}^{(m)}}^{2}) \big\}\\
&+ 2\sum_{t=1}^{\lfloor nu \rfloor}n^{-1/2}\big\{X_{t,n}^{(m)}a_t^{(m)}\epsilon_t-\E(X_{t,n}^{(m)}a_t^{(m)}\epsilon_t)\big\} \\ 
&+ n^{-1}\sum_{t=1}^{\lfloor nu \rfloor}{a_t^{(m)}}^2(\epsilon_t^2-\sigma^2)\\
\equiv& {\rm (I)+(II)+(III)},
\end{align*}
\eqref{ingsep10} is ensured by
\begin{align}
\label{ingsep15}
{\rm (I)}=o_{p}(1),\quad {\rm (II)}=o_{p}(1), \quad {\rm (III)}=o_{p}(1).
\end{align}
In the following, we only prove the first equation of \eqref{ingsep15}
because the other two can be obtained similarly. Write
\begin{align*}
    X_{t,n}^{(m)}=n^{-1/2}\sum_{j=1}^{p_n}V_{n,j}^{(m)}(t)\pi_{t, j},
\end{align*}
where
\begin{align*}
    V_{n,j}^{(m)}(t)=\sum_{s=t+1}^{n}a_{s}^{(m)}\sum_{l=1}^{(s-t)\wedge r_{j}} \beta_{l}^{(j)}p_{s-l-t,j}.
\end{align*}
Let $\bar{V}_{n,j}^{(m)} = \max_{1 \leq t \leq n} |V_{n,j}^{(m)}(t)|$.
By the same argument used in \eqref{ingsep8}, it is not difficult to show that $\sum_{j=1}^{p_{n}}|\bar{V}_{n,j}^{(m)}| \leq C$.
Then, by Assumption (A2), Burkholder's inequality, Minkowski's inequality, and Cauchy-Schwarz inequality,
\begin{align*}
    &\E \left\vert \sum_{t=1}^{\lfloor nu \rfloor}\{ {X_{t,n}^{(m)}}^{2} - \E ({X_{t,n}^{(m)}}^{2}) \} \right\vert^{\eta_{1}} \\
    =& \E\left\vert \frac{1}{n} \sum_{t=1}^{\lfloor nu \rfloor}\sum_{s_{1}=1}^{p_{n}}\sum_{s_{2}=1}^{p_{n}} V_{n,s_{1}}(t) V_{n,s_{2}}(t)(\pi_{t,s_{1}}\pi_{t,s_{2}} - \sigma_{s_{1},s_{2}}) \right\vert^{\eta_{1}} \\
    \leq& \left\{ \frac{1}{n}\sum_{s_{1},s_{2}=1}^{p_{n}} \left\Vert \sum_{t=1}^{\lfloor nu \rfloor} V_{n,s_{1}}(t)V_{n,s_{2}}(t) \sum_{j=0}^{\infty} \bm{\theta}_{j,s_{1},s_{2}}^{\top}\mathbf{e}_{t-j,s_{1},s_{2}} \right\Vert_{\eta_{1}} \right\}^{\eta_{1}} \\
    =& \left\{ \frac{1}{n}\sum_{s_{1},s_{2}=1}^{p_{n}} \left\Vert \sum_{j=-\infty}^{\lfloor nu \rfloor} \left( \sum_{t= 1\vee j}^{\lfloor nu \rfloor} V_{n,s_{1}}(t)V_{n,s_{2}}(t)\bm{\theta}_{t-j,s_{1},s_{2}}^{\top}\right) \mathbf{e}_{j,s_{1},s_{2}} \right\Vert_{\eta_{1}} \right\}^{\eta_{1}} \\
    \leq& C \left\{ \frac{1}{n}\sum_{s_{1},s_{2}=1}^{p_{n}} \left[ \sum_{j=-\infty}^{\lfloor nu \rfloor} \left\Vert \left( \sum_{t= 1\vee j}^{\lfloor nu \rfloor}  V_{n,s_{1}}(t)V_{n,s_{2}}(t)\bm{\theta}_{t-j,s_{1},s_{2}}^{\top}\right) \mathbf{e}_{j,s_{1},s_{2}} \right\Vert_{\eta_{1}}^{2} \right]^{1/2} \right\}^{\eta_{1}} \\
    \leq& C\left\{ \frac{1}{n} \sum_{s_{1},s_{2}=1}^{p_{n}}\left[ \sum_{j=-\infty}^{\lfloor nu \rfloor}\left( \sum_{t=1\vee j}^{\lfloor nu \rfloor} |V_{n,s_{1}}(t)||V_{n,s_{2}}(t)| \Vert \bm{\theta}_{t-j,s_{1},s_{2}}\Vert \right)^{2} \right]^{1/2} \right\}^{\eta_{1}} \\
    \leq& C \left\{ \frac{1}{n} \sum_{s_{1},s_{2}=1}^{p_{n}}|\bar{V}_{n,s_{1}}^{(m)}||\bar{V}_{n,s_{2}}^{(m)}| \left( \left(\sum_{t=0}^{\infty} \Vert \bm{\theta}_{t,s_{1},s_{2}} \Vert\right) \left[ \sum_{t=1}^{\lfloor nu \rfloor}\sum_{j=-\infty}^{t} \Vert \bm{\theta}_{t-j,s_{1},s_{2}} \Vert \right]^{1/2} \right)^{2} \right\}^{\eta_{1}} \\
    \leq& C (\sqrt{\lfloor nu \rfloor} / n )^{\eta_{1}} = o(1).
\end{align*}
Thus, the desired conclusion follows.
\end{proof}

\setcounter{equation}{0}
\section{Proofs of (\ref{ingaug3}), (\ref{ing11}), (\ref{Thm3.2-AppB}), (\ref{Thm4.2-4})--(\ref{Thm4.2-6}), and (\ref{Thm4.2-11}) in Section \ref{theorems}} \label{Appendix_C2}

{\bf Proof of \eqref{ingaug3}.} 
Let's recall $\hat{a}_m$ as defined in \eqref{ingjune231}.
It follows from
\eqref{selectionrule}
that
\begin{align} \label{AppA2-1}
\begin{split}
    \hat{a}_{m} = \hat{a}_{m-1} -\psi^2_{J_{m-1},(j_m,l_m)} \leq \hat{a}_{m-1} - \xi^{2} \max_{(j,l) \notin J_{m-1}}\psi^2_{J_{m-1},(j,l)}.
    \end{split}
\end{align}
Moreover, since for $1\leq m \leq K_n$,
\begin{align*}
\begin{split}
& \hat{a}_{m-1}=
 n^{-1} \bm{\mu}_{n}^{\top}(\I-\OPM_{[q_{n}]\oplus J_{m-1}}) \bm{\mu}_{n} \\
&\geq  
(\sum_{(j, l) \in  {\cal J}_n-J_{m-1}}{\beta_{l}^{(j)}}^2)
\min_{1\leq \sharp(J)\leq K_n}\lambda_{\min}(n^{-1}\sum_{t=\bar{r}_n+1}^{n}\mathbf{w}_t({\cal J}_n\cup J)
\mathbf{w}^{\top}_t({\cal J}_n\cup J)),
\end{split}
\end{align*}
it holds that
\begin{align} \label{AppA2-2}
\begin{split}
   & \hat{a}_{m-1} = \sum_{(j, l) \in {\cal J}_n} \beta_{l}^{(j)} n^{-1} \bm{\mu}_{n}^{\top}(\I-\OPM_{[q_{n}]\oplus J_{m-1}})\bm{x}_{l}^{(j)} \\
    &\leq  \max_{(j,l) \in {\cal J}_n-J_{m-1}} n^{-1}
    | \bm{\mu_{n}}^{\top}(\I-\OPM_{[q_{n}]\oplus J_{m-1}})\bm{x}_{l}^{(j)}| s_0^{1/2}
 \big(\sum_{(j, l) \in  {\cal J}_n-J_{m-1}}{\beta_{l}^{(j)}}^2\big)^{1/2}\\
 &\leq \max_{(j,l)\notin J_{m-1}}|\psi_{J_{m-1},(j,l)}|
\hat{a}_{m-1}^{1/2}s_0^{1/2} D_n^{-1/2},
\end{split}
\end{align}
where
$D_n$ is defined in \eqref{ing16}.
Equations \eqref{AppA2-1} and \eqref{AppA2-2} imply
for $1\leq m \leq K_n$,
\begin{align*}
    \hat{a}_{m} \leq \hat{a}_{m-1}\left(1 - \frac{\xi^{2}D_n}{s_0} \right), 
\end{align*}
noting that $D_n$ is bounded by 1.
Thus, as long as a selection path
obeying \eqref{selectionrule} is chosen, the resultant
noiseless mean squared error satisfies
\begin{align} \label{L1}
    \hat{a}_m \leq \hat{a}_0 {\rm exp}(-\xi^2mD_n/s_0)\leq C_n {\rm exp}(-\xi^2mD_n/s_0), \,\,1\leq m \leq K_n,
\end{align}
where $C_n$ is also defined \eqref{ing16}.

Now since \eqref{ingaug4} ensures that on $\mathcal{A}_n(K_n) \cap \mathcal{B}_{n}(K_n)$,
$\{\hat{J}_{1}, \ldots, \hat{J}_{K_n}\}$ 
obeys \eqref{selectionrule}, with $0 < \xi < 1$ defined after \eqref{ingaug4},
we conclude that
\eqref{L1} holds with
$\hat{a}_m$ replaced by $\hat{s}_m$
on $\mathcal{A}_n(K_n) \cap \mathcal{B}_{n}(K_n)$.
This completes the proof of
\eqref{ingaug3}.
\qed
\medskip

\noindent {\bf Proof of \eqref{ing11}.}
By an argument similar to \eqref{AppA2-2}, one has for $1\leq m \leq K_n$,
\begin{align}
\begin{split} \label{ing17}
    &n^{-1}\bm{\mu}_{n}^{\top}(\I-\OPM_{[q_{n}]\oplus \hat{J}_{m}})\bm{\mu}_{n} \leq
     \min_{0\leq k \leq m-1}n^{-1}\bm{\mu}_{n}^{\top}(\I-\OPM_{[q_{n}]\oplus \hat{J}_{k}})\bm{\mu}_{n}\\
    &\leq \min_{0\leq k \leq m-1}
    \max_{(j,l)\notin \hat{J}_{k}}\psi^2_{\hat{J}_{k},(j,l)} s_0D_n^{-1}.
\end{split}
\end{align}
Consequently, \eqref{ing11} follows from \eqref{ing17} and
\begin{eqnarray*}
    \min_{0\leq k \leq m-1} \max_{(j,l)\notin \hat{J}_{k}}\psi^2_{\hat{J}_{k},(j,l)} \leq \frac{ \tilde{\xi}^2 l_np_{n}^{*^{(q_0+1)/(\eta q_0)}}}{n}
\end{eqnarray*}
on $\mathcal{B}^c_{n}(m)$.
\qed

To prove \eqref{Thm3.2-AppB}, we need an auxiliary lemma.

\begin{lemma} \label{LemmaA2}
Assume that {\rm (A1), (A2), (A4)}, and {\rm (A5)} hold. Then,
\begin{align}
\label{ingjan8}
\begin{split}
& \max_{1 \leq l \leq r_{j}^{(n)}, 1 \leq j \leq p_{n}}
    \left\vert n^{-1}\sum_{t=\bar{r}_{n}+1}^{n} \epsilon_{t}
    x_{t-l, j} \right\vert +
     \max_{1 \leq k \leq q_{n}}
    \left\vert n^{-1}\sum_{t=\bar{r}_{n}+1}^{n} \epsilon_{t}
    z_{t-k} \right\vert \\
   & =
    O_{p}\left(\frac{{p_{n}^{*}}^{(q_0+1)/(2\eta q_0)}}{n^{1/2}}+\frac{q_n^{1/\eta}}{n^{1/2}}\right)=
    O_{p}\left(\frac{{p_{n}^{*}}^{(q_0+1)/(2\eta q_0)}}{n^{1/2}}\right).
\end{split}
\end{align}
\end{lemma}
\noindent {\bf Proof.}
The first identity of \eqref{ingjan8} is ensured by
\begin{align}
\label{ingjan13}
\begin{split}
 \max_{1 \leq l \leq r_{j}^{(n)}, 1\leq j\leq p_n}
\E \left\vert n^{-1/2}\sum_{t=\bar{r}_{n}+1}^{n} \epsilon_{t}
    x_{t-l, j} \right\vert^{\frac{2\eta q_{0}}{q_0+1}}\,+\,
\max_{1 \leq k \leq q_{n}-d}
\E\left\vert n^{-1/2}\sum_{t=\bar{r}_{n}+1}^{n} \epsilon_{t}
    z_{t-k} \right\vert^{\eta} < C,
\end{split}
\end{align}
which can be proved using
Burkholder's inequality,
Jensen's inequality, H\"{o}lder's inequality,
$\mathrm{E}|x_{t-l, j}|^{2\eta q_0} <C$,
and
$\mathrm{E}|z_{t-k}|^{2\eta} <C$
for all $-\infty<t<\infty, 1\leq l \leq r_{j}^{(n)}, 1 \leq j \leq p_{n}$,
and $1\leq k \leq q_n-d$.
The second identity of \eqref{ingjan8} follows from
$q_n=o(n^{1/2})$
and $p_n^{*} \asymp n^{\nu}$ with $\nu \geq 1$.
\qed
\medskip

\noindent {\bf Proof of \eqref{Thm3.2-AppB}}. 
It suffices for \eqref{Thm3.2-AppB} to show that
    \begin{align}
    \label{ingjan11}
    \begin{split}
   & \max_{ \substack{\sharp(J) \leq K_{n} - 1 \\ (i,l) \notin J} } | \psi_{J,(i,l)} - \hat{\psi}_{J,(i,l)}| = \max_{ \substack{\sharp(J) \leq K_{n} - 1 \\ (i,l) \notin J} }
    \frac{n^{-1}|\bm{\varepsilon}_{n}^{\top}(\I - \OPM_{[q_{n}] \oplus J})\bm{x}_{l}^{(i)}|}
    { (n^{-1} {\bm{x}_{l}^{(i)}}^{\top}(\I - \OPM_{[q_{n}] \oplus J})\bm{x}_{l}^{(i)})^{1/2}}\\
  &  = O_{p}\left(\frac{{p_{n}^{*}}^{(q_0+1)/(2\eta q_0)}}{n^{1/2}}\right),
    \end{split}
    \end{align}
which is, in turn, ensured by
\begin{align}
 \label{ingjan9}
     \max_{ \substack{\sharp(J) \leq K_{n} - 1 \\ (i,l) \notin J} }
    n^{-1}|\bm{\varepsilon}_{n}^{\top}(\I - \OPM_{[q_{n}] \oplus J})\bm{x}_{l}^{(i)}|
    = O_{p}\left(\frac{{p_{n}^{*}}^{(q_0+1)/(2\eta q_0)}}{n^{1/2}}\right)
    \end{align}
and
\begin{align}
\label{ingjan10}
     \max_{ \substack{\sharp(J) \leq K_{n} - 1 \\ (i,l) \notin J} }
(n^{-1} {\bm{x}_{l}^{(i)}}^{\top}(\I - \OPM_{[q_{n}] \oplus J})\bm{x}_{l}^{(i)})^{-1/2}
    = O_{p}(1).
\end{align}
Note that \eqref{ingjan10} is an immediate consequence of
\begin{align*}
\max_{\substack{\sharp(J) \leq K_{n} - 1 \\ (i,l) \notin J}}
\vert n^{-1} {\bm{x}_{l}^{(i)}}^{\top}(\I - \OPM_{[q_{n}] \oplus J})\bm{x}_{l}^{(i)} \vert^{-1/2}
\leq
\max_{\sharp(J) \leq K_{n}}
\lambda_{\min}^{-1/2}\left( n^{-1} \sum_{t=\bar{r}_{n}+1}^{n}
    \mathbf{w}_{t}(J)\mathbf{w}_{t}^{\top}(J) \right)
\end{align*}
and Theorem \ref{Theorem_3.1}. Hence, it remains to prove \eqref{ingjan9}.
Since
    \begin{align*}
    \max_{ \substack{\sharp(J) \leq K_{n} - 1 \\ (i,l) \notin J} } \frac{1}{n}|\bm{\varepsilon}_{n}^{\top}(\I - \OPM_{[q_{n}] \oplus J})\bm{x}_{l}^{(i)}|
    \leq& \max_{\substack{1\leq i \leq p_{n} \\ 1 \leq l \leq r_{i}^{(n)}}}\left\vert \frac{1}{n}\sum_{t=\bar{r}_{n}+1}^{n}\epsilon_{t}x_{t-l,i}\right\vert
    + \max_{\substack{\sharp(J)\leq K_{n}-1 \\ (i,j) \notin J}}\left\vert \frac{1}{n}\sum_{t=\bar{r}_{n}+1}^{n}\epsilon_{t}\hat{x}_{t-l,i;J}\right\vert,
    \end{align*}
\eqref{ingjan9} follows from
    \begin{align} \label{AppB1-1}
\max_{\substack{\sharp(J)\leq K_{n}-1 \\ (i,l) \notin J}}\left\vert \frac{1}{n}\sum_{t=\bar{r}_{n}+1}^{n}\epsilon_{t}\hat{x}_{t-l,i;J}\right\vert
    = O_{p}\left(\frac{{p_{n}^{*}}^{(q_0+1)/(2\eta q_0)}}{n^{1/2}}\right)
    \end{align}
in light of Lemma \ref{LemmaA2}.

For $(i, l) \notin J$
    \begin{align} \label{ingjan12}
    \begin{split}
    &\left\vert n^{-1}\sum_{t=\bar{r}_{n}+1}^{n} \epsilon_{t} \hat{x}_{t-l,i;J} \right\vert
    \leq
    \left\Vert n^{-1}\sum_{t=\bar{r}_{n}+1}^{n}\epsilon_{t}\mathbf{s}_{t}(J)\right\Vert \times\\
    & \left\Vert \left( n^{-1}\sum_{t=\bar{r}_{n}+1}^{n} \mathbf{s}_{t}(J)\mathbf{s}_{t}^{\top}(J) \right)^{-1} \right\Vert
     \left\Vert n^{-1}\sum_{t=\bar{r}_{n}+1}^{n}\mathbf{s}_{t}(J)x^{\perp}_{t-l,i; J} \ \right\Vert \\
    &\quad + \left\vert n^{-1}\sum_{t=\bar{r}_{n}+1}^{n} \epsilon_{t}\mathbf{s}_{t}^{\top}(J) \mathbf{b}_{J}(i,l) \right\vert,
    \end{split}
    \end{align}
where
$x^{\perp}_{t-l,i; J}=x_{t-l,i}-\mathbf{s}_{t}^{\top}(J)\mathbf{b}_{J}(i,l)$
and 
$\mathbf{b}_{J}(i,l)=(\underbrace{0,\ldots,0}_{d},
\mathbf{g}^{\top}_{J}(i,l)\mathbf{\Gamma}_{n}^{-1}(J))^{\top}$.
By Lemma \ref{LemmaA2}, \eqref{thm1-asmpt},
\eqref{ingjan13}, and
\begin{eqnarray}
\label{ingfeb19}
\max_{1 \leq k \leq d}
\mathrm{E}\vert n^{-1/2}\sum_{t=\bar{r}_{n}+1}^{n} \epsilon_{t}
    \tilde{y}_{t, k} \vert^{\eta} < C,
\end{eqnarray}
one obtains
\begin{align} \label{AppB1-2}
    \max_{\substack{\sharp(J)\leq K_{n}-1 \\ (i,j) \notin J}}
    \left\vert n^{-1}\sum_{t=\bar{r}_{n}+1}^{n}\epsilon_{t}\mathbf{s}_{t}^{\top}(J)\mathbf{b}_{J}(i,l)\right\vert =
    O_{p}\left(\frac{{p_{n}^{*}}^{\frac{q_0+1}{2\eta q_0}}}{n^{1/2}}\right)
\end{align}
and
 \begin{align}
    \label{ingjan14}
    \begin{split}
    \max_{\sharp(J)\leq K_{n}-1}
    \left\Vert n^{-1}\sum_{t=\bar{r}_{n}+1}^{n}\epsilon_{t}\mathbf{s}_{t}(J)\right\Vert =
    O_{p}\left(\frac{K_n^{1/2}{p_{n}^{*}}^{\frac{q_0+1}{2\eta q_0}}+q_n^{1/2}}{n^{1/2}}\right).
    \end{split}
    \end{align}
Define
\begin{align*}
\begin{split}
& {\rm (I)}=
\left\Vert n^{-1}\sum_{t=\bar{r}_{n}+1}^{n}\mathbf{s}_{t}(J)x_{t-l,i}-(\underbrace{0, \ldots, 0}_{d}, \mathbf{g}^{\top}_{J}(i, l))^{\top} \ \right\Vert,\\
& {\rm (II)}=
\left
\Vert(
n^{-1}\sum_{t=\bar{r}_{n}+1}^{n}\mathbf{s}_{t}(J)\mathbf{s}^{\top}_{t}(J)-\mathbf{S}_{n}(J))\mathbf{b}_{J}(i,l) \ \right\Vert.
\end{split}
\end{align*}
Then,
\begin{align}
\label{ingjan15}
\begin{split}
\left\Vert n^{-1}\sum_{t=\bar{r}_{n}+1}^{n}\mathbf{s}_{t}(J)x^{\perp}_{t-l,i; J} \ \right\Vert \leq
{\rm (I)+(II)}.
\end{split}
\end{align}

It follows from Lemma \ref{LemmaA3} that
\begin{align}
\label{ingfeb1}
\begin{split}
&\max_{\sharp{(J)}\leq K_n-1, (i,l) \notin J} {\rm (I)}\leq
\sqrt{d}\max_{\substack{1\leq l \leq r_{i}^{(n)} \\ 1\leq i \leq p_n, 1 \leq k \leq  d}}
|n^{-1}\sum_{t=\bar{r}_n+1}^{b}\tilde{y}_{t, k}x_{t-l, i}| \\
&+ \left\{
\sum_{k=1}^{q_n-d}
\max_{1\leq l \leq r_{i}^{(n)}, 1\leq i \leq p_n}
\big|n^{-1}\sum_{t=\bar{r}_n+1}^{n}
\{z_{t-k}x_{t-l,i}-\E(z_{t-k}x_{t-l,i})\}
\big|^2
\right\}^{1/2}\\
&+\left\{K_n
\max_{\substack{1\leq l_1 \leq r_{i_1}^{(n)},
1\leq l \leq r_{i}^{(n)} \\ 1\leq i_1, i \leq p_n}}
\big|n^{-1}\sum_{t=\bar{r}_n+1}^{n}
\{x_{t-l_1, i_1}x_{t-l,i}-\E(x_{t-l_1,i_1}x_{t-l,i})\}
\big|^2
\right\}^{1/2}\\
&=O_{p}\left(\frac{p_n^{*^{\frac{q_0+1}{2\eta q_0}}}}{n^{1/2}}\right)
+O_{p}\left(
\frac{q_np_n^{*^{\frac{q_0+1}{2\eta q_0}}}}{n^{1/2}}
\right)+
O_{p}\left(
\frac{K_n^{1/2}p_n^{*^{\frac{2}{\eta q_0}}}}{n^{1/2}}
\right)\\
&=O_{p}\left(
\frac{K_n^{1/2}p_n^{*^{\frac{2}{\eta q_0}}}+
q_np_n^{*^{\frac{q_0+1}{2\eta q_0}}}
}
{n^{1/2}}
\right).
\end{split}
\end{align}
By Lemma \ref{LemmaA3} and \eqref{thm1-asmpt},
we also show below that
\begin{eqnarray}
\label{ingfeb3}
\max_{\sharp{(J)}\leq K_n-1, (i,l) \notin J} {\rm (II)}
=O_{p}
\left(
\frac{
K_n^{1/2}p_n^{*^{\frac{2}{\eta q_0}}}+(K_n^{1/2}+q_n^{1/2})p_n^{*^{\frac{q_0+1}{2\eta q_0}}}
}
{n^{1/2}}
\right).
\end{eqnarray}
According to \eqref{ingjan15}--\eqref{ingfeb3},
\begin{align}
\label{ingfeb2}
\begin{split}
& \max_{\sharp{(J)}\leq K_n-1, (i,l) \notin J}
\left\Vert n^{-1}\sum_{t=\bar{r}_{n}+1}^{n}\mathbf{s}_{t}(J)x^{\perp}_{t-l,i; J} \ \right\Vert \\
&=O_{p}\left(
\frac{
K_n^{1/2}p_n^{*^{\frac{2}{\eta q_0}}}+(K_n^{1/2}+q_n^{1/2})p_n^{*^{\frac{q_0+1}{2\eta q_0}}}}
{n^{1/2}}
\right).
\end{split}
\end{align}
Consequently, \eqref{ingjan12}, \eqref{AppB1-2}, \eqref{ingjan14}, \eqref{ingfeb2}, and \eqref{ing3} imply
\begin{align*}
    \begin{split}
& \max_{\sharp{(J)}\leq K_n-1, (i,l) \notin J}
\left\vert n^{-1}\sum_{t=\bar{r}_{n}+1}^{n} \epsilon_{t} \hat{x}_{t-l,i;J} \right\vert=
O_{p}
\left(
\frac{
K_n^{1/2}p_n^{*^{\frac{2}{\eta q_0}}}+(K_n^{1/2}+q_n^{1/2})p_n^{*^{\frac{q_0+1}{2\eta q_0}}}}
{n^{1/2}}
\right) \\
&\times
O_{p}
\left(
\frac{
K_n^{1/2}p_n^{*^{\frac{q_0+1}{2\eta q_0}}}+q_n^{1/2}}
{n^{1/2}}
\right),
\end{split}
\end{align*}
which, together with \eqref{ingsep18} and (A5), leads to
\eqref{AppB1-1}. Thus, the proof is complete.
\qed
\medskip

\noindent {\bf Proof of \eqref{ingfeb3}}.
Note first that
\begin{align} \label{ingfeb21}
\begin{split}
&\max_{\sharp{(J)}\leq K_n-1, (i,l) \notin J} {\rm (II)} \\
&\leq
\max_{\sharp{(J)}\leq K_n-1, (i,l) \notin J}
\left\{
\sum_{k=1}^{d}\big[
\sum_{s=1}^{q_n-d}
a_{s, J}(i, l)
(n^{-1}\sum_{t=\bar{r}_n+1}^{n}
\tilde{y}_{t, k}z_{t-s}) \big. \right. \\
&+ \left.\big.
\sum_{(i^{*}, l^{*}) \in J}
a_{(i^*,l^*)}(i,l)
(n^{-1}\sum_{t=\bar{r}_n+1}^{n}\tilde{y}_{t, k}
x_{t-l^*, i^*})
\big]^2
\right\}^{1/2}\\
&+
\max_{\sharp{(J)}\leq K_n-1, (i,l) \notin J}
\left\{
\sum_{k=1}^{q_n-d}\big[
\sum_{s=1}^{q_n-d}
a_{s, J}(i, l)
(n^{-1}\sum_{t=\bar{r}_n+1}^{n}
z_{t-k}z_{t-s}-\E(z_{t-k}z_{t-s})) \big. \right. \\
&+ \left.\big.
\sum_{(i^{*}, l^{*}) \in J}
a_{(i^*,l^*)}(i,l)
(n^{-1}\sum_{t=\bar{r}_n+1}^{n}z_{t-k}
x_{t-l^*, i^*}-\E(z_{t-k}
x_{t-l^*, i^*}))
\big]^2
\right\}^{1/2} \\
&+
\max_{\sharp{(J)}\leq K_n-1, (i,l) \notin J}
\left\{
\sum_{(\tilde{i}, \tilde{l}) \in J}\big[
\sum_{s=1}^{q_n-d}
a_{s, J}(i, l)
(n^{-1}\sum_{t=\bar{r}_n+1}^{n}
x_{t-\tilde{l}, \tilde{i}}z_{t-s}-\E(x_{t-\tilde{l}, \tilde{i}}z_{t-s})) \big. \right. \\
&+ \left.\big.
\sum_{(i^{*}, l^{*}) \in J}
a_{(i^*,l^*)}(i,l)
(n^{-1}\sum_{t=\bar{r}_n+1}^{n}x_{t-\tilde{l}, \tilde{i}}
x_{t-l^*, i^*}-\E(x_{t-\tilde{l}, \tilde{i}}
x_{t-l^*, i^*}))
\big]^2
\right\}^{1/2}\\
&\equiv {\rm (III)+(IV)+(V)}.
\end{split}
\end{align}
By Lemma \ref{LemmaA3} and \eqref{thm1-asmpt}, it can be shown that
\begin{eqnarray}
\label{ingfeb22}
{\rm (III)}= O_{p}\left(
\frac{p_n^{*^{\frac{q_0+1}{2\eta q_0}}}}{n^{1/2}}\right)
\end{eqnarray}
and
\begin{eqnarray}
\label{ingfeb23}
{\rm (IV)}= O_{p}\left(
\frac{q_n^{2^{-1}+\eta^{-1}+\delta I_{\{\eta=2\}}}}{n^{1/2}}+
\frac{q_n^{1/2}p_n^{*^{\frac{q_0+1}{2\eta q_0}}}}{n^{1/2}}
\right)=O_{p}\left(
\frac{q_n^{1/2}p_n^{*^{\frac{q_0+1}{2\eta q_0}}}}{n^{1/2}}
\right),
\end{eqnarray}
where $\delta>0$ is arbitrarily small and the second equality is ensured by
(A5). Moreover, it follows that
\begin{align} \label{ingfeb24}
\begin{split}
     {\rm (V)}^{2} \leq& C (K_n-1) \sum_{s_1=1}^{q_n-d} \sum_{s_2=1}^{q_n-d} b_{s_1}b_{s_2}A_{s_1}A_{s_2}\\
    & + C \max_{\sharp{(J)}\leq K_n-1, (i,l) \notin J} \sum_{(i^{*}, l^{*}) \in J}
|a_{(i^*,l^*)}(i,l)| \\
    &\times \max_{\substack{1 \leq \tilde{i} ,i^* \leq p_{n} \\ 1 \leq \tilde{l} \leq r_{\tilde{i}}^{(n)}, \;1 \leq l^* \leq r_{i^*}^{(n)}}} \left|n^{-1}\sum_{t=\bar{r}_n+1}^{n}x_{t-\tilde{l}, \tilde{i}} x_{t-l^*, i^*}-\E(x_{t-\tilde{l}, \tilde{i}} x_{t-l^*, i^*})\right|^{2} \\
    &= O_{p}\left(\frac{K_n(p_n^{*^{\frac{q_0+1}{\eta q_0}}} + p_n^{*^{\frac{4}{\eta q_0}}})}{n}\right),
\end{split}
\end{align}
where
\begin{align*}
\begin{split}
& b_{s}=\max_{\sharp{(J)}\leq K_n-1, (i,l) \notin J}|a_{s, J}(i,l)|,\\
&A_s=\max_{1 \leq i  \leq p_{n}, 1 \leq l \leq r_{i}^{(n)}}
|n^{-1}\sum_{t=\bar{r}_n+1}^{n}
x_{t-l, i}z_{t-s}-\E(x_{t-l, i}z_{t-s})|.
\end{split}
\end{align*}
Combining \eqref{ingfeb21}--\eqref{ingfeb24} yields \eqref{ingfeb3}.
\qed
\medskip

\noindent {\bf Proof of \eqref{Thm4.2-4}--\eqref{Thm4.2-6}}.
Since
\begin{align*}
    & \hat{A}_{n}^{-1} \leq \max_{\substack{\sharp(J) \leq K_{n}-1 \\ (j,l)\notin J}} \{ n^{-1}{\bm{x}_{l}^{(j)}}^{\top}(\I-\OPM_{[q_{n}] \oplus J})\bm{x}_{l}^{(j)} \}^{-1},\\
    & |\hat{B}_{n}| \leq \max_{\substack{\sharp(J)\leq K_{n}-1 \\ (j,l)\notin J}}|n^{-1}{\bm{x}_{l}^{(j)}}^{\top}(\I-\OPM_{[q_{n}] \oplus J})\bm{\varepsilon}_{n}|,
\end{align*}
\eqref{Thm4.2-4} and \eqref{Thm4.2-5} follow directly from \eqref{ingjan9} and \eqref{ingjan10}, respectively.
To show \eqref{Thm4.2-6}, note first that
\begin{align} \label{ingfeb14}
    |\hat{C}_{n}| \leq& \left\vert n^{-1}\sum_{t=\bar{r}_{n}+1}^{n}\epsilon_{t}^{2}-\sigma^{2} \right\vert + n^{-1}\bm{\varepsilon}_{n}^{\top}\OPM_{[q_{n}] \oplus
    \hat{J}_{\tilde{k}}}\bm{\varepsilon}_{n}.
\end{align}
By Assumption (A1), it is easy to show that
\begin{align} \label{ingfeb13}
|n^{-1}\sum_{t=\bar{r}_{n}+1}^{n}\epsilon_{t}^{2}-\sigma^{2}|=O_{p}(n^{-1/2}).
\end{align}
In addition,
\begin{align*}
    n^{-1} \bm{\varepsilon}_{n}^{\top} \OPM_{[q_{n}] \oplus \hat{J}_{\tilde{k}}}\bm{\varepsilon}_{n} \leq& \max_{\sharp(J) \leq m_n^{*}} \lambda_{\min}^{-1} \left( n^{-1} \sum_{t=\bar{r}_{n}+1}^{n} \mathbf{s}_{t}(J) \mathbf{s}_{t}^{\top}(J) \right) \\
    &\times \max_{\sharp(J) \leq m_n^{*}}
    \left\Vert n^{-1}\sum_{t=\bar{r}_{n}+1}^{n}\epsilon_{t}\mathbf{s}_{t}(J)\right\Vert^2 \,\, \mbox{on}\,\, {\cal D}_n,
\end{align*}
which, together with \eqref{ingfeb14}, \eqref{ingfeb13}, \eqref{ingjan14}, \eqref{ing3}, and \eqref{ingfeb15},
gives \eqref{Thm4.2-6}. 
\qed
\medskip

\noindent {\bf Proof of \eqref{Thm4.2-11}}.
Note first that for some $c_1>0$,
\begin{align*}
    \frac{1-\exp(-w_{n, p_n}n^{-1}(\hat{k}-\tilde{k}))}{\hat{k}-\tilde{k}} \geq c_1 \Big\{\frac{w_{n,p_{n}}}{n}\wedge\frac{1}{\hat{k}-\tilde{k}}\Big\} \geq
c_1 \Big\{\frac{w_{n,p_{n}}}{n}\wedge K_n^{-1}\Big\} \,\,\mbox{on}\,\,\{\tilde{k}<\hat{k}\}.
\end{align*}
Define $B_{n, p_n}=(w_{n,p_{n}}/n)\wedge K_n^{-1}$.
Then, it follows from \eqref{ingsep18} and the first part of \eqref{penalty1} that
\begin{align} \label{ingfeb16}
    p_n^{*^{\bar{\theta}}}/n^{1/2}=o(B_{n, p_n}^{1/2}).
\end{align} 
Now, for any $\delta>0$,
\begin{align}
    &P\{(\hat{k}-\tilde{k})(\hat{a}_{n}+\hat{b}_{n}) \geq \delta[1-\exp(-n^{-1}w_{n, p_n}(\hat{k}-\tilde{k})]), \tilde{k}<\hat{k}\} \notag \\
    \leq& P\left( \lambda_{\min}^{-1}\left( n^{-1}\sum_{t=\bar{r}_{n}+1}^{n}\mathbf{s}_{t}(\hat{J}_{\hat{k}})
    \mathbf{s}_{t}^{\top}(\hat{J}_{\hat{k}}) \right) \max_{\substack{1\leq j \leq p_{n} \\ 1\leq l \leq r_{j}^{(n)}}}
    \left\vert n^{-1}\sum_{t=\bar{r}_{n}+1}^{n}x_{t-l,j}\epsilon_{t} \right\vert^{2} \geq
    c_1  \delta B_{n, p_n}\right) \notag \\
    &+P\left( \lambda_{\min}^{-1}\left( n^{-1}\sum_{t=\bar{r}_{n}+1}^{n}\mathbf{s}_{t}(\hat{J}_{\hat{k}})\mathbf{s}_{t}^{\top}(\hat{J}_{\hat{k}}) \right)
    \max_{\substack{\sharp(J)\leq K_{n}-1 \\ (j,l) \notin J}} \left\vert n^{-1}\sum_{t=\bar{r}_{n}+1}^{n} \hat{x}_{t-l,j;J}\epsilon_{t} \right\vert^{2} \geq
     c_1 \delta B_{n, p_n} \right) \notag \\
    :=& {\rm (I)} + {\rm (II)}. \notag
\end{align}
By \eqref{ingjan8}, \eqref{AppB1-1}, Theorem \ref{Theorem_3.1}, and \eqref{ingfeb16}, ${\rm (I)+(II)} = o(1)$.
Thus \eqref{Thm4.2-11} is proved.
\qed

\setcounter{equation}{0}
\section{Some technical details about Examples \ref{egOGA} and \ref{egLASSO}} \label{sec06}

\subsection{Proof of \eqref{ingapr235} in Example \ref{egOGA}}  \label{supp::egOGA}
In this subsection, all summations are understood as summing from $t=3$ to $t=n$. 
Let $z_{t} = y_{t} - y_{t-1}$. Clearly, $z_{t} = a z_{t-1} + \epsilon_{t}$ for $t=1,2,\ldots,n$.
Note that with some algebraic manipulation and using the AR definition, we can express
\begin{align*}
    F_{2,n} =& \frac{(\sum y_{t}y_{t-1})^{2}}{\sum y_{t-1}^{2}} + \frac{(\sum y_{t}y_{t-1})^{2}}{\sum y_{t-1}^{2}} \left( \frac{2 \sum y_{t-2}z_{t-1} + \sum z_{t-1}^{2}}{\sum y_{t-2}^{2}} \right) \\
    &-\frac{2(\sum y_{t}y_{t-1})(\sum y_{t} z_{t-1})}{\sum y_{t-1}^{2}}\left( 1 + \frac{2\sum y_{t-2}z_{t-1} + \sum z_{t-1}^{2}}{\sum y_{t-2}^{2}} \right) \\
    &+ \frac{(\sum y_{t}z_{t-1})^{2}}{\sum y_{t-1}^{2}} \left( 1 + \frac{2\sum y_{t-2}z_{t-1} + \sum z_{t-1}^{2}}{\sum y_{t-2}^{2}} \right).
\end{align*}
By a similar argument used in Lemma \ref{LemmaA3} and Theorem \ref{Theorem_3.1}, we have
\begin{align*}
    \frac{1}{n}(F_{1,n} - F_{2,n}) =& - \frac{\sum y_{t}y_{t-1}}{\sum y_{t-1}^{2}} \frac{\sum y_{t}y_{t-1}}{\sum y_{t-2}^{2}}(2 n^{-1} \sum y_{t-2}z_{t-1} + n^{-1} \sum z_{t-1}^{2}) \\
    &+ 2 \frac{\sum y_{t}y_{t-1}}{\sum y_{t-1}^{2}}(n^{-1}\sum y_{t}z_{t-1})\left(1 + \frac{2\sum y_{t-2}z_{t-1} + \sum z_{t-1}^{2}}{\sum y_{t-2}^{2}} \right) \\
    &- \frac{(\sum y_{t}z_{t-1})^{2}}{\sum y_{t-1}^{2}}\left(\frac{1}{n} + \frac{2\sum y_{t-2}z_{t-1} + \sum z_{t-1}^{2}}{n \sum y_{t-2}^{2}} \right) \\
    =& (-1 + O_{p}(n^{-1})) \left(2 n^{-1} \sum y_{t-2}z_{t-1} + n^{-1} \sum z_{t-1}^{2} \right) \\
    &+ 2(n^{-1}\sum y_{t}z_{t-1})(1+O_{p}(n^{-1})) + O_{p}(n^{-1}) \\
    =& n^{-1}\sum z_{t-1}^{2} + 2n^{-1}\sum z_{t}z_{t-1} + O_{p}(n^{-1}),
\end{align*}
which implies
\begin{align*}
    \frac{1}{n}(F_{1,n} - F_{2,n}) \rightarrow \frac{1}{1-a^{2}} + \frac{2a}{1-a^{2}} \mbox{ in probability}. 
\end{align*}
\qed

\subsection{Proof of \eqref{sch731} in Example \ref{egLASSO}} \label{supp::egLasso}
Note that 
\begin{align}
\label{ingsep003}
\begin{split}
{\cal A}_n&=\{\hat{\bm{\beta}}^{(\lambda_{n})}\,\,
\mbox{selects the correct model}\,\,\}=
\{\hat{\beta}_{1}^{(\lambda_{n})} \neq 0,
\hat{\beta}_{3}^{(\lambda_{n})} \neq 0,
\hat{\beta}_{2}^{(\lambda_{n})}=0\}\\
&= \big\{\bm{s}_n(1)=(\mbox{sign}(\hat{\beta}_{1}^{(\lambda_{n})}),
\mbox{sign}(\hat{\beta}_{3}^{(\lambda_{n})})
)^{\top} \in \{\bm{a}_1,\ldots, \bm{a}_4\} \,\,\mbox{and} \,\, \hat{\beta}_{2}^{(\lambda_{n})} = 0 \big\},
\end{split}
\end{align}
where $\bm{a}_1^{\top}=(1, 1)$,
$\bm{a}_2^{\top}=(1, -1)$,
$\bm{a}_3^{\top}=(-1, 1)$, and 
$\bm{a}_4^{\top}=(-1, -1)$.
Define
\begin{align*}
    \mathbf{C}_{11} =  \sum_{t=3}^{n} \begin{pmatrix}
        y_{t-1} \\
        x_{t-1}
    \end{pmatrix} \begin{pmatrix}
        y_{t-1} & x_{t-1}
    \end{pmatrix}, \quad \mathbf{c}_{21} =  \sum_{t=3}^{n} \begin{pmatrix}
        y_{t-1}y_{t-2} & x_{t-1}y_{t-2}
    \end{pmatrix}, \\ 
    \hat{\mathbf{u}}_{n} = \begin{pmatrix}
        \hat{\beta}_{1}^{(\lambda_{n})} - 1 \\
        \hat{\beta}_{3}^{(\lambda_{n})} - 1
    \end{pmatrix}, \quad
    \mathbf{w}_{n}(1) = \sum_{t=3}^{n} \epsilon_{t} \begin{pmatrix}
        y_{t-1} \\
        x_{t-1}
    \end{pmatrix}, \quad
    w_{n}(2) = \sum_{t=3}^{n} \epsilon_{t}y_{t-2}.
\end{align*}
Then by an argument used in \cite{Zhao2006},
\begin{eqnarray}
\label{ingsep004}
{\cal A}_n \subseteq
\bigcup_{i=1}^{4} {\cal E}_{n}(i),
\end{eqnarray}
where 
\begin{eqnarray*}
{\cal E}_{n}(i)=\{
\mathbf{C}_{11} \hat{\mathbf{u}}_{n}
- \mathbf{w}_{n}(1)=-\frac{\lambda_{n}\bm{a}_i}{2}, \,\,
-\frac{\lambda_{n}}{2} \leq
\mathbf{c}_{21}\hat{\mathbf{u}}_{n} - w_{n}(2) \leq \frac{\lambda_{n}}{2}, \,\,\bm{s}_{n}(1)=\bm{a}_i\}.
\end{eqnarray*}
In the following, we will show that regardless of whether $\{\lambda_n\}$ satisfies (a') $\lambda_n/n \to \infty$, (b') $\lambda_n/n \to 0$, or (c') $0<\lim_{n \to \infty}
\lambda_n/n =d^* <\infty$,
\begin{align} \label{ingsep001}
    \limsup_{n \to \infty} P({\cal E}_{n}(1))\leq 1/2,
\end{align}
and
\begin{align} \label{ingsep002}
    \lim_{n \to \infty} P({\cal E}_{n}(i))=0, \quad i=2, 3, 4.
\end{align}
By \eqref{ingsep003}--\eqref{ingsep002},
the desired conclusion \eqref{sch731} follows.

We commence by proving \eqref{ingsep001}. 
Straightforward calculations give
\begin{align}
\label{ingsep005}
   {\cal E}_{n}(1) \subseteq \left\{ \mathbf{c}_{21}\mathbf{C}_{11}^{-1}\mathbf{w}_{n}(1) - w_{n}(2) \geq -\frac{\lambda_n}{2}\left(1 - \mathbf{c}_{21}\mathbf{C}_{11}^{-1}\bm{a}_1 \right) \right\},
\end{align}
\begin{align*}
    \mathbf{c}_{21}\mathbf{C}_{11}^{-1} = \begin{pmatrix}
        1 + O_{p}(n^{-1}) & -n^{-1} \sum_{t=3}^{n}x_{t-1}\delta_{t-1} + O_{p}(n^{-1})
    \end{pmatrix},
\end{align*}
where $\delta_{t} = x_{t-1} + \epsilon_{t}$, 
\begin{align} \label{eglasso3}
\mathbf{c}_{21}\mathbf{C}_{11}^{-1}\mathbf{w}_{n}(1) - w_{n}(2) = \sum_{t=3}^{n} \delta_{t-1}\epsilon_{t} + O_{p}(1),
\end{align}
\begin{align} \label{eglasso4}
1- \mathbf{c}_{21}\mathbf{C}_{11}^{-1}\bm{a}_1 = \frac{1}{n}
 \sum_{t=3}^{n} x_{t-1}\delta_{t-1} + O_{p}(1/n),
\end{align}
and
\begin{align}
\label{ingsep006}
    \begin{pmatrix}
        \frac{1}{\sqrt{n}}\sum_{t=3}^{n} \delta_{t-1}\epsilon_{t} \\
        \frac{1}{\sqrt{n}}\sum_{t=3}^{n} x_{t-1}\delta_{t-1}
    \end{pmatrix} \Rightarrow \begin{pmatrix}
        N_{1} \\
        N_{2}
    \end{pmatrix},
\end{align}
where $N_{1}$ and $N_{2}$ are two independent Gaussian random variables with mean zero and variance 2. 
By \eqref{ingsep005}--\eqref{ingsep006}, it holds that
\begin{align}
\label{ingsep007}
\begin{split}
P({\cal E}_{n}(1)) &\leq 
P\left( \frac{1}{\sqrt{n}}\sum_{t=3}^{n} \delta_{t-1}x_{t-1} \geq o_{p}(1) \right) \to 1/2\,\,
\mbox{in case}\,\,(a'),
\\
P({\cal E}_{n}(1)) &\leq 
P\left( \frac{1}{\sqrt{n}}\sum_{t=3}^{n} \delta_{t-1}\epsilon_{t} \geq o_{p}(1) \right)\to 1/2\,\,
\mbox{in case}\,\,(b'),\\
P({\cal E}_{n}(1)) &\leq 
P\left( \frac{1}{\sqrt{n}}\sum_{t=3}^{n} \delta_{t-1}\epsilon_{t} \geq
\frac{-\lambda_n}{2n}
\frac{1}{\sqrt{n}}\sum_{t=3}^{n} \delta_{t-1}x_{t-1}+
o_{p}(1) \right) \\
& \to 
P\left(N_{1} \geq -\frac{d^*}{2}N_{2}\right)=1/2 
\,\,
\mbox{in case}\,\,(c').
\end{split}
\end{align}
Thus, \eqref{ingsep001} follows.

For $i=2, 3$, and 4,
\begin{align}
\label{ingsep008}
\begin{split}
{\cal E}_{n}(i)
& \subseteq \{ \hat{\mathbf{u}}_{n}
= \mathbf{C}_{11}^{-1}[\mathbf{w}_{n}(1)-(\lambda_{n}\bm{a}_i)/2],  \,\,
\bm{s}_{n}(1)=\bm{a}_i
\}\\
&= \{ (\hat{\beta}_{1}^{(\lambda_{n})},
\hat{\beta}_{3}^{(\lambda_{n})})^{\top}=(1,1)^{\top}
+ \mathbf{C}_{11}^{-1}[\mathbf{w}_{n}(1)-(\lambda_{n}\bm{a}_i)/2],  \,\,
\bm{s}_{n}(1)=\bm{a}_i
\}.
\end{split}
\end{align}
By an argument similar to that used to prove \eqref{eglasso3} and \eqref{eglasso4},
\begin{align}
\label{ingsep009}
\begin{split}
& \mathbf{C}_{11}^{-1}[\mathbf{w}_{n}(1)-(\lambda_{n}\bm{a}_i)/2]\\
&=\left(
O_{p}\big(n^{-1}+\frac{\lambda_n}{n^2}\big),
O_{p}(n^{-1/2})+
\frac{\lambda_{n}}{2n}
(I_{\{i=2, 4\}}-I_{\{i=3\}})
(1+o_{p}(1)).
\right)^{\top}
\end{split}
\end{align}
Combining 
\eqref{ingsep008}
and \eqref{ingsep009} yields that
there exist an arbitrarily small positive constant $\varepsilon>0$ and
an arbitrarily large positive constant
$M<\infty$
such that for all sufficiently large $n$,
\begin{align} \label{ingsep010}
\begin{split}
    P({\cal E}_{n}(j)) &\leq P(\hat{\beta}_{3}^{(\lambda_{n})}>1-\varepsilon, \hat{\beta}_{3}^{(\lambda_{n})}<0)+o(1) \\
    &=o(1) \,\,\mbox{in all cases of}\,\, (a'), (b'),\,\,\mbox{and}\,\, (c'),
\end{split}
\end{align}
where $j=2, 4$, and
\begin{align} \label{ingsep011}
\begin{split}
    &P({\cal E}_{n}(3)) \leq P(\hat{\beta}_{3}^{(\lambda_{n})}<-M, \hat{\beta}_{3}^{(\lambda_{n})}>0)+o(1) = o(1) \,\,\mbox{in case}\,\, (a'),\\
    & P({\cal E}_{n}(3)) \leq P(\hat{\beta}_{1}^{(\lambda_{n})}>1-\varepsilon, \hat{\beta}_{1}^{(\lambda_{n})}<0)+o(1)= o(1) \,\,\mbox{in cases}\,\, (b')\,\,\mbox{and}\,\,(c').
\end{split}
\end{align}
Consequently, \eqref{ingsep002} is ensured by \eqref{ingsep010} and \eqref{ingsep011}. This completes the proof of \eqref{sch731}.  \qed

\setcounter{equation}{0}
\section{Complementary simulation results} \label{supp::sim}
We generate data from
\begin{align}
\label{ingapr15}
    (1 + 0.4B)(1 - B)^{2}y_{t} = \sum_{j=1}^{2}\sum_{l=1}^{4}\beta_{l}^{(j)}x_{t-l,j} + \epsilon_{t},
\end{align}
where $\{\epsilon_{t}\}$ is a GARCH(1,1) model,
\begin{align*}
    \epsilon_{t} =& \sigma_{t}Z_{t}, \\
    \sigma_{t}^{2} =& 5\times 10^{-2} + 0.5\epsilon_{t-1}^{2} + 0.1\sigma_{t-1}^{2},
\end{align*}
in which $\{Z_{t}\}$ is a sequence of i.i.d. standard Gaussian random variables. Let
$\{\pi_{t,1}\}$ and $\{\pi_{t,2}\}$ be two independent ARCH(1) processes such that
for $j=1$ and 2,
\begin{align*}
    \pi_{t, j} =& h_{t,j}G_{t, j}, \\
    h_{t, j}^{2} =& 1 + 0.2\pi_{t-1,j}^{2},
\end{align*}
where $\{G_{t,1}\}$ and $\{G_{t,2}\}$ have the same probabilistic structure as that
of $\{Z_t\}$ and these three sequences are independent of each other. Also let
$v_{t,j}$, $1 \leq t \leq n, 1 \leq j \leq p_n$, be independent standard Gaussian
random variables and independent of $\{G_{t,1}\}$, $\{G_{t,2}\}$, and $\{Z_t\}$.
Define $w_{t,j} = \pi_{t,1} + v_{t,j}$ if $j$ is odd, $w_{t,j} = \pi_{t,2} + v_{t,j}$ if $j$ is
even. Then, $x_{t,j}$ are MA(2) processes satisfying $x_{t,j} = 0.8w_{t,j} +
0.1w_{t-1,j}$ if $j$ is odd and $x_{t,j} = 0.2w_{t,j} + 0.6w_{t-1,j}$ otherwise. The
coefficients are set to
\begin{align*}
    (\beta_{1}^{(1)}, \beta_{2}^{(1)}, \beta_{3}^{(1)}, \beta_{4}^{(1)}) =& (-7.62, 6.72, -5.55, 3.77), \\
    (\beta_{1}^{(2)}, \beta_{2}^{(2)}, \beta_{3}^{(2)}, \beta_{4}^{(2)}) =& (6.89, -6.18, 4.47, -3.10).
\end{align*}
Using Theorem 2.2 of \cite{LING2002109} again, one can verify that $\epsilon_{t}$
only has a finite fourth moment and $x_{t,j}$ has a finite twelfth moment.
Moreover, it is easy to show that (A1) and (A2) in Section 3.1 are fulfilled by the above model specification.

One distinct feature of this example is that the error term and all candidate
covariates are conditionally heteroscedastic. Table \ref{table:3} records the
performance of the methods introduced in Section 4
based on 1000 replications and $(n, p_n, r^{(n)})=(800, 250, 4), (1000, 275, 5)$, and (1500, 300, 6).
The table reveals that FHTD is the
only method that efficiently identifies the correct ARX model. More specifically, it
successfully chooses the correct ARX model over 89\% of the time, in all cases of
$n$ considered in this example.

\begin{table}[h!]
    \centering
    \caption{Values of E, SS, TP, and FP in Example \ref{ingapr15}} \label{table:3}
    \begin{tabular}{crrrrrr}
    \hline
    & LASSO & ALasso & OGA-3 & AR-ALasso & AR-OGA-3 & FHTD  \\
    \multicolumn{7}{l}{\footnotesize $(n, p_{n}^{*}, p_n, r^{(n)}, q_n) = (800, 1000, 250, 4, 10)$} \\
    E  & 0    & 0    & 0    & 0    & 137   & 926 \\
    SS & 0    & 0    & 0    & 0    & 138   & 1000 \\
    TP & 1.08 & 1.00 & 3.81 & 1.00 & 9.27  & 11.00 \\
    FP & 0.13 & 0.00 & 0.17 & 0.00 & 5.39  & 0.09 \\ \hline
    \multicolumn{7}{l}{\footnotesize $(n, p_{n}^{*}, p_n, r^{(n)}, q_n) = (1000, 1375, 275, 5, 11)$} \\
    E  & 0     & 0    & 0    & 0    & 181   & 891 \\
    SS & 0     & 0    & 0    & 0    & 183   & 932 \\
    TP & 1.05  & 1.00 & 3.56 & 1.00 & 9.32  & 10.83 \\
    FP & 0.17  & 0.00 & 0.35 & 0.00 & 6.31  & 0.32 \\ \hline
    \multicolumn{7}{l}{\footnotesize $(n, p_{n}^{*}, p_n, r^{(n)}, q_n) = (1500, 1800, 300, 6, 12)$} \\
    E  & 0     & 0    & 0    & 0    & 299   & 960 \\
    SS & 0     & 0    & 0    & 0    & 301   & 989 \\
    TP & 1.02  & 1.00 & 3.19 & 1.00 & 9.57  & 10.99 \\
    FP & 0.19  & 0.01 & 0.24 & 0.00 & 5.83  & 0.08 \\ \hline
    \end{tabular}
\end{table}
\end{supplement}


\bibliographystyle{imsart-number} 
\bibliography{mainbib}       

@article{ing2011,
  title={A stepwise regression method and consistent model selection for high-dimensional sparse linear models},
  author={Ing, Ching-Kang and Lai, Tze Leung},
  journal={Statistica Sinica},
  pages={1473--1513},
  year={2011}
}

@article{Ing2012,
   author = {Ching-Kang Ing and Chor-Yiu Sin and Shu-Hui Yu},
   issn = {0047259X},
   journal = {Journal of Multivariate Analysis},
   month = {4},
   pages = {57--71},
   title = {Model selection for integrated autoregressive processes of infinite order},
   volume = {106},
   year = {2012}
}

@article{Ing2014,
   author = {Ching Kang Ing and Chiao Yi Yang},
   issue = {505},
   journal = {Journal of the American Statistical Association},
   pages = {243--253},
   publisher = {American Statistical Association},
   title = {Predictor selection for positive autoregressive processes},
   volume = {109},
   year = {2014}
}

@article{Ing2001,
   author = {C. K. Ing},
   issue = {6},
   journal = {Journal of Time Series Analysis},
   pages = {711--724},
   title = {A note on mean-squared prediction errors of the least squares predictors in random walk models},
   volume = {22},
   year = {2001}
}

@article{Ing2010,
   author = {Ching Kang Ing and Chor Yiu Sin and Shu Hui Yu},
   issue = {3},
   journal = {Econometric Theory},
   month = {6},
   pages = {774--803},
   title = {Prediction errors in nonstationary autoregressions of infinite order},
   volume = {26},
   year = {2010}
}

@article{tib1996,
    author = {Robert Tibshirani},
    journal = {Journal of the Royal Statistical Society. Series B (Methodological)},
    number = {1},
    pages = {267--288},
    title = {Regression shrinkage and selection via the Lasso},
    volume = {58},
    year = {1996}
}

@article{zou2006,
    author = {Hui Zou},
    title = {The adaptive Lasso and its oracle properties},
    journal = {Journal of the American Statistical Association},
    volume = {101},
    number = {476},
    pages = {1418--1429},
    year  = {2006},
    publisher = {Taylor & Francis}
}

@article{medeiros2016,
    title = {$\ell_{1}$-regularization of high-dimensional time-series models with non-Gaussian and heteroskedastic errors},
    journal = {Journal of Econometrics},
    volume = {191},
    number = {1},
    pages = {255--271},
    year = {2016},
    author = {Marcelo C. Medeiros and Eduardo F. Mendes}
}

@article{Han2020,
    author = {Yuefeng Han and Ruey S. Tsay},
    journal = {Statistica Sinica},
    number = {4},
    pages = {1797--1827},
    publisher = {Institute of Statistical Science, Academia Sinica},
    title = {High-dimensional linear regression for dependent data with applications to nowcasting},
    volume = {30},
    year = {2020}
}

@article{ing2019p,
    author = {Ching-Kang Ing},
    title = {Model selection for high-dimensional linear regression with dependent observations},
    volume = {48},
    journal = {The Annals of Statistics},
    number = {4},
    publisher = {Institute of Mathematical Statistics},
    pages = {1959--1980},
    year = {2020}
}

@article{kock2016, 
    title={Consistent and conservative model selection with the adaptive Lasso in stationary and nonstationary autoregressions}, 
    volume={32}, 
    number={1}, 
    journal={Econometric Theory}, 
    publisher={Cambridge University Press}, 
    author={Kock, Anders Bredahl}, 
    year={2016}, 
    pages={243--259}
}

@article{chan1988,
    author = "Chan, N. H. and Wei, C. Z.",
    journal = "The Annals of Statistics",
    month = "03",
    number = "1",
    pages = "367--401",
    publisher = "The Institute of Mathematical Statistics",
    title = "Limiting distributions of least squares estimates of unstable autoregressive processes",
    volume = "16",
    year = "1988"
}

@article{ling1998,
    author = "Ling, Shiqing and Li, W. K.",
    journal = "The Annals of Statistics",
    month = "02",
    number = "1",
    pages = "84--125",
    publisher = "The Institute of Mathematical Statistics",
    title = "Limiting distributions of maximum likelihood estimators for unstable autoregressive moving-average time series with general autoregressive heteroscedastic errors",
    volume = "26",
    year = "1998"
}

@article{bierens2001,
    author = {Herman J. Bierens},
    journal = {Econometric Theory},
    number = {5},
    pages = {962--983},
    publisher = {Cambridge University Press},
    title = {Complex unit roots and business cycles: Are they real?},
    volume = {17},
    year = {2001}
}

@article{Montgomery1998,
    author = {Alan L. Montgomery and Victor Zarnowitz and Ruey S. Tsay and George C. Tiao},
    title = {Forecasting the U.S. unemployment rate},
    journal = {Journal of the American Statistical Association},
    volume = {93},
    number = {442},
    pages = {478--493},
    year  = {1998},
    publisher = {Taylor & Francis}
}

@BOOK{billingsley1999,
  title={Convergence of Probability Measures},
  author={Billingsley, Patrick},
  year={1999},
  publisher={Wiley}
}

@article{helland1982,
  title={Central limit theorems for martingales with discrete or continuous time},
  author={Helland, Inge S},
  journal={Scandinavian Journal of Statistics},
  pages={79--94},
  year={1982}
}

@article{findley1993,
  title={Moment bounds for deriving time series {CLT}'s and model selection procedures},
  author={Findley, David F and Wei, Ching-Zong},
  journal={Statistica Sinica},
  pages={453--480},
  year={1993}
}

@article{LING2002109,
    title = "Stationarity and the existence of moments of a family of GARCH processes",
    journal = "Journal of Econometrics",
    volume = "106",
    number = "1",
    pages = "109--117",
    year = "2002",
    issn = "0304-4076",
    author = "Shiqing Ling and Michael McAleer"
}

@book{brillinger1974,
  title={Time Series: Data Analysis and Theory},
  author={Brillinger, D.R.},
  year={1975},
  publisher={Holt, Rinehart, and Winston, New York}
}

@article{Buhlmann2016,
    author = {Peter B\"{u}hlmann},
    title = {{Boosting for high-dimensional linear models}},
    volume = {34},
    journal = {The Annals of Statistics},
    number = {2},
    publisher = {Institute of Mathematical Statistics},
    pages = {559--583},
    year = {2006}
}

@article{wang2009,
    author = {Hansheng Wang},
    title = {Forward regression for ultra-high dimensional variable screening},
    journal = {Journal of the American Statistical Association},
    volume = {104},
    number = {488},
    pages = {1512--1524},
    year  = {2009}
}

@article{Fan2008,
    author = {Fan, Jianqing and Lv, Jinchi},
    title = {Sure independence screening for ultrahigh dimensional feature space},
    journal = {Journal of the Royal Statistical Society: Series B (Statistical Methodology)},
    volume = {70},
    number = {5},
    pages = {849--911},
    year = {2008}
}

@article{AlZoubi2008,
    title = {The long swings in the spot exchange rates and the complex unit roots hypothesis},
    journal = {Journal of International Financial Markets, Institutions and Money},
    volume = {18},
    number = {3},
    pages = {236--244},
    year = {2008},
    issn = {1042--4431},
    author = {Haitham A. Al-Zoubi}
}

@article{Faria2009,
    title = {Unemployment and entrepreneurship: A cyclical relation?},
    journal = {Economics Letters},
    volume = {105},
    number = {3},
    pages = {318--320},
    year = {2009},
    author = {João Ricardo Faria and Juan Carlos Cuestas and Luis A. Gil-Alana}
}

@article{Tsay1984,
    author = {Ruey S. Tsay},
    title = {Order selection in nonstationary autoregressive models},
    volume = {12},
    journal = {The Annals of Statistics},
    number = {4},
    publisher = {Institute of Mathematical Statistics},
    pages = {1425--1433},
    year = {1984}
}

@article{Zhao2006,
  author  = {Peng Zhao and Bin Yu},
  title   = {On model selection consistency of lasso},
  journal = {Journal of Machine Learning Research},
  year    = {2006},
  volume  = {7},
  number  = {90},
  pages   = {2541--2563}
}

@article{candes2007,
    author = {Emmanuel Candes and Terence Tao},
    title = {The Dantzig selector: Statistical estimation when p is much larger than n},
    volume = {35},
    journal = {The Annals of Statistics},
    number = {6},
    publisher = {Institute of Mathematical Statistics},
    pages = {2313--2351},
    year = {2007}
}

@article{Zheng2014,
    author = {Zheng, Zemin and Fan, Yingying and Lv, Jinchi},
    title = {High dimensional thresholded regression and shrinkage effect},
    journal = {Journal of the Royal Statistical Society: Series B (Statistical Methodology)},
    volume = {76},
    number = {3},
    pages = {627--649},
    year = {2014}
}

@article{Chudik2018,
    author = {Chudik, A. and Kapetanios, G. and Pesaran, M. Hashem},
    title = {A one covariate at a time, multiple testing approach to variable selection in high-dimensional linear regression models},
    journal = {Econometrica},
    volume = {86},
    number = {4},
    pages = {1479--1512},
    year = {2018}
}

@article{huang2022,
    author = {Hsueh-Han Huang and Ngai Hang Chan and Kun Chen and Ching-Kang Ing},
    title = {Consistent order selection for ARFIMA processes},
    journal = {The Annals of Statistics},
    year = {2022}
}

@article{Bollerslev1988,
    author = {Tim Bollerslev and Robert F. Engle and Jeffrey M. Wooldridge},
    journal = {Journal of Political Economy},
    number = {1},
    pages = {116--131},
    publisher = {University of Chicago Press},
    title = {A capital asset pricing model with time-varying covariances},
    volume = {96},
    year = {1988}
}

@ARTICLE{Tropp2004,  
    author={Tropp, J.A.},  
    journal={IEEE Transactions on Information Theory},   
    title={Greed is good: Algorithmic results for sparse approximation},   
    year={2004},  
    volume={50},  
    number={10}, 
    pages={2231--2242}
}

@article{Bickel2009,
    author = {Peter J. Bickel and Ya’acov Ritov and Alexandre B. Tsybakov},
    title = {{Simultaneous analysis of Lasso and Dantzig selector}},
    volume = {37},
    journal = {The Annals of Statistics},
    number = {4},
    pages = {1705--1732},
    year = {2009}
}

@article{Ing2016,
    author = {Ching-Kang Ing and Hai-Tang Chiou and Meihui Guo},
    title = {{Estimation of inverse autocovariance matrices for long memory processes}},
    volume = {22},
    journal = {Bernoulli},
    number = {3},
    publisher = {Bernoulli Society for Mathematical Statistics and Probability},
    pages = {1301--1330},
    year = {2016}
}

@article{Baxter1962,
    author = {Glen Baxter},
    journal = {Mathematica Scandinavica},
    pages = {137--144},
    title = {An asymptotic result for the finite predictor},
    volume = {10},
    year = {1962}
}

@article{lai1982,
  title={Asymptotic properties of projections with applications to stochastic regression problems},
  author={Lai, T. L. and Wei, C. Z.},
  journal={Journal of Multivariate Analysis},
  volume={12},
  number={3},
  pages={346--370},
  year={1982}
}

@article{Zhang2010,
    author = {Cun-Hui Zhang},
    title = {{Nearly unbiased variable selection under minimax concave penalty}},
    volume = {38},
    journal = {The Annals of Statistics},
    number = {2},
    publisher = {Institute of Mathematical Statistics},
    pages = {894--942},
    year = {2010}
}

@article{Wei1987,
    author = {C. Z. Wei},
    title = {{Adaptive prediction by least squares predictors in stochastic regression models with applications to time series}},
    volume = {15},
    journal = {The Annals of Statistics},
    number = {4},
    publisher = {Institute of Mathematical Statistics},
    pages = {1667--1682},
    year = {1987}
}

@article{Wei1992,
    author = {C. Z. Wei},
    title = {{On predictive least squares principles}},
    volume = {20},
    journal = {The Annals of Statistics},
    number = {1},
    publisher = {Institute of Mathematical Statistics},
    pages = {1--42},
    year = {1992}
}

@article{McC2016,
    author = {Michael W. McCracken and Serena Ng},
    title = {FRED-MD: A monthly database for macroeconomic research},
    journal = {Journal of Business and Economic Statistics},
    volume = {34},
    number = {4},
    pages = {574--589},
    year  = {2016},
    publisher = {Taylor & Francis}
}

@article{AlZoubi18,
    author = {Al-Zoubi, Haitham A. and O'Sullivan, Jennifer A. and Alwathnani, Abdulaziz M.},
    title = {Business cycles, financial cycles and capital structure},
    journal = {Annals of Finance},
    volume = {14},
    number = {1},
    pages = {105--123},
    year = {2018}
}

@article{Maddanu22,
    author = {Maddanu, Federico and Proietti, Tommaso},
    title = {Modelling persistent cycles in solar activity},
    journal = {Solar Physics},
    volume = {297},
    number = {13},
    year = {2022}
}

@article{Proietti22,
    title = {Modelling cycles in climate series: The fractional sinusoidal waveform process},
    journal = {Journal of Econometrics},
    volume = {239},
    number = {1},
    pages = {105299},
    year = {2024},
    author = {Tommaso Proietti and Federico Maddanu}
}

@article{Gilalana09,
    author = {Luis A. Gil-Alana},
    title = {Time series modeling of sunspot numbers using long-range cyclical dependence},
    journal = {Solar Physics},
    volume = {257},
    pages = {371--381},
    year = {2009}
}

@article{Gilalana14,
    author = {Luis A. Gil-Alana and Rangan Gupta},
    title = {Persistence and cycles in historical oil price data},
    journal = {Energy Economics},
    volume = {45},
    pages = {511--516},
    year = {2014}
}

@article{Castro22,
    author = {del Barrio Castro, Tomás and Cubadda, Gianluca and Osborn, Denise R.},
    title = {On cointegration for processes integrated at different frequencies},
    journal = {Journal of Time Series Analysis},
    volume = {43},
    number = {3},
    pages = {412--435},
    year = {2022}
}

@article{Lee2022,
    title = {On LASSO for predictive regression},
    journal = {Journal of Econometrics},
    volume = {229},
    number = {2},
    pages = {322--349},
    year = {2022},
    author = {Ji Hyung Lee and Zhentao Shi and Zhan Gao}
}

@article{Georgiev2018,
    title = {Testing for parameter instability in predictive regression models},
    journal = {Journal of Econometrics},
    volume = {204},
    number = {1},
    pages = {101--118},
    year = {2018},
    author = {Iliyan Georgiev and David I. Harvey and Stephen J. Leybourne and A.M. Robert Taylor}
}

@article{Welch2007,
    author = {Welch, Ivo and Goyal, Amit},
    title = {A comprehensive look at the empirical performance of equity premium prediction},
    journal = {The Review of Financial Studies},
    volume = {21},
    number = {4},
    pages = {1455--1508},
    year = {2007}
}

@article{Diebold2002,
    author = {Francis X Diebold and Robert S Mariano},
    title = {Comparing predictive accuracy},
    journal = {Journal of Business and Economic Statistics},
    volume = {20},
    number = {1},
    pages = {134--144},
    year = {2002}
}

@article{Chan2014,
    author = {Ngai Hang Chan and Chun Yip Yau and Rong-Mao Zhang},
    title = {Group LASSO for Structural Break Time Series},
    journal = {Journal of the American Statistical Association},
    volume = {109},
    number = {506},
    pages = {590--599},
    year = {2014}
}

@article{GJR-GARCH,
    author = {Lawrence R. Glosten and Ravi Jagannathan and David E. Runkle},
    journal = {The Journal of Finance},
    number = {5},
    pages = {1779--1801},
    title = {On the relation between the expected value and the volatility of the nominal excess return on stocks},
    volume = {48},
    year = {1993}
}

\end{document}